\documentclass[12pt]{article}
\usepackage [utf8]{inputenc}
\usepackage [english]{babel}
\usepackage{graphicx}
\usepackage {amsmath}
\usepackage {amsfonts}
\usepackage {amssymb}
\usepackage {mathrsfs}
\usepackage {longtable}
\usepackage{multicol}
\usepackage{dcolumn}

\title{Magnetorotational and convective instabilities in a thin layer of electrically conductive nanofluid under an external helical magnetic field}
\author{\textbf{M. I. Kopp}$^{1}$, \; \textbf{V. V. Yanovsky}$^{1,2}$ }

\begin{document}

\maketitle
$^{1}$\textit{Institute for Single Crystals, NAS of Ukraine, Nauky Ave. 60, Kharkiv 61072, Ukraine}

$^{2}$\textit{V. N. Karazin Kharkiv National University 4 Svobody Sq., Kharkiv 61022, Ukraine}

\bigskip

\begin{abstract}
This review summarizes recent advances in the theoretical analysis of the stability of magnetized flows in a non-uniformly rotating layer of electrically conductive nanofluid, incorporating the effects of Brownian diffusion and thermophoresis. In the absence of temperature gradients, different forms of magnetorotational instability (MRI): standard (SMRI), azimuthal (AMRI), and helical (HMRI) are investigated for nanofluid layers subjected to axial, azimuthal, and helical magnetic fields. The corresponding growth rates and instability regions are analyzed in relation to the rotation profile (quantified by the Rossby number $\textrm{Ro}$) and the radial wave number $k$. When temperature gradients and nanoparticle concentration effects are present, stationary convective modes in both axial and helical magnetic fields are examined under conditions of non-uniform rotation. Analytical expressions for the critical Rayleigh number $\textrm{Ra}_{st}$ are derived, and neutral stability curves are constructed as functions of the angular velocity profile, the azimuthal magnetic field inhomogeneity (magnetic Rossby number $\textrm{Rb}$), and the wave number $k$. The study identifies and discusses key mechanisms responsible for the stabilization or destabilization of stationary convection in axial and spiral magnetic field configurations, highlighting the role of nanoparticle-driven effects in modifying classical magnetoconvective behavior.
\end{abstract}

\section{Introduction}
The concept of nanofluid, introduced by Choi and Eastman in 1995 \cite{1s}, refers to a suspension of nanoparticles (typically less than 100 nm in diameter) in a conventional base fluid such as water, ethylene glycol, or oil. Nanofluids have attracted significant attention due to their enhanced thermal and electrical properties, which make them promising candidates for advanced heat and mass transfer applications, especially under the influence of external fields. The incorporation of nanoparticles into a base fluid significantly alters its rheological behavior, thermal conductivity, and electrical conductivity, offering new opportunities for controlling instabilities in fluid systems.
The dispersed nanoparticles, although larger than individual atoms (ranging from 1 to 100 nm), are still small enough for Brownian motion to play a significant role. A major theoretical contribution to the understanding of transport phenomena in nanofluids was made by Buongiorno \cite{2s}, who systematically analyzed various mechanisms influencing nanoparticle movement, including Brownian diffusion, the Magnus effect, particle inertia, thermophoresis, fluid drainage, and diffusiophoresis. His findings indicated that, in laminar regimes without turbulent eddies, Brownian diffusion and thermophoresis are the predominant mechanisms. As a result, investigating how these two effects influence forced convection in nanofluid-based heat transfer systems remains a critical area of research.

There has been growing interest in understanding heat transfer mechanisms in electrically conductive nanofluids subjected to magnetic fields, particularly when the effects of Brownian diffusion and thermophoresis are taken into account \cite{3s, 4s}. Convective instabilities in nanofluid layers under vertical magnetic fields have been examined for various boundary conditions free-free, rigid-rigid, and rigid-free in \cite{5s}. That study demonstrated that increasing the strength of the magnetic field enhances the stability of the nanofluid, whereas higher nanoparticle concentrations promote an earlier onset of convection. The conditions for the onset of convection in pure (single-component) fluids under gravitational fields, uniform rotation, and external magnetic fields have been extensively analyzed in prior works \cite{6s}-\cite{8s}. This background motivates further investigation into convective instability in a rotating nanofluid layer influenced by a magnetic field. The first study addressing this combined scenario appeared in \cite{9s}, where the authors explored how rotation and magnetization jointly affect the onset of convection in a horizontal layer of electrically conductive nanofluid, while also considering the influences of Brownian motion and thermophoresis. It was shown in \cite{9s} that the critical Rayleigh number $Ra_c$ for nanofluids is lower than that for conventional fluids under identical values of the Chandrasekhar number $Q$ (quantifying magnetic/Lorentz force effects) and the Taylor number $Ta$ (representing rotational/Coriolis force effects). Furthermore, the study found that increasing nanoparticle concentration has a destabilizing impact on convective onset, while fluid rotation contributes to stabilization.

The convective instability of a nonuniformly rotating layer of an electrically conducting fluid in the presence of a constant vertical magnetic field ($\mathbf{H}_0 = \text{const}$) was first investigated in \cite{10s,11s}, and later extended to the case of a helical magnetic field in \cite{12s}. The results presented in \cite{12s} demonstrate that, in the absence of thermal forcing ($\text{Ra} = 0$), i.e., when there is no imposed temperature gradient, the criteria for the onset of convective instability reduce to the classical conditions for the development of the standard magnetorotational instability (MRI) and the helical MRI in a dissipative, electrically conducting medium. In these studies, the nonuniform rotation of the conducting medium was modeled using the Couette flow configuration.
The Couette flow, particularly its rotational configuration between coaxial cylinders (Taylor-Couette flow), is one of the fundamental models in fluid dynamics, widely used to study linear and nonlinear stability and transition to turbulence \cite{13s}. This setup allows precise control of system parameters and the observation of complex flow regimes, including Taylor vortices, quasiperiodic oscillations, and chaotic behavior \cite{14s, 15s}. Moreover, Taylor-Couette flow is extensively used as a model system for studying magnetorotational instability (MRI), which plays a crucial role in astrophysical contexts, particularly in accretion disk dynamics \cite{16s}. 

The fundamental significance of the accretion process in astrophysics has driven extensive efforts to reproduce and study analogous phenomena under laboratory conditions. As rotating electrically conducting media, liquid metals such as sodium and gallium \cite{17s}, as well as low-temperature plasma \cite{18s}, have been employed. In particular, liquid metal flows subjected to helical magnetic fields are actively investigated within the framework of the PROMISE experiment (Rossendorf, Germany) \cite{19s}. 

To the best of our knowledge, there have been no experimental studies on the interaction between helical magnetic fields and nanofluid flows. Considering that the thermophysical properties of nanofluids can be tuned experimentally by synthesizing nanoparticles from a variety of materials \cite{20s}, we propose utilizing electrically conductive nanofluids (including hybrid formulations) in laboratory experiments on hydromagnetic instabilities under the influence of helical magnetic fields \cite{20sa}. We believe that the introduction of helical magnetic field concepts into such systems could contribute to the development of novel nanomaterials and nanotechnologies.
In light of the importance of understanding the influence of helical magnetic fields on nanofluid behavior, the present study explores new forms of hydromagnetic instabilities in a rotating layer of electrically conductive nanofluid, incorporating the effects of Brownian diffusion and thermophoresis. In addition to the general theoretical interest, the results of this study suggest that electrically conductive nanofluids may serve as a promising medium for laboratory modeling of various types of magnetorotational instability (MRI), including standard (SMRI), azimuthal (AMRI), and helical (HMRI) forms. Compared to conventional liquid metals, nanofluids offer greater tunability of key physical parameters such as the magnetic Prandtl number, modified diffusion coefficients, and nanoparticle Rayleigh numbers allowing access to a broader range of dynamical regimes. This enhanced flexibility makes it possible to reproduce experimental conditions that more closely approximate astrophysical and geophysical environments, thereby opening new avenues for analog modeling of magnetohydrodynamic phenomena relevant to disks, planetary interiors, and plasma systems.

\begin{figure}
  \centering
	\includegraphics[width=10 cm, height=8 cm]{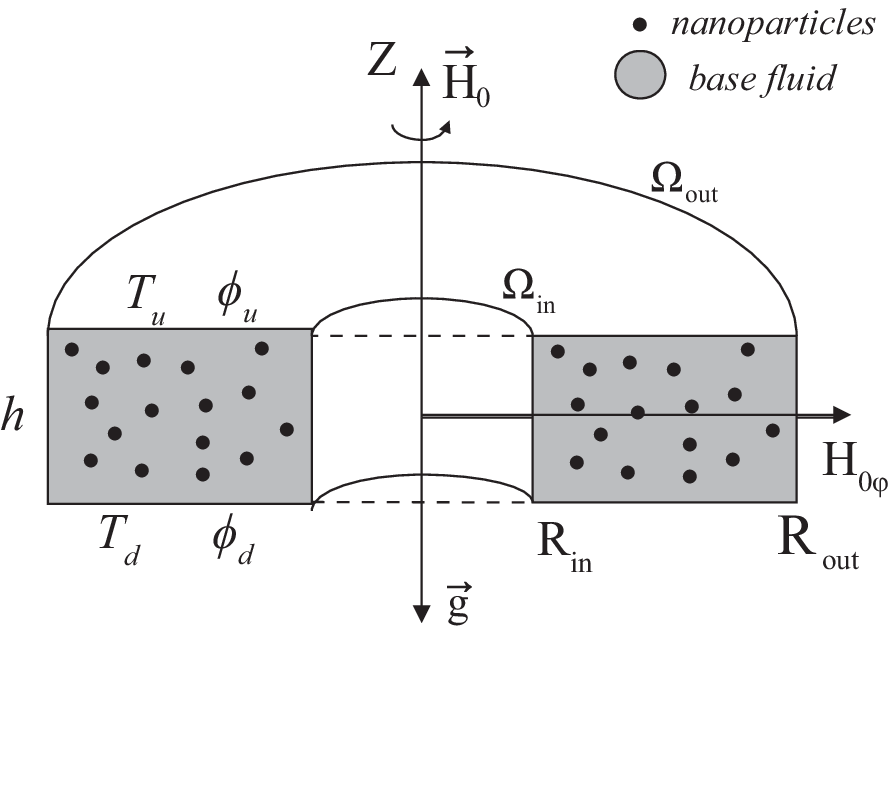} \\
\caption{\small Schematic of a rotating nanofluid layer confined between two coaxial cylinders in a helical magnetic field. The lower and upper surfaces are maintained at temperatures \( T_d > T_u \) and nanoparticle volume fractions \( \phi_d < \phi_u \), respectively.}\label{fg1}
\end{figure}

  \section{Problem formulation and basic equations} 

Let us consider a layer of incompressible viscous electrically conducting nanofluid of thickness \( h \), confined between two rotating cylinders with inner radius \( R_{\textrm{in}} \) and outer radius \( R_{\textrm{out}} \), such that \( h \ll (R_{\textrm{out}} - R_{\textrm{in}}) \).
The nanofluid is bounded by two parallel planes at \( z = 0 \) and \( z = h \), where the temperature and the nanoparticle volume fraction are maintained constant: \( T = T_d, \; \phi = \phi_d \) at \( z = 0 \), and \( T = T_u, \; \phi = \phi_u \) at \( z = h \), with \( T_d > T_u \) and \( \phi_u > \phi_d \) (see Fig.~\ref{fg1}). The fluid is assumed to be under the influence of a constant gravitational field \( \mathbf{g} \), directed vertically downward along the \( OZ \)-axis: \( \mathbf{g} = (0, 0, -g) \). The electrically conducting fluid rotates with an angular velocity \( \boldsymbol{\Omega} \), directed vertically upward along the \( OZ \)-axis. This rotation induces a steady azimuthal flow:
\[\mathbf{V}_0 = \mathbf{e}_\phi \, \Omega(R) R, \]
where \( \Omega(R) \) is the angular velocity with an arbitrary dependence on the radial coordinate \( R \).
In addition, the nanofluid is subjected to a helical magnetic field \( \mathbf{H}_0 \), which is represented as the sum of an inhomogeneous azimuthal component \( H_{0\phi}(R) \) and a uniform axial component \( H_{0z} \):
\begin{equation} \label{eq1a}
\mathbf{H}_0 = H_{0\phi}(R) \, \mathbf{e}_\phi + H_{0z} \, \mathbf{e}_z, \quad H_{0z} = \text{const}.
\end{equation}
A topological characteristic of the magnetic field lines of the helical field~\eqref{eq1a} is the pseudoscalar quantity known as the magnetic helicity (see, e.g.,~\cite{21s}):
\begin{equation} \label{eq2a}
\mathbf{H}_0 \cdot \operatorname{rot} \mathbf{H}_0 = \frac{H_{0z}}{R} \frac{\partial}{\partial R}(R H_{0\phi}).
\end{equation}
It is evident that for analyzing this type of flow, it is convenient to employ a cylindrical coordinate system \( (R, \phi, z) \), which is motivated by the potential practical applications of the theory developed herein.
The azimuthal magnetic field is assumed to be radially inhomogeneous. To quantify this inhomogeneity, we introduce the magnetic Rossby number:
\[\textrm{Rb} = \frac{R}{2 H_{0\phi} R^{-1}} \frac{d}{dR} \left( H_{0\phi} R^{-1} \right),\]
as defined in~\cite{22s}. For instance, if the azimuthal component of the magnetic field is given by \( H_{0\phi}(R) = 2I/R \), which corresponds to an external axial current \( I \) isolated from the fluid, then \( H_{0\phi} \sim R^{-1} \), and the magnetic Rossby number becomes \( \textrm{Rb} = -1 \). In this case, the helicity of the magnetic field vanishes: \( \mathbf{H}_0 \cdot \operatorname{rot} \mathbf{H}_0 = 0 \).
When \( H_{0\phi}(R) \sim R \), the magnetic Rossby number is zero, \( \textrm{Rb} = 0 \), but the magnetic helicity does not vanish:
\[\mathbf{H}_0 \cdot \operatorname{rot} \mathbf{H}_0 = \frac{2 H_{0z} H_{0\phi}}{R}.
\]
For a quadratic radial dependence of the azimuthal magnetic field, \( H_{0\phi}(R) \sim R^2 \), the magnetic Rossby number takes the value \( \textrm{Rb} = 1/2 \), and the helicity becomes:
\[
\mathbf{H}_0 \cdot \operatorname{rot} \mathbf{H}_0 = \frac{3 H_{0z} H_{0\phi}}{R}.
\]
It is worth noting that even in the case of a uniform azimuthal magnetic field, \( H_{0\phi} = \text{const} \), the helicity remains non-zero:
\[
\mathbf{H}_0 \cdot \operatorname{rot} \mathbf{H}_0 = \frac{H_{0z} H_{0\phi}}{R},
\]
while the magnetic Rossby number is \( \textrm{Rb} = -1/2 \).
We consider the following assumptions that underlie the mathematical formulation of the physical model:
\begin{enumerate}
  \item All thermophysical properties are assumed to be constant, except for the fluid density in the buoyancy term, which varies according to the Boussinesq approximation.
  \item The fluid and nanoparticles are in thermal equilibrium, allowing the use of a single-temperature model to describe heat transfer.
  \item The nanofluid is incompressible, viscous, electrically conducting, and exhibits laminar flow; the suspended nanoparticles are spherical, non-magnetic, and uniformly dispersed.
  \item Both boundary walls are assumed to be impermeable and perfectly thermally conducting.
  \item Radiative heat transfer between the boundary walls is considered negligible compared to other modes of heat transfer.
\end{enumerate}
To describe the convective processes, we employ the Boussinesq-Oberbeck approximation for an incompressible, electrically conducting nanofluid~\cite{5s}. The governing hydrodynamic equations are formulated in a cylindrical coordinate system:
\begin{equation} \label{eq1} \rho_0\left(\frac{\partial V_{R} }{\partial t} +({\bf V}\nabla)V_R-\frac{V_\phi^2}{R}\right) -\frac{\mu_e}{4\pi} \left(({\bf H}\nabla)H_R-\frac{H_\phi^2}{R}\right) =-\frac{\partial}{\partial R}\left(P+\frac{\mu_e H^2}{8\pi}\right) +$$
$$+\mu \left(\nabla^2 V_{R} -\frac{2}{R^2}\frac{\partial V_\phi}{\partial \phi} - \frac{V_R}{R^2} \right)\end{equation}
\begin{equation} \label{eq2} \rho_0 \left(\frac{\partial V_{\phi } }{\partial t} + ({\bf V}\nabla)V_\phi+\frac{V_\phi V_R}{R}\right)-\frac{\mu_e}{4\pi}\left( ({\bf H}\nabla)H_\phi+ \frac{H_\phi H_R}{R}\right)=-\frac{1}{R} \frac{\partial}{\partial \phi }\left(P+\frac{\mu_eH^2}{8\pi}\right) +$$
$$+\mu \left(\nabla^2 V_{\phi} +\frac{2}{R^2}\frac{\partial V_R}{\partial \phi} - \frac{V_\phi}{R^2} \right)  \end{equation} 
\begin{equation} \label{eq3} \rho_0\left(\frac{\partial V_{z}}{\partial t}+({\bf V}\nabla)V_z\right) -\frac{\mu_e}{4\pi}({\bf H}\nabla)H_z=-\frac{\partial }{\partial z}\left(P+\frac{\mu_e H^2}{8\pi}\right)+$$
$$+\mu \nabla^2 V_{z}-[\phi\rho_{p}+(1-\phi)\rho_{0}(1-\beta(T-T_u))]g  \end{equation}
\begin{equation} \label{eq4} 
(\rho c)_{f}\left(\frac{\partial T}{\partial t}+({\bf V}\nabla)T\right)=k_f\nabla^2 T+(\rho c)_{p}\left(D_B \nabla\phi\cdot\nabla T+D_T\frac{\nabla T\cdot\nabla T}{T_u}\right)
\end{equation}
\begin{equation} \label{eq5}
\frac{\partial \phi}{\partial t}+({\bf V}\nabla)\phi=D_B\nabla^2\phi+\frac{D_T}{T_u}\nabla^2 T
\end{equation}
\begin{equation} \label{eq6} \frac{\partial H_{R} }{\partial t}+({\bf V}\nabla)H_R-({\bf H}\nabla)V_R=\eta \left(\nabla^2 H_{R}-\frac{2}{R^2}\frac{\partial H_\phi}{\partial \phi}-\frac{H_R}{R^2} \right)  \end{equation} 
\begin{equation} \label{eq7} \frac{\partial H_{\phi } }{\partial t}+({\bf V}\nabla)H_\phi-({\bf H}\nabla)V_\phi +\frac{1}{R}\left(V_\phi H_R-V_RH_\phi\right)-=$$
$$=\eta\left(\nabla^2 H_{\phi} +\frac{2}{R^2}\frac{\partial H_R}{\partial \phi} - \frac{H_\phi}{R^2} \right)   \end{equation} 
\begin{equation} \label{eq8} \frac{\partial H_{z} }{\partial t} +({\bf V}\nabla)H_z-({\bf H}\nabla)V_z  =\eta \nabla^2 H_{z}  \end{equation} 
\begin{equation} \label{eq9} \frac{\partial V_{R} }{\partial R} +\frac{V_{R} }{R} +\frac{1}{R} \frac{\partial V_{\phi } }{\partial \phi } +\frac{\partial V_{z} }{\partial z} =0, $$
$$ \frac{\partial H_{R} }{\partial R} +\frac{H_{R} }{R} +\frac{1}{R} \frac{\partial H_{\phi } }{\partial \phi } +\frac{\partial H_{z} }{\partial z} =0,  \end{equation} 
Here, the scalar product \( (\mathbf{A} \nabla) \) and the Laplacian \( \Delta \equiv \nabla^2 \) in cylindrical coordinates are given by:
\[
(\mathbf{A} \nabla) = A_R \frac{\partial}{\partial R} + \frac{A_\phi}{R} \frac{\partial}{\partial \phi} + A_z \frac{\partial}{\partial z}, \quad
\Delta = \frac{\partial^2}{\partial R^2} + \frac{1}{R} \frac{\partial}{\partial R} + \frac{1}{R^2} \frac{\partial^2}{\partial \phi^2} + \frac{\partial^2}{\partial z^2}.
\]
Here, \( \rho_0 = \phi \rho_p + (1 - \phi) \rho_f \) is the nanofluid density evaluated at the reference temperature \( T_u \), where \( \rho_p \) and \( \rho_f \) are the densities of the nanoparticles and the base fluid at temperature \( T_u \), respectively, and \( \phi \) is the nanoparticle volume fraction. The parameter \( \beta \) denotes the thermal expansion coefficient. The terms \( (\rho c)_f \) and \( (\rho c)_p \) represent the effective heat capacities of the base fluid and the nanoparticles, respectively. The coefficients \( D_B \) and \( D_T \) are the Brownian diffusion and thermophoretic diffusion coefficients, respectively, both positive and given by:
\[
D_B = \frac{k_B T}{3 \pi \mu d_p}, \quad
D_T = \left( \frac{\mu}{\rho_f} \right) \left( \frac{0.26\,k_f}{2k_f + k_p} \right),
\]
where \( d_p \) is the diameter of the nanoparticles, \( k_B \) is the Boltzmann constant, \( k_f \) and \( k_p \) are the thermal conductivities of the base fluid and the nanoparticles, respectively, and \( \mu \) is the dynamic viscosity of the base fluid.
The magnetic permeability \( \mu_e \), magnetic diffusivity \( \eta \), and electrical conductivity \( \sigma \) of the nanofluid are defined as:
\[
\mu_e = \phi \mu_{ep} + (1 - \phi) \mu_{ef}, \quad
\eta = \frac{1}{4 \pi \mu_e \sigma} = \phi \eta_p + (1 - \phi) \eta_f, \quad
\sigma = \phi \sigma_p + (1 - \phi) \sigma_f,
\]
where \( \mu_{ep} \) and \( \mu_{ef} \) are the magnetic permeabilities, \( \eta_p \) and \( \eta_f \) are the magnetic diffusivities, and \( \sigma_p \) and \( \sigma_f \) are the electrical conductivities of the nanoparticles and the base fluid, respectively.
We represent all physical quantities in equations~(\ref{eq1})--(\ref{eq9}) as the sum of a steady-state component and a perturbation:
\begin{equation} \label{eq10}
\begin{pmatrix} V_R \\ V_\phi \\ V_z \end{pmatrix} =
\begin{pmatrix} 0 \\ \Omega(R)R \\ 0 \end{pmatrix} +
\begin{pmatrix} u_R \\ u_\phi \\ u_z \end{pmatrix}, \;
\begin{pmatrix} H_R \\ H_\phi \\ H_z \end{pmatrix} =
\begin{pmatrix} 0 \\ H_{0\phi} \\ H_{0z} \end{pmatrix} +
\begin{pmatrix} b_R \\ b_\phi \\ b_z \end{pmatrix}, \;
P = P_b + p, \; T = T_b + T', \; \phi = \phi_b + \phi'.
\end{equation}
The equilibrium of the base (stationary) flow is governed by the radial force balance:
\begin{equation} \label{eq11}
\Omega^2 R = \frac{1}{\rho_0} \frac{dP_b}{dR} + \frac{\mu_e H_{0\phi}}{4\pi \rho_0 R} \frac{d}{dR}(R H_{0\phi}).
\end{equation}
In the vertical direction, the hydrostatic balance for the stationary state is described by:
\begin{equation} \label{eq12}
\frac{dP_b}{dz} = -g \left[ \phi_b(\rho_p - \rho_0) + \rho_0 - \rho_0 \beta (T_b - T_u) \right].
\end{equation}
The steady-state profiles of temperature \( T_b = T_b(z) \) and nanoparticle volume fraction \( \phi_b = \phi_b(z) \) are determined from the following equations:
\begin{equation} \label{eq13}
0 = \frac{k_f}{(\rho c)_f} \frac{d^2 T_b}{dz^2}
+ \frac{(\rho c)_p}{(\rho c)_f} \left( D_B \frac{d\phi_b}{dz} \frac{dT_b}{dz}
+ \frac{D_T}{T_u} \left( \frac{dT_b}{dz} \right)^2 \right), \qquad
0 = D_B \frac{d^2 \phi_b}{dz^2} + \frac{D_T}{T_u} \frac{d^2 T_b}{dz^2}.
\end{equation}
Taking into account the boundary conditions, we find linear dependence on \( z \) for \( T_b(z) \) and \( \phi_b(z) \):
\[
T_b(z) = T_d - \frac{T_d - T_u}{h} z, \quad
\phi_b(z) = \phi_d - \frac{\phi_d - \phi_u}{h} z.
\]
We now focus on the stability of small perturbations in physical quantities
\( (\mathbf{u} = (u_R, u_\phi, u_z), \ \mathbf{b} = (b_R, b_\phi, b_z),\ p,\ T',\ \phi') \)
superimposed on the base state. Substituting expressions~(\ref{eq10}) into the governing equations~(\ref{eq1})--(\ref{eq9}), we obtain the linearized evolution equations for small perturbations:
\begin{equation} \label{eq14} \frac{\partial u_{R} }{\partial t} +\Omega \frac{\partial u_{R} }{\partial \varphi } -2\Omega u_{\varphi } -\frac{\mu_e}{4\pi \rho_{0}}\left(\frac{H_{0\varphi}}{R}\frac{\partial b_R}{\partial \varphi}-\frac{2H_{0\varphi}b_\varphi}{R}+ H_{0z} \frac{\partial b_{R} }{\partial z}\right) =$$
$$=-\frac{1}{\rho_{0}} \frac{\partial \widetilde p}{\partial R} +\nu \left(\nabla^2 u_{R} -\frac{2}{R^2}\frac{\partial u_\varphi}{\partial \varphi} - \frac{u_R}{R^2} \right),\; \nu=\mu/\rho_0   \end{equation} 
\begin{equation} \label{eq15} \frac{\partial u_{\varphi } }{\partial t} +\Omega \frac{\partial u_{\varphi } }{\partial \phi } +2\Omega(1+\textrm{Ro})u_{R} -\frac{1}{4\pi \rho _{0} }\left(\frac{H_{0\varphi}}{R}\frac{\partial b_R}{\partial \phi}+\frac{2H_{0\varphi}}{R}\left(1+\textrm{Rb}\right)+H_{0z} \frac{\partial b_{\varphi}}{\partial z}\right) =$$
$$=-\frac{1}{\rho_{0} R} \frac{\partial \widetilde p}{\partial\varphi} +\nu\left(\nabla^2 u_{\varphi}+\frac{2}{R^2}\frac{\partial u_R}{\partial \varphi} - \frac{u_\varphi}{R^2} \right)  \end{equation} 
\begin{equation} \label{eq16} \frac{\partial u_{z} }{\partial t} +\Omega \frac{\partial u_{z} }{\partial \varphi } -\frac{\mu_e}{4\pi\rho_{0} }\left(\frac{H_{0\varphi}}{R}\frac{\partial b_z}{\partial \varphi}+ H_{0z} \frac{\partial b_{z} }{\partial z}\right) =-\frac{1}{\rho_{0}} \frac{\partial \widetilde p}{\partial z}+\nu \nabla^2 u_{z}-$$
$$-\frac{g}{\rho_0}[\phi^{'}(\rho_p-\rho_0)-\rho_0\beta T^{'}] \end{equation} 
\begin{equation} \label{eq17} \frac{\partial T^{'} }{\partial t} +\Omega \frac{\partial T^{'}}{\partial \varphi }-u_{z}\left(\frac{T_d-T_u}{h}\right)=\chi_f \nabla^2 T^{'}-$$
$$-\frac{(\rho c)_p}{(\rho c)_f}\left[D_B\left(\frac{T_d-T_u}{h}\right)\frac{d \phi^{'}}{dz}+D_B\left(\frac{\phi_d-\phi_u}{h}\right)\frac{d T^{'}}{dz}+\frac{2D_T}{T_u}\left(\frac{T_d-T_u}{h}\right)\frac{d T^{'}}{dz}\right],\;\chi_f=\frac{k_f}{(\rho c)_f} \end{equation} 	
\begin{equation} \label{eq18} \frac{\partial \phi^{'} }{\partial t} +\Omega \frac{\partial\phi^{'}}{\partial\varphi }+u_{z}\left(\frac{\phi_u-\phi_d}{h}\right)=D_B\nabla^2\phi^{'}+\frac{D_T}{T_u}\nabla^2T^{'} \end{equation} 
\begin{equation} \label{eq19} \frac{\partial b_{R} }{\partial t} +\Omega \frac{\partial b_{R} }{\partial \varphi }-\frac{H_{0\varphi}}{R}\frac{\partial u_R}{\partial \varphi}-H_{0z} \frac{\partial u_{R} }{\partial z} =\eta \left(\nabla^2 b_{R} -\frac{2}{R^2}\frac{\partial b_\varphi}{\partial \varphi}-\frac{b_R}{R^2} \right)  \end{equation} 
\begin{equation} \label{eq20} \frac{\partial b_{\varphi } }{\partial t} +\Omega \frac{\partial b_{\varphi } }{\partial \varphi }-\frac{H_{0\varphi}}{R}\frac{\partial u_\varphi}{\partial \varphi}-H_{0z} \frac{\partial u_{\varphi } }{\partial z}-2\Omega \textrm{Ro} b_{R} + \frac{2H_{0\varphi}}{R}\textrm{Rb} u_R=$$
$$=\eta\left(\nabla^2 b_{\varphi}+\frac{2}{R^2}\frac{\partial b_R}{\partial \varphi} - \frac{b_\phi}{R^2} \right)   \end{equation} 
\begin{equation} \label{eq21} \frac{\partial b_{z}}{\partial t} +\Omega \frac{\partial b_{z} }{\partial\varphi}-\frac{H_{0\varphi}}{R}\frac{\partial u_z}{\partial \varphi}-H_{0z}\frac{\partial u_{z} }{\partial z}=\eta \nabla^2 b_{z}  \end{equation}	
\begin{equation} \label{eq22} \frac{\partial u_{R} }{\partial R} +\frac{u_{R} }{R} +\frac{1}{R} \frac{\partial u_{\varphi } }{\partial \varphi } +\frac{\partial u_{z} }{\partial z} =0, $$
$$ \frac{\partial b_{R} }{\partial R} +\frac{b_{R} }{R} +\frac{1}{R} \frac{\partial b_{\varphi } }{\partial \varphi } +\frac{\partial b_{z} }{\partial z} =0, 
\end{equation} 		
Here, \( \widetilde{p} = p + \frac{1}{4\pi} (\mathbf{H}_0 \cdot \mathbf{b}) \) is the total perturbed pressure, and
\[
\textrm{Ro} = \frac{R}{2\Omega} \frac{\partial \Omega}{\partial R}
\]
is the hydrodynamic Rossby number, which characterizes the radial inhomogeneity of the rotation rate of the medium. 
Note that for solid-body rotation, the Rossby number is zero, \( \textrm{Ro} = 0 \). In the case of Keplerian rotation, where \( \Omega(R) \sim R^{-3/2} \), the Rossby number is \( \textrm{Ro} = -\frac{3}{4} \). For the Rayleigh profile of angular velocity \( \Omega(R) \sim R^{-2} \), the Rossby number is \( \textrm{Ro} = -1 \).

\section{Local WKB approximation and dispersion relation}

The system of equations (\ref{eq14})-(\ref{eq22}) will be used to analyze the stability of small axisymmetric perturbations. Since the characteristic scale of inhomogeneity in the horizontal plane is much larger than in the vertical direction, i.e., \( L_{R} \gg L_{h} \), we can apply the local WKB (Wentzel-Kramers-Brillouin) method to perturbations that depend on the radial coordinate \( R \).
To this end, all physical quantities are expanded in a Taylor series in the vicinity of a fixed point \( R_0 \), retaining only the zeroth-order terms with respect to the local coordinate \( \widetilde{R} = R - R_0 \). As a result, we obtain a system of differential equations with constant coefficients, taking into account the following relations:
$$\Omega_0=\Omega(R_0),\quad \nabla^2 \rightarrow \widehat{D}^2+\frac{\partial^2}{\partial\widetilde{R}^2}+\frac{1}{R_0}\frac{\partial}{\partial \widetilde{R}}, \quad \widehat D \equiv \frac{\partial}{\partial z} ,  $$
$$ \left(\nabla^2 \begin{pmatrix} \bf u \\ \bf b \end{pmatrix}\right)_{R}= \nabla^2 \begin{pmatrix}  u_R \\ b_R \end{pmatrix}-\frac{1}{R_0^2}\begin{pmatrix}  u_R \\ b_R \end{pmatrix}, $$
$$ \left(\nabla^2 \begin{pmatrix} \bf u \\ \bf b \end{pmatrix}\right)_{\phi}= \nabla^2 \begin{pmatrix}  u_\phi \\ b_\phi \end{pmatrix}-\frac{1}{R_0^2}\begin{pmatrix}  u_\phi \\ b_\phi \end{pmatrix} . $$
All perturbations in the system of equations (\ref{eq14})-(\ref{eq22}) are represented as plane waves:
\begin{equation} \label{eq23}
\begin{pmatrix} \mathbf{u} \\ \mathbf{b} \\ T^{'} \\ \phi^{'} \\ \widetilde{p} \end{pmatrix}
=
\begin{pmatrix} \mathbf{U}(z) \\ \mathbf{B}(z) \\ \Theta(z) \\ \Phi(z) \\ \widetilde{P}(z) \end{pmatrix}
\exp\left(\gamma t + ik \widetilde{R}\right)
\end{equation}
After substituting expression (\ref{eq23}) into the system of equations (\ref{eq14})--(\ref{eq22}) and applying the short-wavelength approximation \( k \gg \frac{1}{R_0} \), neglecting terms of the order \( \frac{ik}{R_0} \) and \( \frac{1}{R_0^2} \), we obtain:
\begin{equation} \label{eq24}
\widehat L_{\nu} U_R+\frac{{2\Omega }}{\nu }U_\varphi-\frac{{\mu_eH_{0\varphi}}}{{2\pi\rho_0\nu R_0 }}B_\varphi+\frac{{\mu_eH_{0z}}}{{4\pi\rho_0\nu }}\widehat D B_R-\frac{{ik}}{{\nu\rho_0}}\widetilde P = 0 
\end{equation}
 \begin{equation} \label{eq25}
\widehat L_{\nu} U_\varphi-\frac{{2\Omega}}{\nu}(1+\textrm{Ro})U_R+\frac{{\mu_eH_{0\varphi}}}{{2\pi\rho_0\nu R_0}}(1+\textrm{Rb})B_R+\frac{{\mu_eH_{0z}}}{{4\pi\rho_0\nu}}\widehat D B_\varphi = 0 
\end{equation}
\begin{equation} \label{eq26}
\widehat L_{\nu} U_z +\frac{{\mu_eH_{0z}}}{{4\pi\rho_0\nu}}\widehat D H_z-\frac{{\widehat D}}{{\nu\rho_0}}\widetilde P-\frac{g}{\rho_0\nu}\left(\Phi(\rho_p-\rho_0)-\rho_0\beta\Theta\right) = 0 
\end{equation}	
\begin{equation} \label{eq27}
\widehat L_{\chi}\Theta+\left(\frac{T_d-T_u}{\chi_f h}\right)U_z-\frac{(\rho c)_p D_B}{(\rho c)_f\chi_f}\left(\frac{T_d-T_u}{h}\right)\widehat D\Phi-$$
$$-\frac{(\rho c)_p}{(\rho c)_f\chi_f}\left[D_B\left(\frac{\phi_d-\phi_u}{ h}\right)+\frac{2D_T}{T_u}\left(\frac{T_d-T_u}{ h}\right)\right]\widehat D\Theta=0
\end{equation}
\begin{equation} \label{eq28}
\widehat L_{\phi}\Phi+\left(\frac{\phi_d-\phi_u}{D_B h}\right)U_z+\frac{D_T}{D_B T_u}\left(\widehat D^2-k^2\right)\Theta=0
\end{equation}
\begin{equation} \label{eq29}
\widehat L_{\eta} B_R+\frac{{H_{0z}}}{{\eta }}\widehat D U_R = 0 
\end{equation}
\begin{equation} \label{eq30}
\widehat L_{\eta} B_\varphi+\frac{{H_{0z}}}{{\eta}}\widehat D U_\varphi-\frac{2H_{0\varphi}}{\eta R_0} \textrm{Rb} U_R+ \frac{2\Omega}{\eta} \textrm{Ro} B_R= 0 
\end{equation}
\begin{equation} \label{eq31}
\widehat L_{\eta} B_z +\frac{{H_{0z}}}{\eta}\widehat D U_z = 0 
\end{equation}
The following notations are introduced for the differential operators:
\[
\widehat{L}_{\nu} = \widehat{D}^2 - \left( \frac{\gamma}{\nu} + k^2 \right), \quad
\widehat{L}_{\eta} = \widehat{D}^2 - \left( \frac{\gamma}{\eta} + k^2 \right),
\]
\[
\widehat{L}_{\chi} = \widehat{D}^2 - \left( \frac{\gamma}{\chi_f} + k^2 \right), \quad
\widehat{L}_{\phi} = \widehat{D}^2 - \left( \frac{\gamma}{D_B} + k^2 \right).
\]
For the subsequent analysis of the system of equations (\ref{eq24})-(\ref{eq31}), it is convenient to rewrite it in dimensionless form by introducing the following nondimensional variables, denoted with an asterisk:
\[
(R_0^*, z^*) = \frac{1}{h}(R_0, z), \quad 
(U_R^*, U_\phi^*, U_z^*) = \frac{h}{\chi_f}(U_R, U_\phi, U_z),
\]
\[
(B_R^*, B_\phi^*, B_z^*) = \frac{1}{H_0}(B_R, B_\phi, B_z), \quad H_0 = H_{0z},
\]
\[
\varphi^* = \varphi, \quad 
\Theta^* = \frac{\Theta}{T_d - T_u}, \quad 
\Phi^* = \frac{\Phi}{\phi_u - \phi_d}, \quad 
\widetilde{P}^* = \widetilde{P} \left( \frac{h^2}{\rho_0 \nu \chi_f} \right),
\]
\[
t^* = t \left( \frac{\nu}{h^2} \right), \quad 
\frac{\partial}{\partial t^*} = \frac{h^2}{\nu} \frac{\partial}{\partial t}.
\]
Omitting the asterisks for simplicity, we obtain the following system of dimensionless equations:
\begin{equation} \label{eq32}
\widehat L_\nu  U_R  + \sqrt {\textrm{Ta}} U_\varphi-2\Pr\textrm{Pm}^{-1}\textrm{Q}\xi B_\varphi + \Pr\textrm{Pm}^{-1}\textrm{Q}\widehat DB_R-ik\widetilde P = 0 \end{equation}
\begin{equation} \label{eq33}
\widehat L_\nu U_\varphi-\sqrt {\textrm{Ta}}(1+\textrm{Ro})U_R+2\Pr\textrm {Pm}^{-1}\textrm{Q}\xi(1+\textrm{Rb})B_R+\Pr\textrm{Pm}^{-1}\textrm{Q}\widehat DB_\varphi= 0 \end{equation}
\begin{equation} \label{eq34}
\widehat L_\nu U_z +\Pr\textrm{Pm}^{-1}\textrm{Q}\widehat DB_z-\widehat D\widetilde P +\textrm{Ra}\Theta-R_n\Phi = 0 \end{equation}
\begin{equation} \label{eq35}
\widehat L_\chi\Theta+U_z+\frac{N_B}{L_e}\left(\widehat D\Theta- \widehat D\Phi\right)-\frac{2N_AN_B}{L_e}\widehat D\Theta=0  \end{equation}
\begin{equation} \label{eq36}
\widehat L_\phi\Phi-L_e U_z+N_A\left(\widehat D^2-k^2\right)\Theta=0  \end{equation}
\begin{equation} \label{eq37}
\widehat L_\eta B_R+{\Pr}^{-1}\textrm{Pm}\widehat DU_R  = 0 \end{equation}
\begin{equation} \label{eq38}
\widehat L_\eta B_\varphi+{\Pr}^{-1}\textrm{Pm}\widehat DU_\varphi-2{\Pr}^{-1}\textrm{Pm}\xi\textrm{Rb}U_R  +\textrm{PmRo}\sqrt{\textrm{Ta}}B_R=0 \end{equation}
\begin{equation} \label{eq39}
\widehat L_\eta B_z+{\Pr}^{-1}\textrm{Pm}\widehat DU_z= 0, \end{equation}
where the operators \(\widehat{L}_{\nu}\), \(\widehat{L}_{\eta}\), \(\widehat{L}_{\chi}\), and \(\widehat{L}_{\phi}\) in terms of the dimensionless variables take the following forms:
\[\widehat L_\nu=\widehat D^2-\gamma-k^2, \; \widehat L_\eta=\widehat D^2-\textrm{Pm}\gamma-k^2,\]
\[\widehat L_\chi=\widehat D^2-\Pr\gamma-k^2, \; \widehat L_{\phi}=\widehat D^2-\Pr L_e\gamma-k^2. \]		
In Eqs. (\ref{eq32})-(\ref{eq39}), we present the following dimensionless parameters:
$\textrm{Ta}=\frac{4{\Omega}^2 h^4}{\nu^2}$ is the Taylor number; $\textrm{Pr}=\nu/\chi_f$ is  the Prandtl number; $\textrm{Pm}=\nu/\eta$ is the magnetic Prandtl number; $\textrm{Q}=\frac{\mu_e H_0^2 h^2}{4\pi\rho_0\nu\eta}$ is the Chadrasekhar number; $\xi= \frac{H_{0\varphi}h}{R_0 H_0}$  is the ratio of the azimuthal to the axial magnetic fields; $\textrm{Ra}=(T_d-T_u)\rho_{0}g\beta h^3/\mu\chi_f$ is the Rayleigh number; $\textrm{R}_n=(\rho_p-\rho_f)(\phi_u-\phi_d)gh^3/\mu\chi_f$ is the concentration Rayleigh number; $L_e=\chi_f/D_B$ is the Lewis number; 
$$N_B=(\phi_u-\phi_d)\cdot\frac{(\rho c)_p}{(\rho c)_f}$$ -- a coefficient characterizing the increment of the density of nanoparticles; $$N_A=D_T(T_d-T_u)/D_BT_u(\phi_u-\phi_d)$$ -- a modified diffusion coefficient.

Equations (\ref{eq32})-(\ref{eq39}) are accompanied by the solenoidality equations for the fields ${\bf u}$ and ${\bf b}$:
\begin{equation} \label{eq40}
\widehat D U_z+ikU_R=0,\quad \widehat D B_z+ikB_R=0. \end{equation}
Using condition (\ref{eq40}), we eliminate pressure $\widetilde P$ from Eqs. (\ref{eq32}) and (\ref{eq34}): 
\begin{equation} \label{eq41} 
\widetilde P=\frac{1}{\widehat D^2-k^2}\left[\sqrt {\textrm{Ta}}ikU_\phi-2 \textrm{Q}\xi ik\left(-\frac{\widehat D U_\phi}{\widehat L_\eta}+2\xi\textrm{Rb}\frac{U_R}{\widehat L_\eta}+\textrm{PmRo}\sqrt{\textrm{Ta}}\frac{\widehat D}{\widehat L_\eta^2}U_R\right)-\right.$$
$$\left. - \textrm{Ra}\cdot\frac{\widehat D\left(\widehat{L}_\phi-\frac{N_B}{L_e}\widehat{D}\right)}{\widehat L} U_z-\textrm{R}_n\cdot\frac{\widehat{D}}{\widehat L\widehat L_\phi}\left(\widehat L L_e+N_A\left(\widehat D^2-k^2\right)\left(\widehat L_\phi-\frac{N_B}{L_e}\widehat D\right)\right)U_z \right],
\end{equation}
where $$\widehat L=\widehat L_\chi\widehat L_\phi+\frac{N_B}{L_e}(1-2N_A)\widehat L_\phi\widehat{D}+\frac{N_AN_B}{L_e}(\widehat D^2-k^2)\widehat D .$$
 Substituting expression (\ref{eq41}) into the system of linear equations (\ref{eq32})-(\ref{eq39}), after straightforward but cumbersome transformations, we obtain a single differential equation for $U_z$:
 \begin{equation} \label{eq42}
\left[ {\widehat a_{33}\left({\widehat a_{11} \widehat a_{22}-\widehat a_{21}\widehat a_{12} } \right)+\widehat a_{13}\left({\widehat a_{21}\widehat a_{32}-\widehat a_{31}\widehat a_{22}}\right)}\right]U_z  = 0. \end{equation}
Here  \[\widehat a_{11}= \widehat L_\nu-\textrm{Q} \frac{{\widehat D^2 }}{{\widehat L_\eta}}-2\textrm {Q}\xi\left(2\xi\textrm{Rb}+\sqrt{\textrm{Ta} }\textrm{RoPm}\frac{\widehat D}{\widehat L_\eta}\right) \cdot \frac{\widehat D^2}{\widehat L_\eta(\widehat D^2- k^2)},\; \widehat a_{12}=\frac{\widehat D^2 }{\widehat D^2-k^2}\cdot\left( \sqrt {\textrm{Ta}}+2\textrm{Q}\xi\frac{\widehat D}{\widehat L_\eta} \right),\]
\[  \widehat a_{13}=\frac{ik\widehat D}{\widehat D^2-k^2}\left[\textrm{Ra}\cdot\frac{\widehat L_\phi-\frac{N_B}{L_e}\widehat D}{\widehat L}+\frac{\textrm{R}_n}{\widehat L\widehat L_\phi} \left(\widehat L L_e+N_A(\widehat D^2-k^2)\left(\widehat L_\phi-\frac{N_B}{L_e}\widehat D\right)\right)\right], \]
\[\widehat a_{21}=-\sqrt{\textrm{Ta}}(1 +\textrm{Ro})+\textrm{QRoPm}\sqrt {\textrm{Ta}}\frac{\widehat D^2}{\widehat L_\eta^2}-2\textrm{Q}\xi\frac{\widehat D}{\widehat L_\eta}, \; \widehat a_{22}=\widehat L_\nu-\textrm{Q}\frac{\widehat D^2}{\widehat L_\eta},  \]
\[\widehat a_{31}=\frac{2\textrm {Q}\xi ik}{\widehat D^2-k^2}\cdot\left(2\xi\textrm{Rb}\frac{\widehat D}{\widehat L_\eta}+\sqrt{\textrm{Ta}}\textrm{RoPm}\frac{\widehat D^2}{\widehat L_\eta^2}\right), \; \widehat a_{32}=-\frac{ik \widehat D}{\widehat D^2-k^2}\cdot\left(\sqrt{\textrm{Ta}}+2\textrm {Q}\xi\frac{\widehat D}{\widehat L_\eta}\right),\]
\[\widehat a_{33}=\widehat L_\nu-\textrm {Q}\frac{\widehat D^2}{\widehat L_\eta}+\frac{k^2\textrm{Ra}}{\widehat L(\widehat D^2-k^2)}\left(\widehat L_\phi-\frac{N_B}{L_e}\widehat D\right)+\]
\[+\frac{k^2\textrm{R}_n}{\widehat L\widehat L_\phi(\widehat D^2-k^2)}\left(\widehat L L_e+N_A(\widehat D^2-k^2)\left(\widehat L_\phi-\frac{N_B}{L_e}\widehat D\right)\right) .\]
Equation (\ref{eq42}) is subject to boundary conditions imposed solely along the \( z \)-direction:
\begin{equation} \label{eq43}
U_z=\frac{d^2 U_z}{dz^2}=0, \quad \textrm{at} \quad z=0,\quad z=1. \end{equation}
Equation~(\ref{eq42}) governs convective processes in a thin layer of an electrically conducting nanofluid undergoing nonuniform rotation in the presence of an external helical magnetic field. To satisfy the free boundary conditions~(\ref{eq43}), the function \( U_z \) is chosen in the form:
\begin{equation} \label{eq44} U_z=U_{0z} \sin n\pi z \quad (n=1,2,3 \ldots), \end{equation}
where $U_{0z}=\textrm{const}$ is the amplitude of perturbations of the $z$ component of velocity. 
Substituting expression~(\ref{eq44}) into Eq.~(\ref{eq42}) and integrating over the layer thickness \( z \in (0, 1) \), we derive the following dispersion relation under the single-mode approximation (\( n = 1 \)):
\begin{equation} \label{eq45} 
\textrm{Ra} = \frac{a^2\Gamma_A^4\Gamma_\eta(a^2\Gamma_A^2\Gamma_\phi^2\Gamma_\chi-m_0)+a^2\Gamma_A^2\Gamma_\phi^2\Gamma_\chi m_1-\pi^2 a^2\Gamma_A^2\Gamma_\phi\left(\frac{N_B}{L_e}(1-2N_A)\Gamma_\phi+\frac{N_AN_B}{L_e}a^2\right)\cdot m_2}{k^2\Gamma_\eta\Gamma_\phi(a^2\Gamma_\eta\Gamma_\phi\Gamma_A^4+\frac{\pi^2}{L_e}(N_B-N_A)\cdot m_2)}
\end{equation}
where
\[\Gamma_A^2=(\gamma+a^2)(\gamma\textrm{Pm}+a^2)+\pi^2\textrm{Q},\; \Gamma_\chi=\gamma \Pr+a^2, \; \Gamma_\eta = \gamma\textrm{Pm}+a^2,\;\Gamma_\phi = \gamma\textrm{Pr}L_e+a^2,\]
\[ \quad a^2=\pi^2+k^2,\; m_0=k^2\textrm{R}_n\Gamma_\eta\Gamma_\phi(L_e\Gamma_\chi+a^2N_A), \; m_1=\pi^2\textrm{Ta}(1+\textrm{Ro})\Gamma_\eta^3+\pi^4\textrm{QTaRoPm}\Gamma_\eta-\]
\[-4\pi^4\textrm{Q}^2\xi^2\Gamma_\eta-4\pi^2\textrm{Q}\xi^2\Gamma_A^2\textrm{Rb}\Gamma_\eta, \;m_2=2\pi^2\textrm{Q}\xi\sqrt{\textrm{Ta}}[(2+\textrm{Ro})\Gamma_\eta^2+\textrm{PmRo}(\pi^2\textrm{Q}-\Gamma_A^2)].   \]
When both the azimuthal magnetic field is absent (\( \xi = 0 \)) and the influence of nanoparticles is neglected (\( \mathrm{R}_n = N_B = 0 \)), the dispersion equation~(\ref{eq45}) reproduces the result presented in~\cite{10s}. In contrast, in the presence of an azimuthal magnetic field (\( \xi \neq 0 \)) but still without nanoparticle effects (\( \mathrm{R}_n = N_B = 0 \)), the result coincides with the findings reported in~\cite{12s}.

\section{Standard MRI in thin nanofluid layers }

We now examine the simplified scenario in which the nanofluid layer has equal boundary temperatures (\( \mathrm{Ra} = N_A = 0 \)) and is subjected to a purely axial magnetic field (\( \xi = 0 \)). Under these conditions, the dispersion relation (\ref{eq45}) simplifies to a sixth-order polynomial in the growth rate \( \gamma \), corresponding to the classical form of the magnetorotational instability (MRI) in a thin layer of nanofluid:
\begin{equation} \label{eq46}
\mathscr P(\gamma)\equiv a_0\gamma^6+a_1\gamma^5+a_2\gamma^4+a_3\gamma^3+a_4\gamma^2+a_5\gamma+a_6=0,
\end{equation}
where the coefficients $a_j$ $(j=0 \ldots 6)$ are given by
\[ a_0={\Pr}^2\textrm{Pm}^2L_ea^2,\; a_1=2{\Pr}^2\textrm{Pm}L_ea^4(1+\textrm{Pm})+{\Pr}(1+L_e)a^4\textrm{Pm}^2, \]
\[a_2={\Pr}^2L_e[a^6(1+4\textrm{Pm}+\textrm{Pm}^2)+2a^2\textrm{Pm}\pi^2\textrm{Q}+\pi^2\textrm{Pm}^2\textrm{Ta}(1+\textrm{Ro})]+2a^6(1+L_e)\Pr\textrm{Pm}(1+\textrm{Pm})+\]
\[+a^6\textrm{Pm}^2-k^2\textrm{R}_nL_e\textrm{Pm}^2,\; a_3={\Pr}^2L_e[2a^8(1+\textrm{Pm})+2\pi^2a^4\textrm{Q}(1+\textrm{Pm})+2\pi^2a^2\textrm{Pm Ta}(1+\textrm{Ro})]+ \]
\[+a^2(1+L_e)\Pr[a^6(1+4\textrm{Pm}+\textrm{Pm}^2)+2a^2\textrm{Pm}\pi^2\textrm{Q}+\pi^2\textrm{Pm}^2\textrm{Ta}(1+\textrm{Ro})]+2a^8\textrm{Pm}(1+\textrm{Pm})- \]
\[-k^2\textrm{R}_nL_e({\Pr}\textrm{Pm}a^2(2+\textrm{Pm})+a^2\textrm{Pm}^2),\]
\[ a_4=a^2\Pr(1+L_e)[2a^8(1+\textrm{Pm})+2\pi^2a^4\textrm{Q}(1+\textrm{Pm})+2\pi^2a^2\textrm{Pm Ta}(1+\textrm{Ro})]+\]
\[+{\Pr}^2L_e[a^2(a^4+\pi^2\textrm{Q})^2+\pi^2a^4\textrm{Ta}(1+\textrm{Ro})+\pi^4\textrm{PmRoTaQ}]+\]
\[+a^4[a^6(1+4\textrm{Pm}+\textrm{Pm}^2)+2a^2\textrm{Pm}\pi^2\textrm{Q}+\pi^2\textrm{Pm}^2\textrm{Ta}(1+\textrm{Ro})]-k^2\textrm{R}_nL_e({\Pr}(a^4(1+2\textrm{Pm})+\pi^2\textrm{QPm})+\]
\[+a^4\textrm{Pm}(2+\textrm{Pm})), \; a_5=a^2(1+L_e)\Pr[a^2(a^4+\pi^2\textrm{Q})^2+\pi^2a^4\textrm{Ta}(1+\textrm{Ro})+\]
\[+\pi^4\textrm{PmRoTaQ}]+a^4[2a^8(1+\textrm{Pm})+2\pi^2a^4\textrm{Q}(1+\textrm{Pm})+2\pi^2a^2\textrm{Pm Ta}(1+\textrm{Ro})]-\]
\[-k^2\textrm{R}_nL_e(a^2{\Pr}(a^4+\pi^2\textrm{Q})+a^6(1+2\textrm{Pm})+\pi^2a^2\textrm{QPm}),\]
\[a_6=a^4[a^2(a^4+\pi^2\textrm{Q})^2+\pi^2a^4\textrm{Ta}(1+\textrm{Ro})+\pi^4\textrm{PmRoTaQ}]-k^2a^4(a^4+\pi^2\textrm{Q})\textrm{R}_nL_e.\]
In the special case where nanoparticles are absent that is, for a "pure" electrically conducting fluid the dispersion equation (\ref{eq46}) reduces to the classical result obtained in \cite{10s}. Although Eq.~(\ref{eq46}) is analytically unsolvable in general due to its algebraic complexity, the stability of the corresponding perturbations can still be assessed without explicitly solving the equation. This can be accomplished by examining the signs and relationships among the polynomial coefficients using algebraic stability criteria, such as those of Routh-Hurwitz or Lienard-Shepard \cite{23s}. The Lienard-Shepard criterion is particularly advantageous in this case, as it involves fewer determinant conditions than the Routh-Hurwitz approach, thereby simplifying the stability analysis. According to this criterion, a necessary and sufficient condition for all roots of the polynomial in \( \gamma \) to lie in the left half of the complex plane ensuring asymptotic stability is the fulfillment of the following system of determinant inequalities:
\begin{enumerate}
\item All coefficients \( a_j \) of the polynomial in \( \gamma \) must be strictly positive for \( j = 0, 1, \ldots, 6 \);
\item The following sequence of Hurwitz determinants must satisfy the inequalities:
\[
\Delta_1 > 0,\quad \Delta_3 > 0,\quad \Delta_5 > 0,\ldots,
\]
where \( \Delta_m \) denotes the Hurwitz determinant of order \( m \), constructed from the coefficients \( a_j \) of the characteristic polynomial:
\[\Delta _{m} =\left|\begin{array}{cccc} {\begin{array}{c} {a_{1} } \\ {a_{0} } \\ {0} \\ {\begin{array}{c} {0} \\ {\cdot } \end{array}} \end{array}} & {\begin{array}{c} {a_{3} } \\ {a_{2} } \\ {a_{1} } \\ {\begin{array}{c} {a_{0} } \\ {\cdot } \end{array}} \end{array}} & {\begin{array}{c} {a_{5} } \\ {a_{4} } \\ {a_{3} } \\ {\begin{array}{c} {a_{2} } \\ {\cdot } \end{array}} \end{array}} & {\begin{array}{c} {\begin{array}{cc} {\cdot } & {\cdot } \end{array}} \\ {\begin{array}{cc} {\cdot } & {\cdot } \end{array}} \\ {\begin{array}{cc} {\cdot } & {\cdot } \end{array}} \\ {\begin{array}{c} {\begin{array}{cc} {\cdot } & {\cdot } \end{array}} \\ {\begin{array}{cc} {\cdot } & {a_{m} } \end{array}} \end{array}} \end{array}} \end{array}\right|\] 
\end{enumerate}
Using the Lienard-Shepard algorithm, we obtain the necessary and sufficient conditions for stability:
\begin{equation} \label{eq47}
a_{j} > 0,\quad j = 0, \ldots, 6,\quad \Delta_{3} > 0,\quad \Delta_{5} > 0
\end{equation}
Substituting the expressions for the coefficients \( a_j \) into conditions (\ref{eq47}), we arrive at the following inequalities:
\begin{enumerate}
\item  ($a_{0}>0$) $\Rightarrow $ ${\Pr}^2\textrm{Pm}^2L_ea^2 >0$, ($a_{1} > 0$) $\Rightarrow $ $2{\Pr}^2\textrm{Pm}L_ea^4(1+\textrm{Pm})+{\Pr}(1+L_e)a^4\textrm{Pm}^2 >0$ \\ These inequalities are satisfied automatically.
\item  ($a_{2}>0$)$\Rightarrow $ ${\Pr}^2L_e[a^6(1+4\textrm{Pm}+\textrm{Pm}^2)+2a^2\textrm{Pm}\pi^2\textrm{Q}+\pi^2\textrm{Pm}^2\textrm{Ta}(1+\textrm{Ro})]+2a^6(1+L_e)\Pr\textrm{Pm}(1+\textrm{Pm})+a^6\textrm{Pm}^2>k^2\textrm{R}_nL_e\textrm{Pm}^2$ \\ It follows that dissipative processes naturally contribute to the stabilization of the nanofluid flow. Additional stabilizing factors include the uniform magnetic field and differential rotation when the angular velocity profile corresponds to positive Rossby numbers ($\textrm{Ro} > 0$). In contrast, the presence of nanoparticles acts as a destabilizing factor for the nanofluid flow.
\item  ($a_{3}>0$)$\Rightarrow$ ${\Pr}^2L_e[2a^8(1+\textrm{Pm})+2\pi^2a^4\textrm{Q}(1+\textrm{Pm})+2\pi^2a^2\textrm{Pm Ta}(1+\textrm{Ro})]+a^2(1+L_e)\Pr[a^6(1+4\textrm{Pm}+\textrm{Pm}^2)+2a^2\textrm{Pm}\pi^2\textrm{Q}+\pi^2\textrm{Pm}^2\textrm{Ta}(1+\textrm{Ro})]+2a^8\textrm{Pm}(1+\textrm{Pm})>k^2\textrm{R}_nL_e({\Pr}\textrm{Pm}a^2(2+\textrm{Pm})+a^2\textrm{Pm}^2)$ \\
From this, we observe that a uniform magnetic field and differential rotation with positive Rossby numbers ($\textrm{Ro} > 0$) also have a stabilizing effect, while the concentration of nanoparticles (terms involving the Rayleigh concentration number $\textrm{R}_n$) exerts a destabilizing influence.
\item The inequalities $a_{4} > 0$ and $a_{5} > 0$ do not provide new conditions for the stabilization of perturbations.
\item The inequality $a_{6} > 0$ can be written as:
\begin{equation} \label{eq48}
 \textrm{Ro}>-\frac{a^2(a^4+\pi^2\textrm{Q})^2+\pi^2a^4\textrm{Ta}}{\pi^2\textrm{Ta}(a^4+\pi^2\textrm{QPm})}+\frac{k^2(a^4+\pi^2\textrm{Q})\textrm{R}_nL_e}{\pi^2\textrm{Ta}(a^4+\pi^2\textrm{QPm})}= \textrm{Ro}_{\textrm{cr}},\end{equation} 
where the parameter $\textrm{Ro}_{\textrm{cr}}$ is the critical Rossby number at the stability boundary, which corresponds to the neutral state $\gamma=0$. 
\end{enumerate}
Now, let's proceed to the stability conditions consisting of inequalities with Hurwitz determinants (\ref{eq47}). For the determinant $\Delta_{3}$:
$$\Delta_{3}=\begin{vmatrix} a_1 & a_3 & a_5 \\ a_0 & a_2 & a_4 \\ 0 & a_1 & a_3 \end{vmatrix} = a_1a_2a_3 + a_0a_1a_5 - a_1^2a_4 - a_0a_3^2$$
the stability criterion takes the form:
\begin{equation} \label{eq48a}
 a_1a_2a_3 + a_0a_1a_5 > a_1^2a_4 + a_0a_3^2
\end{equation}
For the second Hurwitz determinant, from condition (\ref{eq47}):
$$\Delta_{5}=\begin{vmatrix} a_1 & a_3 &  a_5 & 0 & 0 \\
a_0 & a_2 & a_4 & a_6 & 0 \\ 0 & a_1 & a_3 & a_5 & 0 \\ 0 & a_0 & a_2 & a_4 & 0 \\  0 & 0 & a_1 & a_3 & a_5 \\ \end{vmatrix}=a_1a_2(a_3a_4a_5-a_2a_5^2)-a_1a_4(a_1a_4a_5-a_0a_5^2)+$$
$$+a_1a_6(a_1a_2a_5-a_0a_3a_5)-a_3a_0(a_3a_4a_5-a_2a_5^2)+a_5a_0(a_1a_4a_5-a_0a_5^2)$$
\begin{figure}
  \centering
	\includegraphics[width=5.5 cm, height=5.5 cm]{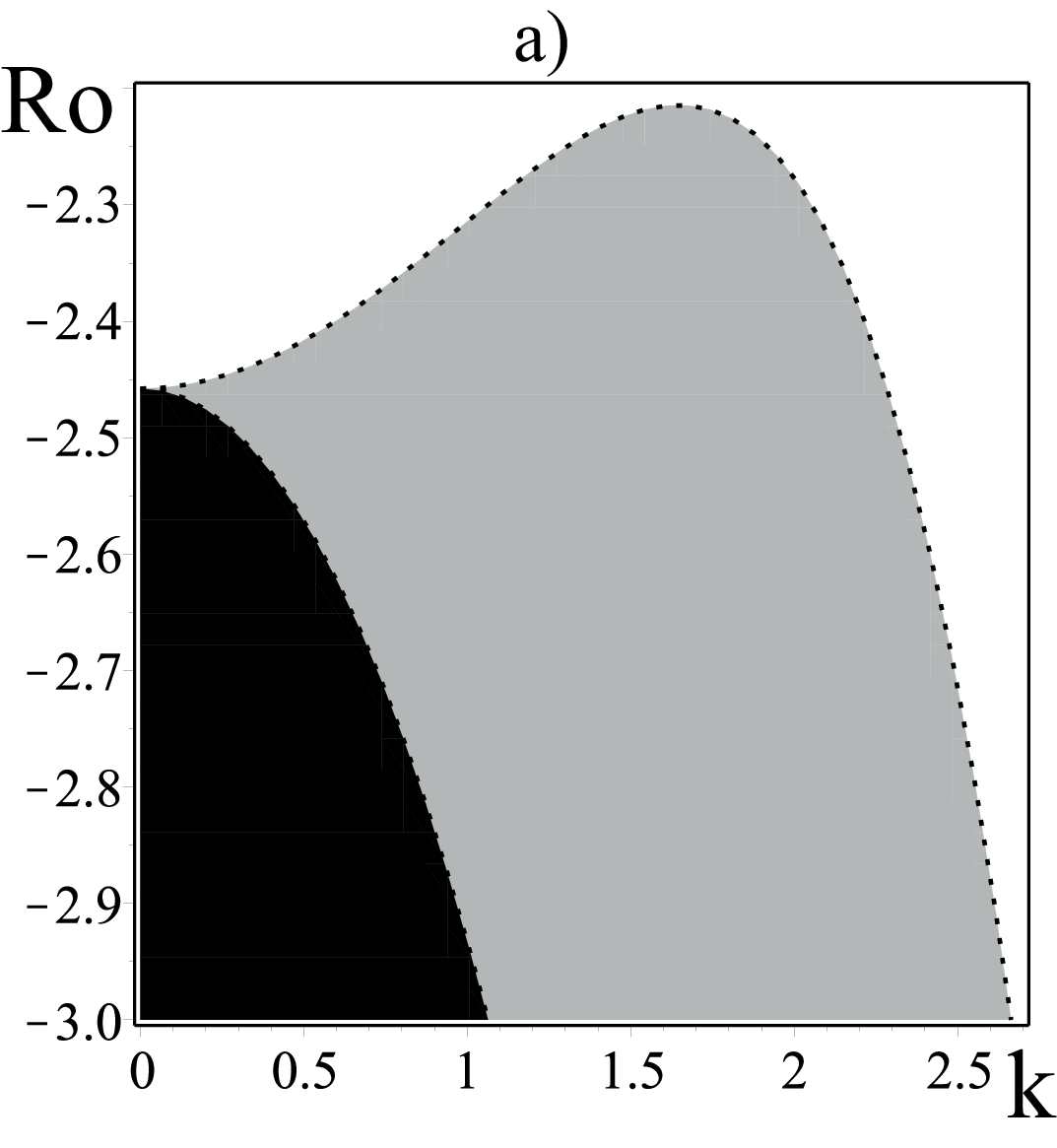}
	\includegraphics[width=5.5 cm, height=5.5 cm]{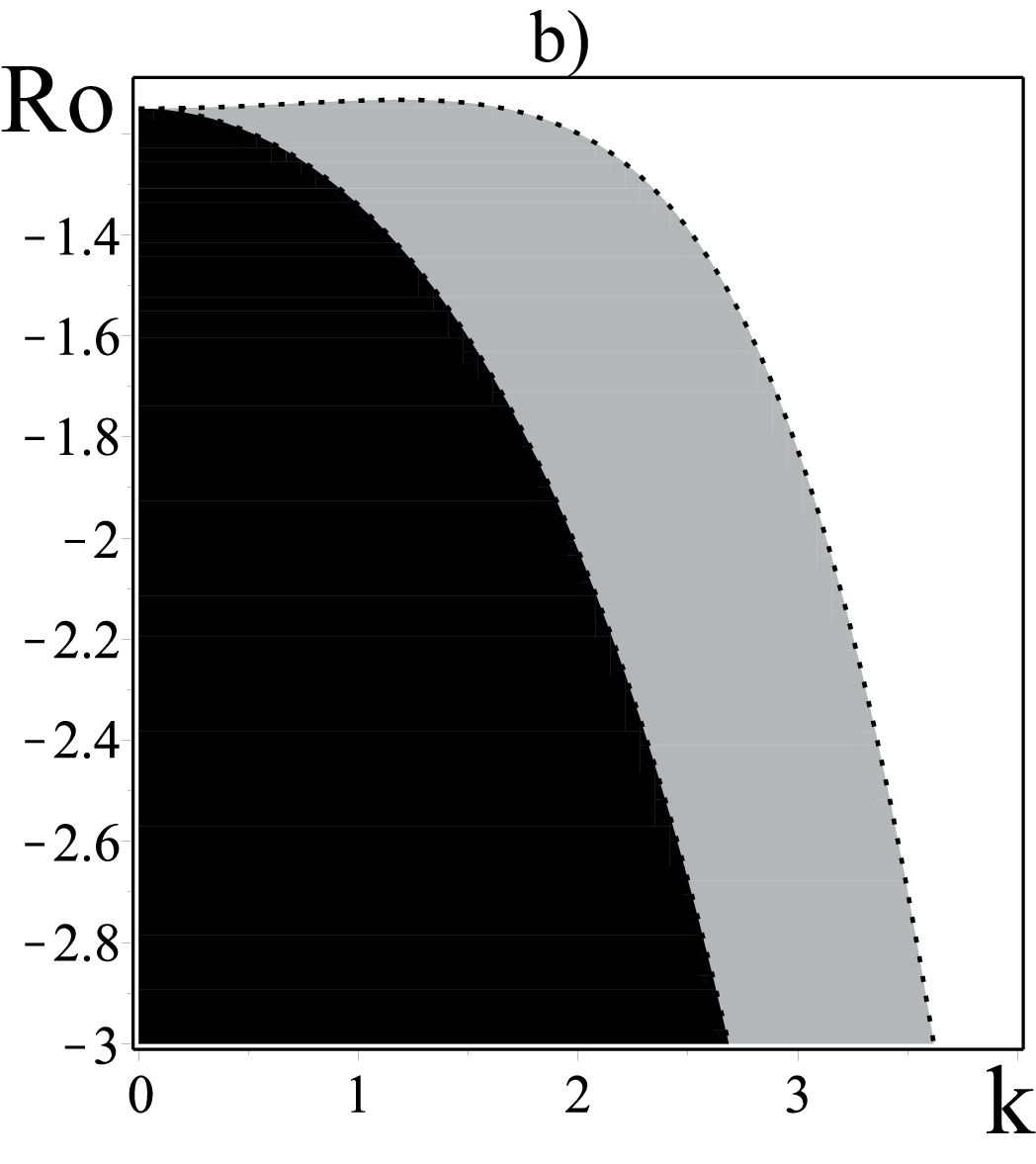}
	\includegraphics[width=5.5 cm, height=5.5 cm]{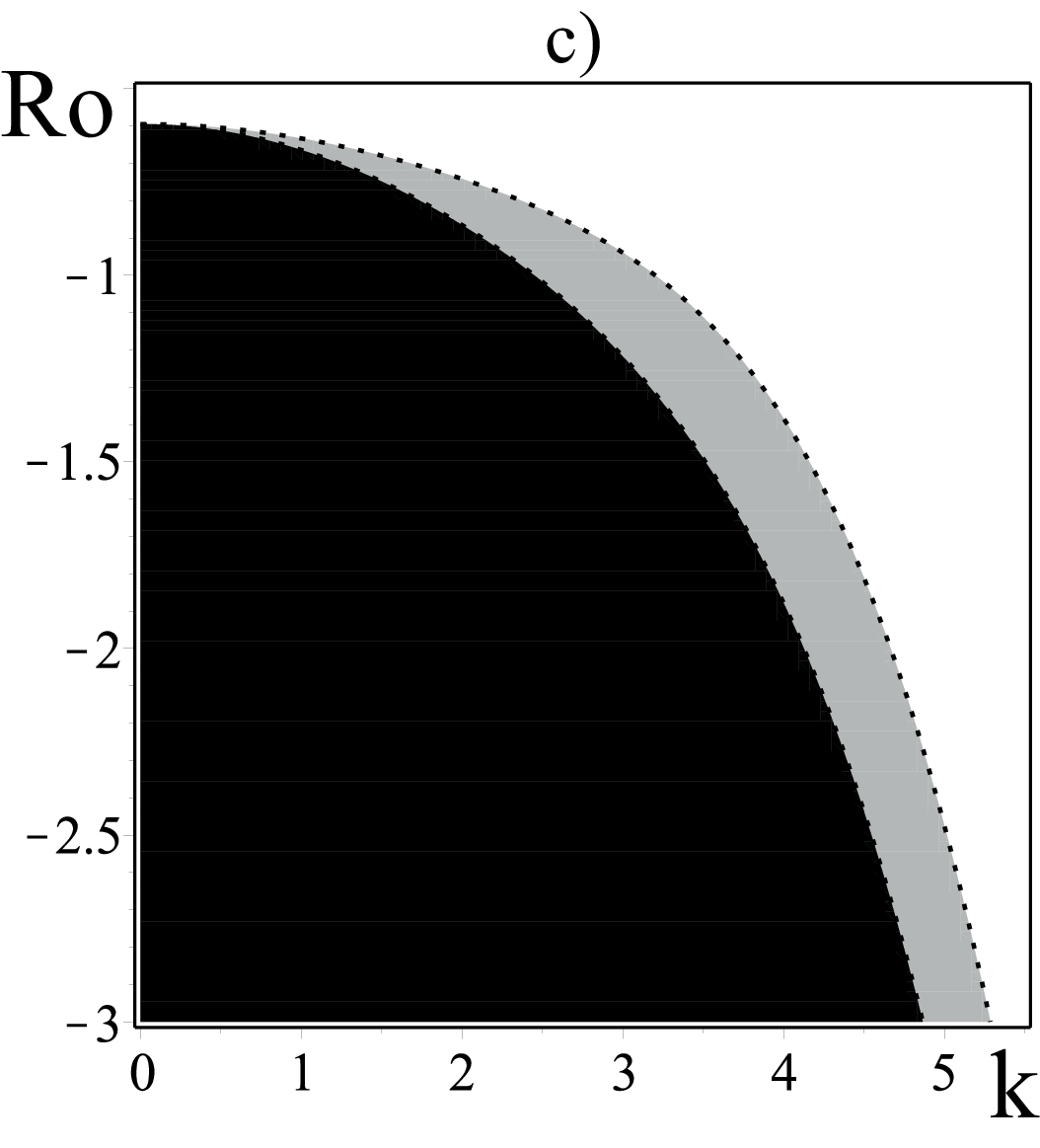} \\ 
	\caption{\small The black color shows the regions in which the standard MRI arises in a pure fluid, and the gray color, in a nanofluid with parameters: $\textrm{Q}=10$, $\textrm{Pm}=1$, $\textrm{R}_n=0.122, L_e=5000$ and Taylor numbers: a) $\textrm{Ta}=100$, b) $\textrm{Ta}=300$, c) $\textrm{Ta}=2000$.}\label{fg2}
\end{figure}
we obtain the following stability criterion:
\begin{equation} \label{eq48b}
a_4a_5(a_1a_2a_3 + a_0a_1a_5 - a_1^2a_4 - a_0a_3^2) + a_1^2a_2a_5a_6 + a_0a_5^2(a_2a_3 + a_1a_4) > a_0a_1a_3a_5a_6 + a_5^2(a_1a_2^2 + a_0^2a_5)
\end{equation}
The stability criteria (\ref{eq48a}) and (\ref{eq48b}) indicate that the concentration of nanoparticles has a destabilizing effect on the stability of axisymmetric perturbations.

Using the critical Rossby number expression (\ref{eq48}), we will calculate the regions of standard MRI development for both "pure" fluid and nanofluid. In the plots of Fig. \ref{fg2}, the instability regions for Rossby numbers $\textrm{Ro} < \textrm{Ro}_{\textrm{cr}}$ are highlighted, illustrating the variation of the rotation parameter $\textrm{Ta}$ (Taylor numbers) in the $(\textrm{k}, \textrm{Ro})$ plane, where $\textrm{k}$ is the dimensionless radial wave number. The black areas in Fig. \ref{fg2} represent the instability zones for the "pure" fluid, while the grey areas indicate the instability zones for the nanofluid. From the graphs in Fig. \ref{fg2}, it is apparent that for small Taylor numbers $\textrm{Ta} = 100, 300$, the region of standard MRI development in the nanofluid (Fig. \ref{fg2}a-b)) is considerably larger than the instability region for the "pure" fluid with $\textrm{R}_n = 0$. In contrast, for larger Taylor numbers $\textrm{Ta} = 2000$, the instability regions for the nanofluid and "pure" fluid become much more similar. The nanofluid parameters (e.g., $\textrm{Al}_2\textrm{O}_3$-water) $ \textrm{R}_n = 0.122, L_e = 5000, \textrm{Pr} = 5, N_A = 5, N_B = 7.5 \cdot 10^{-4}$ were obtained from the work \cite{9s}.

\begin{figure}
  \centering
	\includegraphics[width=5.5 cm, height=5.0 cm]{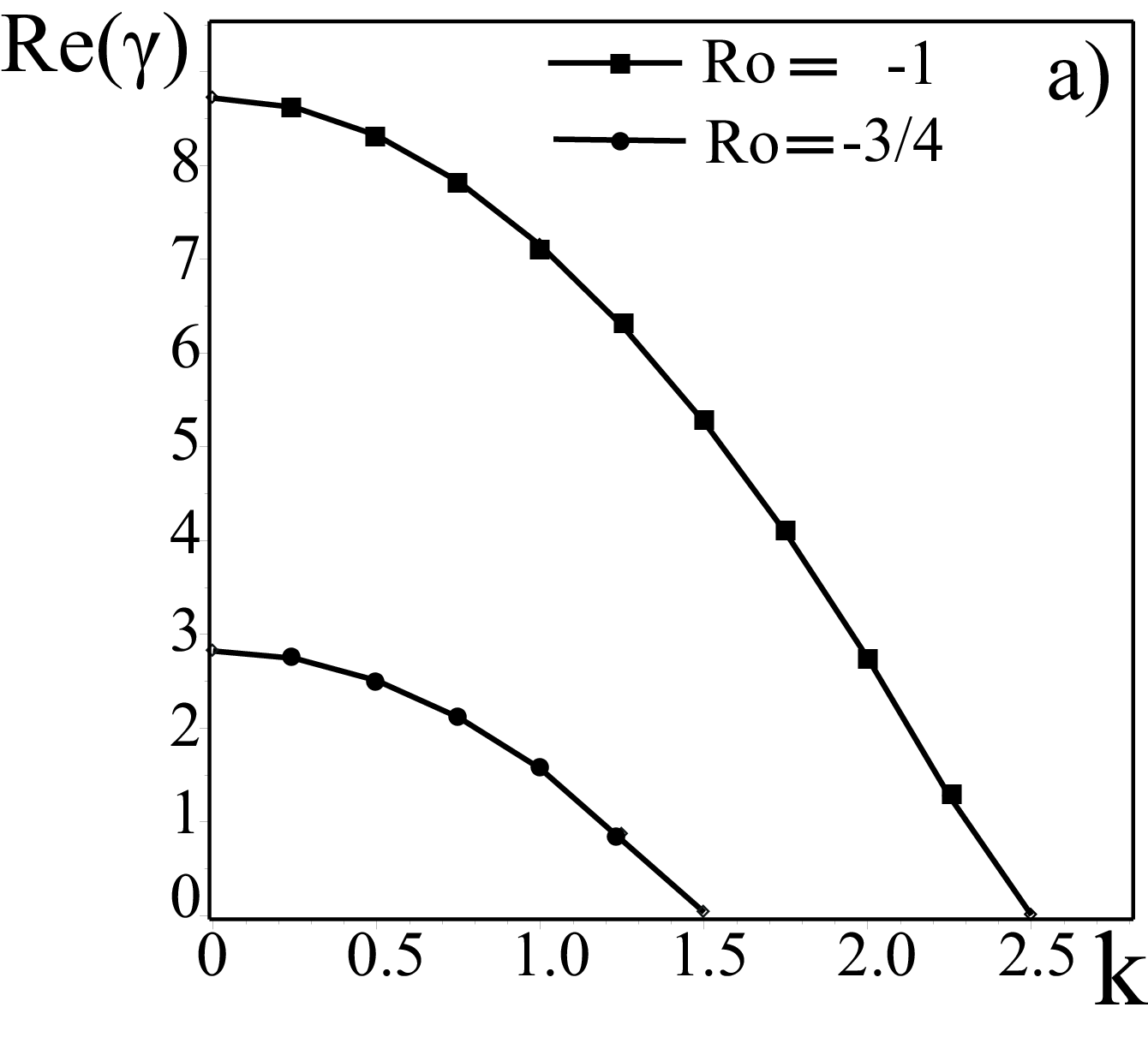}
	\includegraphics[width=5.5 cm, height=5.0 cm]{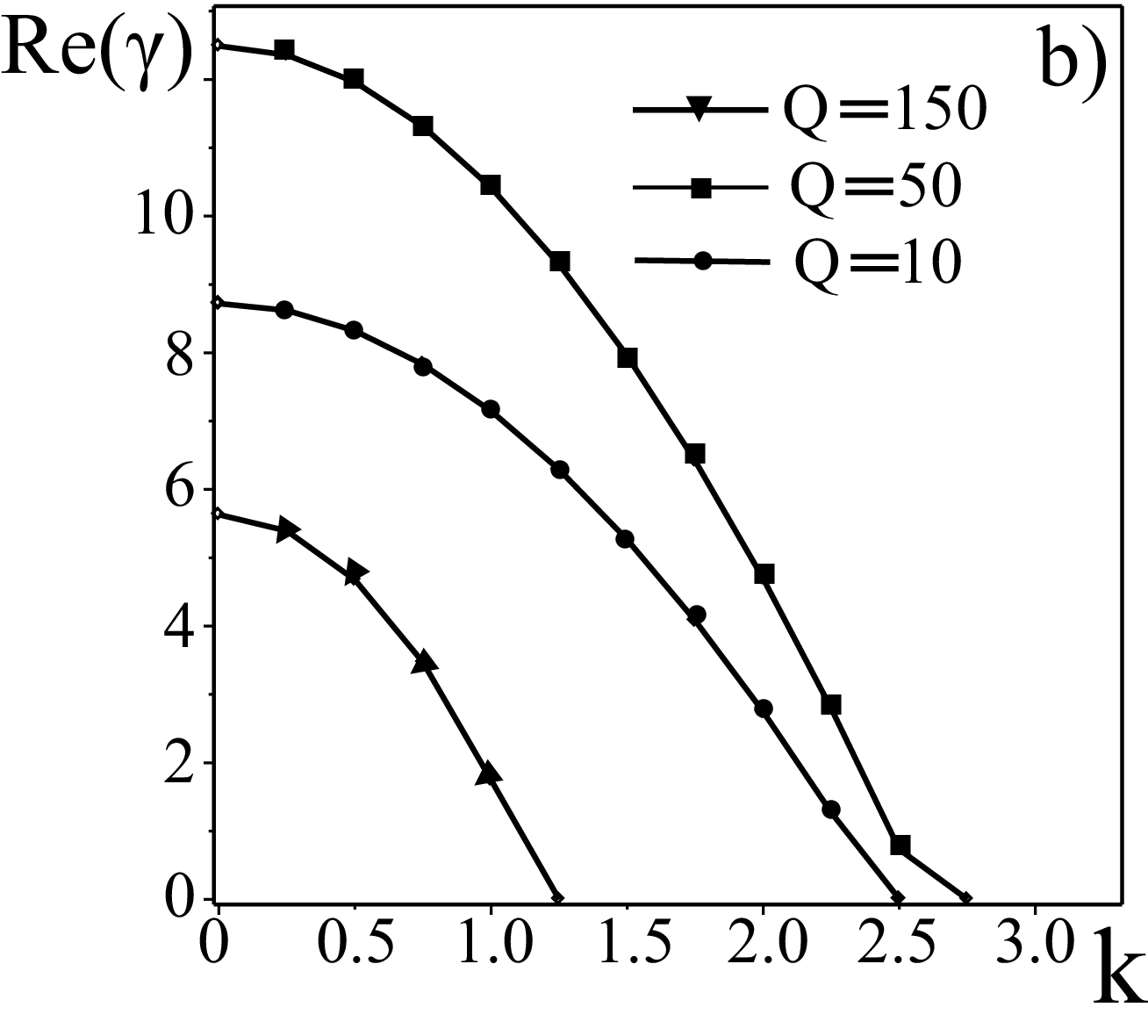}
	\includegraphics[width=5.5 cm, height=5.0 cm]{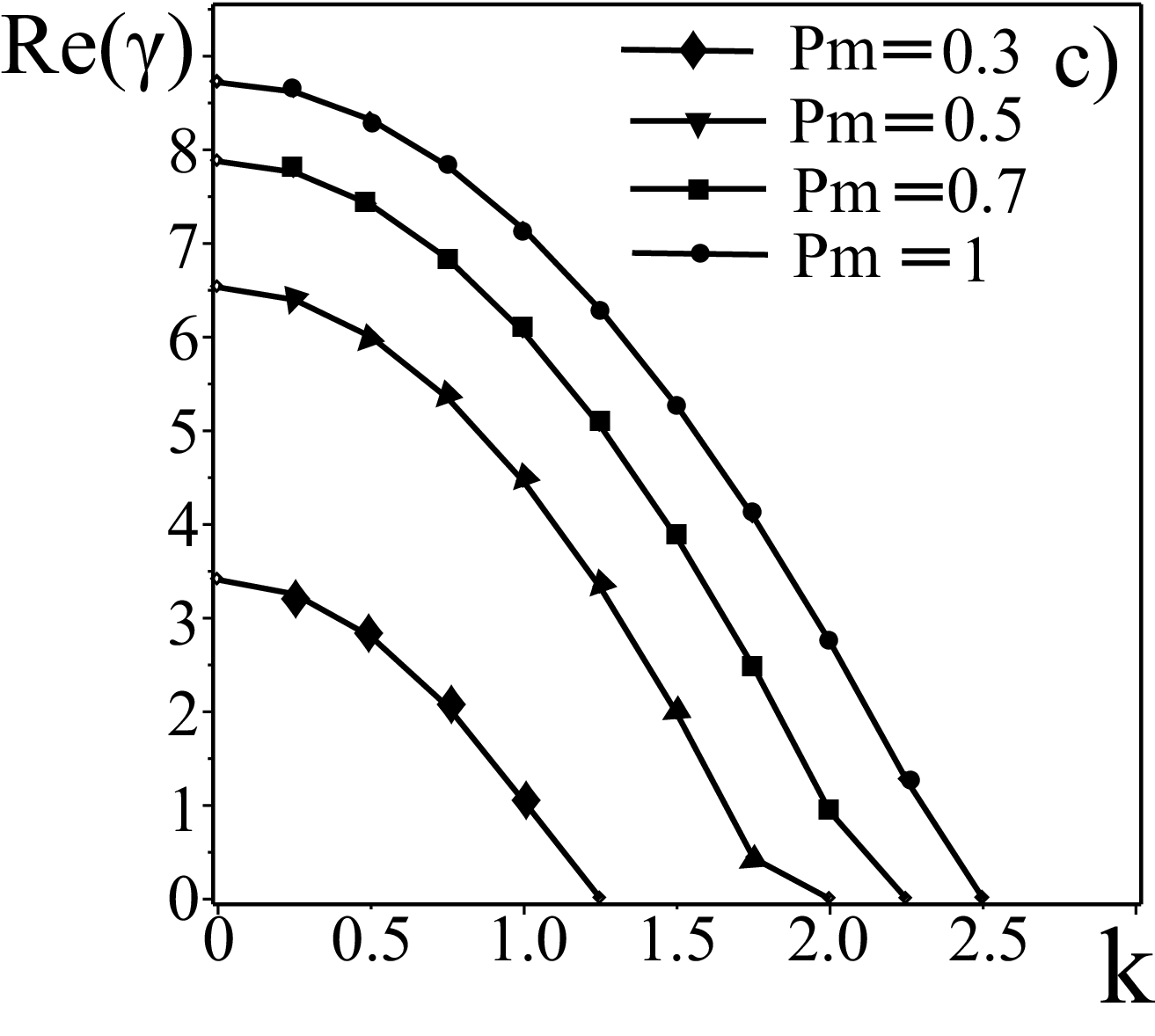} \\
\caption{\small Dependence of the standard MRI growth rate ($\textrm{Re}(\gamma)>0$) in a nanofluid on the radial wave number $\textrm{k}$. The plots show: a) growth rates of perturbations for the Rayleigh ($\textrm{Ro} = -1$) and Keplerian ($\textrm{Ro} = -3/4$) rotation profiles;
b) growth rates of perturbations for different values of the axial magnetic field (Chandrasekhar number) $\textrm{Q} = 10, 50, 150$;
c) growth rates of perturbations for various magnetic Prandtl numbers $\textrm{Pm} = 0.3, 0.5, 0.7, 1$.} \label{fg3}
\end{figure}

The plots in Fig. \ref{fg3} present numerical results for the growth rate of the standard MRI, corresponding to the positive real root of the dispersion relation (\ref{eq46}) with $\textrm{Re}(\gamma)>0$, as a function of the radial wave number $\textrm{k}$. For fixed nanofluid parameters: $\textrm{Pr} = 5$, $\textrm{R}n = 0.122$, and $L_e = 5000$, Fig. \ref{fg3}a) shows that at a Taylor number of $\textrm{Ta} = 2000$, the growth rates of axisymmetric perturbations are higher for the Rayleigh rotation profile ($\textrm{Ro} = -1$) than for the Keplerian rotation profile ($\textrm{Ro} = -3/4$). Fig. \ref{fg3}b) illustrates how the growth rate of standard MRI depends on the strength of the axial magnetic field for $\textrm{Q} = 10, 50, 150$, considering $\textrm{Ro} = -1$ and $\textrm{Ta} = 2000$. It can be seen that increasing the magnetic field intensity $H{0z}$ initially leads to an increase in the instability growth rate ($\textrm{Q} = 10 \rightarrow 50$), but a further increase ($\textrm{Q} = 50 \rightarrow 150$) results in a decrease in the growth rate. Variations in the magnetic 
Prandtl number $\textrm{Pm}$ also significantly affect the development of standard MRI in the nanofluid, as shown in Fig. \ref{fg3}c). For fixed parameters $\textrm{Ta} = 2000$, $\textrm{Q} = 10$, and $\textrm{Ro} = -1$, the growth rates of perturbations are significantly lower at small Prandtl numbers $\textrm{Pm} \ll 1$.

It is worth noting that by varying the physical properties of the nanofluid such as electrical conductivity $\sigma$, thermal conductivity $\chi$, and viscosity $\nu$, which determine the dimensionless parameters $\textrm{Q}, \textrm{Pr}, \textrm{Pm}, \textrm{Ta}, \textrm{R}_n$ the development of standard MRI is indeed possible under the Rayleigh rotation profile ($\textrm{Ro} = -1$).

\subsection{Unmagnetized Couette flow in a nanofluid}

\begin{figure}
  \centering
	\includegraphics[width=5.5 cm, height=5.5 cm]{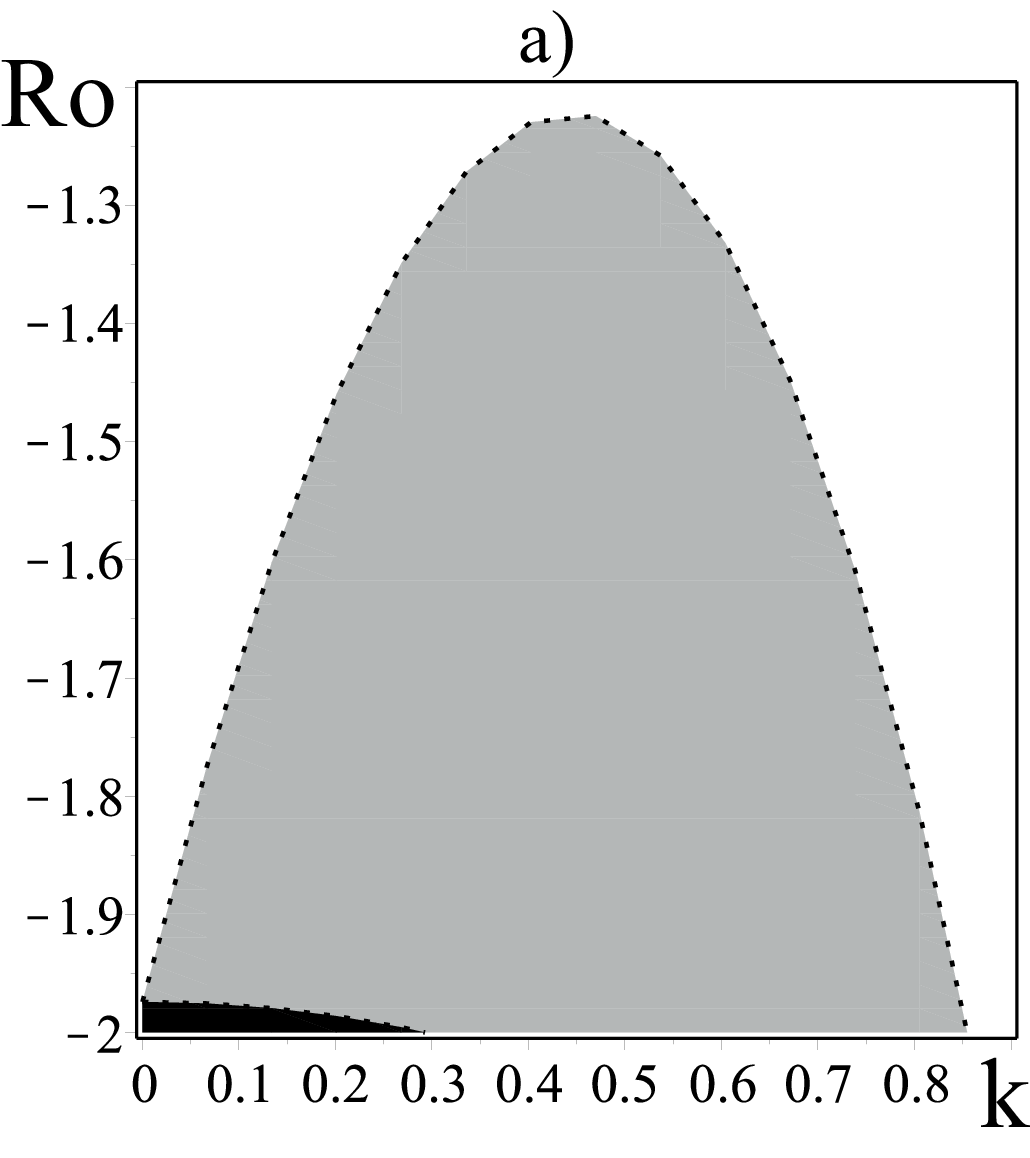}
	\includegraphics[width=5.5 cm, height=5.5 cm]{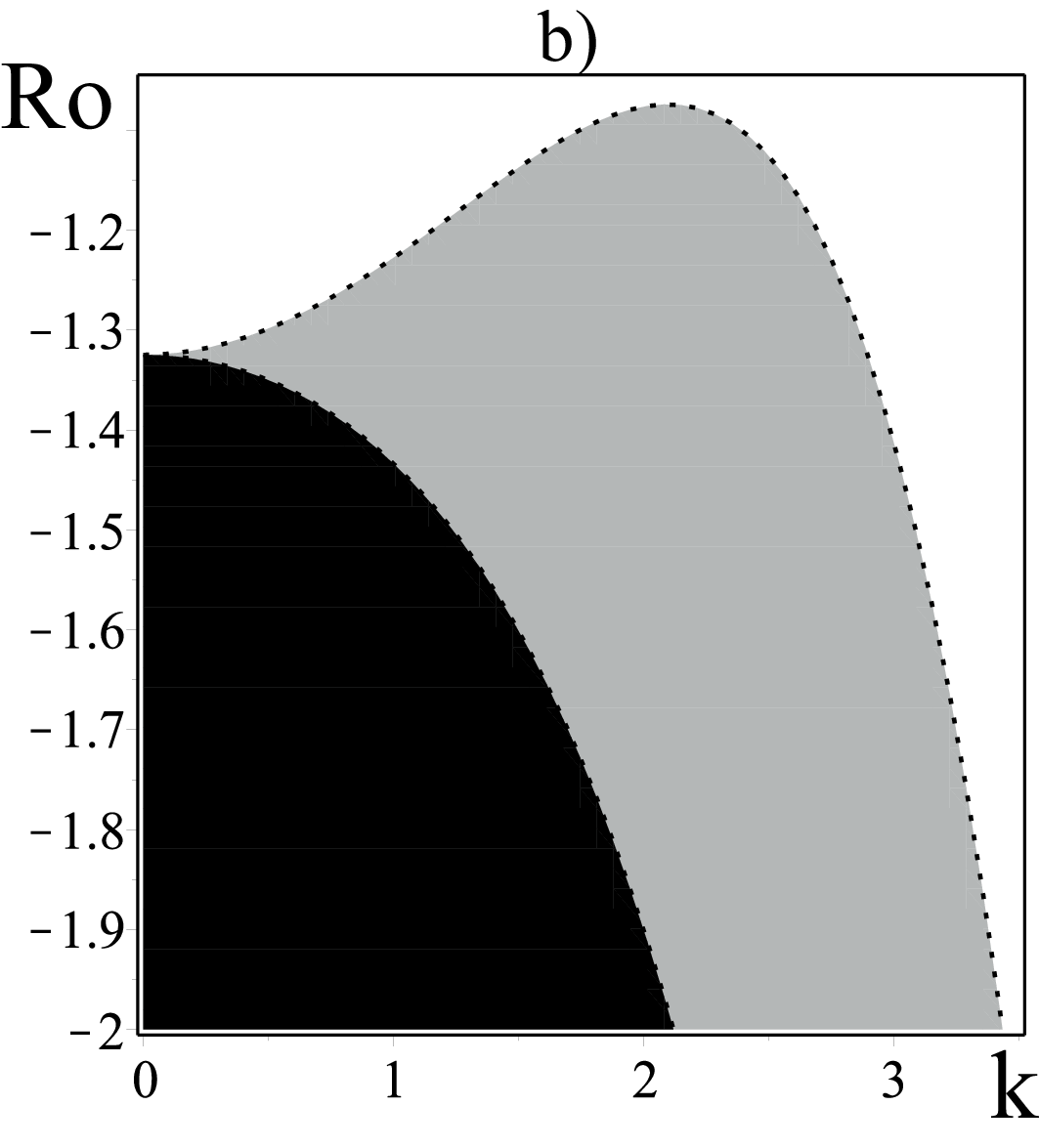}
	\includegraphics[width=5.5 cm, height=5.5 cm]{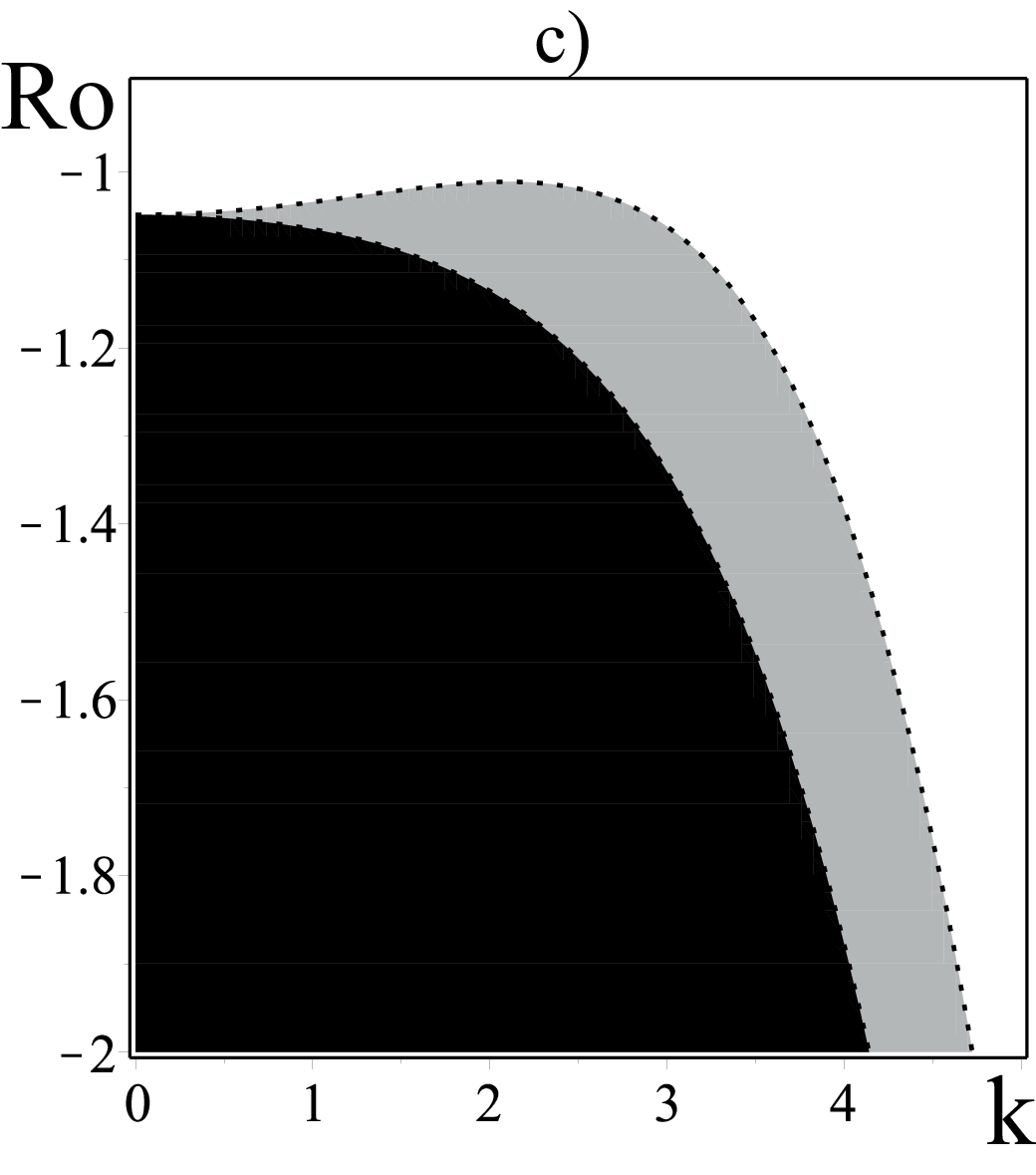}\\
\caption{\small The regions of rotational instability for unmagnetized Couette flow are shown: black areas correspond to the "pure" fluid, while gray areas indicate instability regions in the nanofluid with parameters $\textrm{R}_n = 0.122$, $L_e = 5000$. The plots are presented for different Taylor numbers: a) $\textrm{Ta} = 100$, b) $\textrm{Ta} = 300$, c) $\textrm{Ta} = 2000$. }\label{fg4}
\end{figure}

In the absence of a magnetic field $(\textrm{Q} = 0)$, the stability criterion (\ref{eq48}) takes the form
\begin{equation} \label{eq49}
\textrm{Ro} > -1 - \frac{a^6}{\pi^2 \textrm{Ta}} + \frac{k^2 \textrm{R}n L_e}{\pi^2 \textrm{Ta}} = \textrm{Ro}{\textrm{cr}}
\end{equation}
From this, it follows that for an inviscid nanofluid $(\nu \rightarrow 0)$, the critical Rossby number $\textrm{Ro}_{\textrm{cr}}=-1$ corresponds to the Rayleigh rotation profile. Clearly, an ideal nanofluid flow with a Keplerian rotation profile is stable with respect to axisymmetric perturbations, since $\textrm{Ro} = -3/4 > \textrm{Ro}_{\textrm{cr}} = -1$. In the limiting case where nanoparticles are absent $(\textrm{R}_n = 0)$, the stability criterion (\ref{eq49}) was studied in \cite{24s}.
We now turn to the question of the development of rotational instability (RI) for Rossby numbers $\textrm{Ro} < \textrm{Ro}_{\textrm{cr}}$. To this end, we examine how the instability region of nanofluid Couette flow evolves in the $(\textrm{k}, \textrm{Ro})$ plane for different values of the rotation parameter $\textrm{Ta}$ (Taylor number). Fig. \ref{fg4} shows the results of numerical calculations for $\textrm{Ta} = 100, 300, 2000$. These results indicate that for small Taylor numbers $\textrm{Ta} = 100, 300$, the region of RI development in the nanofluid (shaded in gray in Fig. \ref{fg4}) is significantly larger than the instability region for the "pure" fluid with $\textrm{R}_n = 0$ (shown in black in Fig. \ref{fg4}). In contrast, for large Taylor numbers $\textrm{Ta} = 2000$, the instability regions for the nanofluid and the "pure" fluid differ much less.

\subsection{Chandrasekhar's paradox for electrically conductive nanofluid}

Chandrasekhar's paradox for "pure" electrically conductive fluids lies in the fact that for an ideal conductor $(\eta\rightarrow 0)$, in the "unmagnetized" limit $(H\_0 \rightarrow 0)$, the critical Rossby number becomes zero, $\textrm{Ro}_{\textrm{cr}} = 0$. In contrast, from the stability criterion for an "unmagnetized" fluid (\ref{eq49}) (with $\textrm{R}_n = 0$), we see that $\textrm{Ro}_{\textrm{cr}} \neq 0$; for an ideal fluid, the critical value is $\textrm{Ro}_{\textrm{cr}} = -1$. In other words, the threshold value depends on the order of performing the two limiting transitions.
A similar situation arises with Chandrasekhar's paradox in an ideally electrically conductive nanofluid $\eta \rightarrow 0$. In this case, from expression (\ref{eq48}) for an ideally electrically conductive nanofluid, we obtain the following expression for the critical Rossby number:
$$\textrm{Ro}_{\textrm{cr}} \approx a^2\left(\frac{\mu_e H_0^2}{16\pi\rho_0\Omega^2h^2}\right)$$
This shows that for very small magnetic field values $H_0 \rightarrow 0$, the critical Rossby number tends to zero: $\textrm{Ro}_{\textrm{cr}} \rightarrow 0$, whereas from the stability criterion (\ref{eq49}) in the "ideal" nanofluid limit, $\textrm{Ro}_{\textrm{cr}} \rightarrow -1$. This contradiction is resolved with a more detailed description of the limiting transition. Let's express the critical Rossby number from equation (\ref{eq48}) using the dimensional variables introduced earlier:
$$\textrm{Ro}_{\textrm{cr}}=-\frac{(\omega _A^2+\omega_\nu\omega_\eta)^2+4\alpha^2\Omega^2 \omega_\eta^2}{4\Omega^2 \alpha^2 (\omega_A^2+\omega_\eta^2)}+(1-\alpha^2)\frac{(\omega_A^2+\omega_\nu\omega_\eta)\textrm{R}_nL_e\omega_\nu\omega_\eta}{4\Omega^2 \alpha^2 (\omega_A^2+\omega_\eta^2)(|{\bf k}|h)^4}$$
Next, let's express $\omega_A$ and $\omega_\eta$ through the parameters $(\varepsilon, \varphi)$: $\omega\_A = \varepsilon\cos\varphi$, $\omega\_\eta = \varepsilon\sin\varphi$. The critical Rossby number $\textrm{Ro}_{\textrm{cr}}$ can then be written in terms of $(\varepsilon, \varphi)$:
\begin{equation} \label{eq50}
\textrm{Ro}_{\textrm{cr}} = -\frac{(\varepsilon\cos^2\varphi + \omega_\nu\sin\varphi)^2 + 4\alpha^2\Omega^2\sin^2\varphi}{4\alpha^2\Omega^2} + (1-\alpha^2)\frac{(\varepsilon\cos^2\varphi + \omega\_\nu\sin\varphi)\omega_\nu\sin\varphi}{4\alpha^2\Omega^2(|{\bf k}|h)^4} \textrm{R}_nL\_e
\end{equation}
Now, let's take the limit $\varepsilon \rightarrow 0$ in (\ref{eq50}), which corresponds to the case of both a weak magnetic field $(H_0 \rightarrow 0)$ and an ideally conductive nanofluid $(\eta \rightarrow 0)$. In this case, (\ref{eq50}) reduces to 
\begin{figure}
  \centering
	\includegraphics[width=13 cm, height=6 cm]{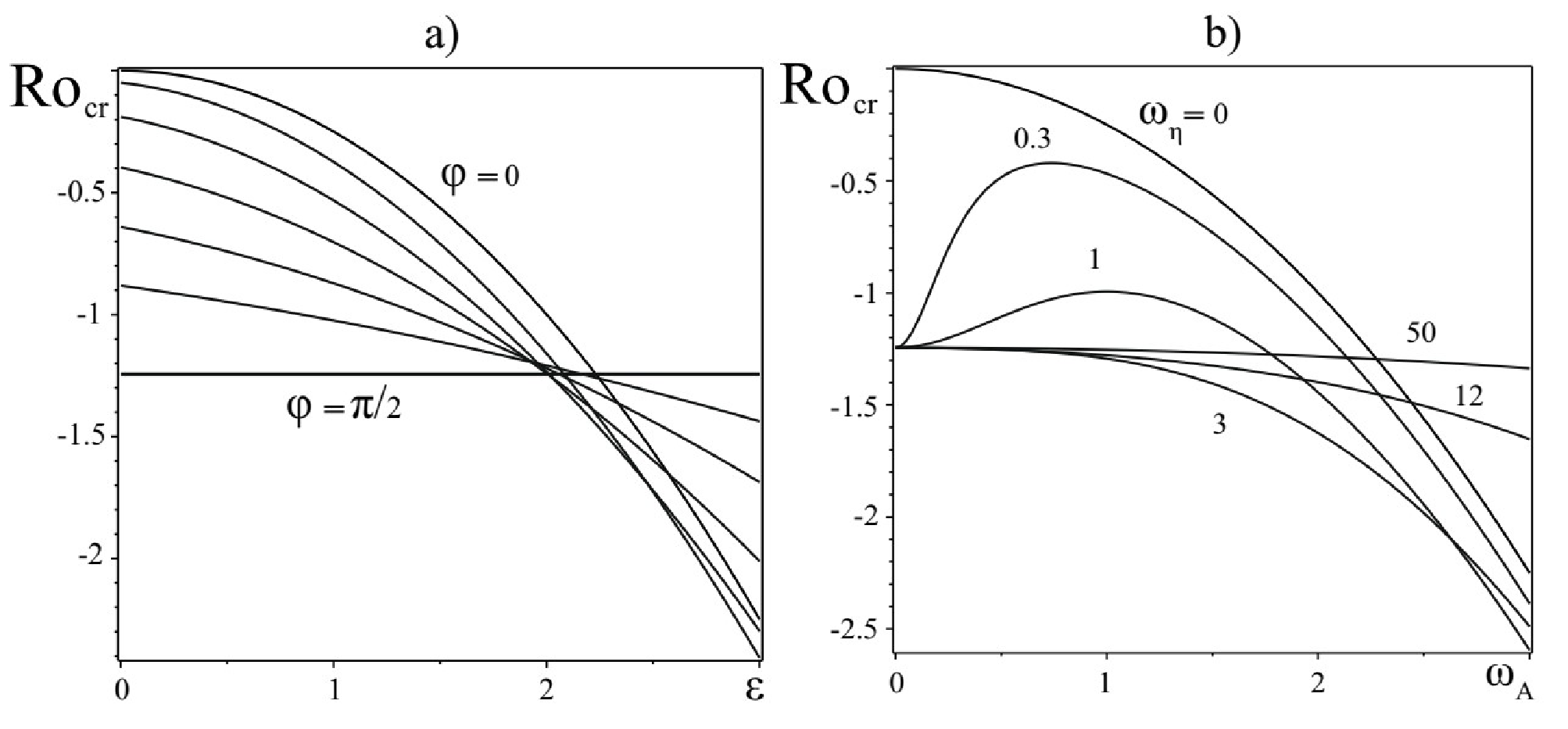} \\
\caption{\small Shown are the variations of the critical Rossby number $\textrm{Ro}_{\textrm{cr}}$ during the transition from a highly conductive to a poorly conductive nanofluid.}\label{fg5}
\end{figure}
the following form:
\begin{equation} \label{eq51}
\textrm{Ro}_{\textrm{cr}}^{(0)} = -\frac{1 + (2Re)^{-2}}{1 + Lu^2} + (1-\alpha^2)\frac{\textrm{R}_nL\_e}{(2Re)^2(|{\bf k}|h)^4(1 + Lu^2)}
\end{equation}
where $Re = \alpha\Omega/\omega_\nu$ is the Reynolds number, and $Lu = \omega_A/\omega_\eta$ is the Lundquist number, which characterizes the degree of "locking" of the magnetic field lines in the electrically conductive fluid. For a highly conductive fluid, $Lu = \infty$, the magnetic field does not "diffuse" out of the fluid as it is "frozen" into it.
For clarity, let's analyze expressions (\ref{eq50})-(\ref{eq51}) using the following values: $Re = 1$, $\alpha\Omega = 1$, $\alpha = 0.8$, $\omega_\nu = 1$, $\textrm{R}_n = 0.122$, $L_e = 5000$. In Fig. \ref{fg5}a, the graph shows the dependence of the critical Rossby number $\textrm{Ro}_{\textrm{cr}}$ on the parameter $\varepsilon$ for varying angles $\varphi \in [0, \pi/2]$ with a step $\Delta\varphi = 0.2$. Here, we see that when both limits $\varepsilon \rightarrow 0$ and $\varphi \rightarrow 0$ are taken simultaneously, the critical Rossby number $\textrm{Ro}_{\textrm{cr}} = 0$, while for $\varepsilon \rightarrow 0$ and $\varphi \rightarrow \pi/2$, the critical Rossby number $\textrm{Ro}_{\textrm{cr}} < -1$, which is less than the critical value for the Rayleigh profile $\textrm{Ro}_{\textrm{cr}} = -1$. Variations of the critical Rossby number $\textrm{Ro}_{\textrm{cr}}$ are also shown in the graph in Fig. \ref{fg5}b) for different values of $\omega_A$ and $\omega_\eta$. It follows from this that for an ideally electrically conductive nanofluid $Lu \rightarrow \infty$ (or when $\varphi = 0$), the critical Rossby number $\textrm{Ro}_{\textrm{cr}} = 0$, while for a poorly conductive fluid, $Lu \rightarrow 0$ (or when $\varphi = \pi/2$), it matches the result of (\ref{eq49}), i.e., resolving the Chandrasekhar paradox.

\section{Azimuthal MRI in thin nanofluid layers}

Let us consider a non-uniformly rotating layer of nanofluid with a constant and uniform temperature at the boundaries, subjected to an external azimuthal magnetic field $H_{0\varphi}$. In this case, the axial magnetic field is zero, $H_{0z}=0$, and $\textrm{Ra}=N_A=0$. The steady flow of the nanofluid aligns with the direction of the magnetic field. Thus, the geometry of the problem corresponds to the azimuthal magnetorotational instability (MRI) \cite{25s}. The dispersion relation for azimuthal MRI in thin nanofluid layers is obtained from equation (\ref{eq45}) by setting $H_{0z}=0$:
\begin{equation} \label{eq52}
\mathscr P(\gamma)\equiv a_0\gamma^6+a_1\gamma^5+a_2\gamma^4+a_3\gamma^3+a_4\gamma^2+a_5\gamma+a_6=0,
\end{equation}
\begin{figure}
  \centering
	\includegraphics[width=5.5 cm, height=5.5 cm]{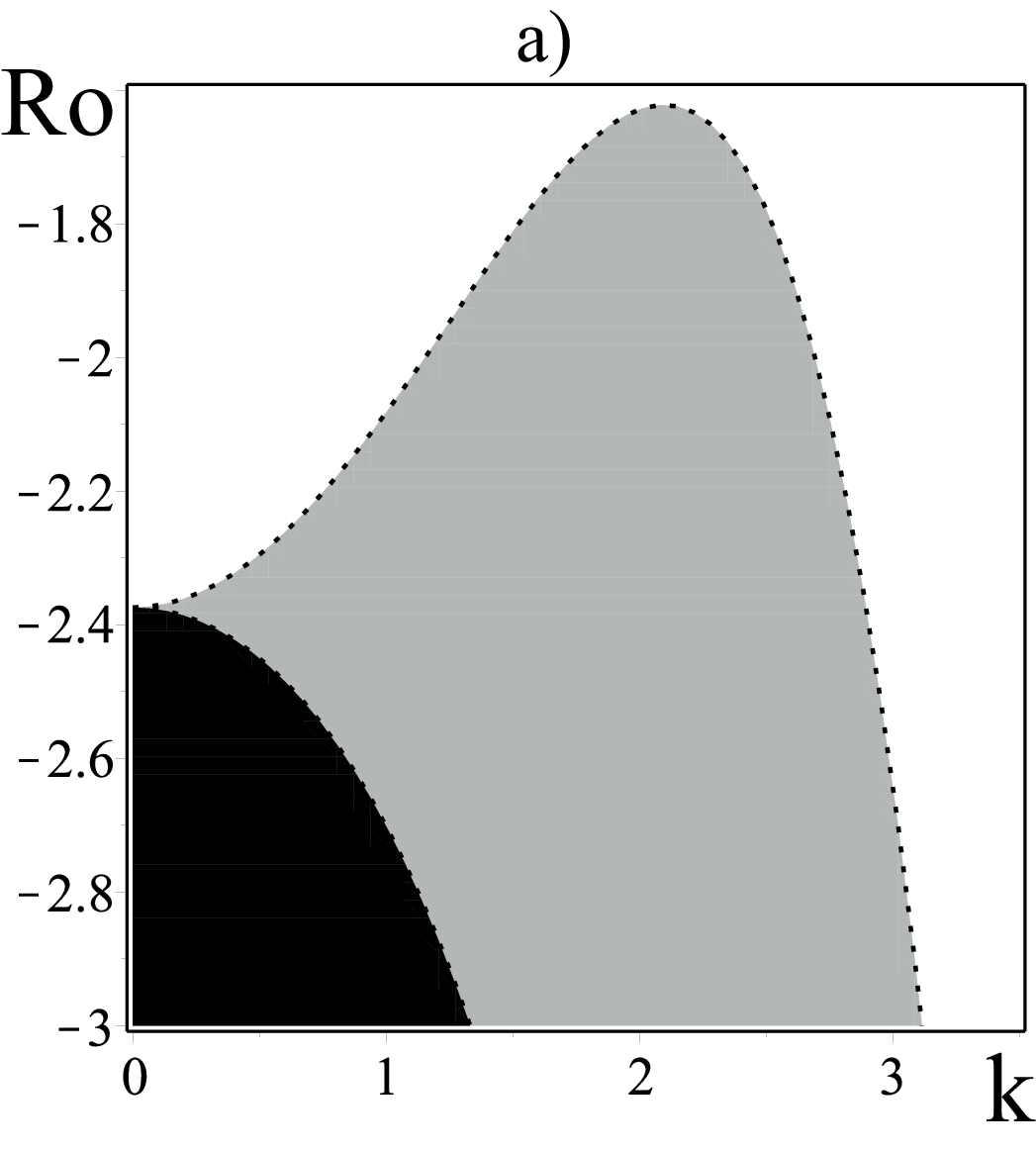}
	\includegraphics[width=5.5 cm, height=5.5 cm]{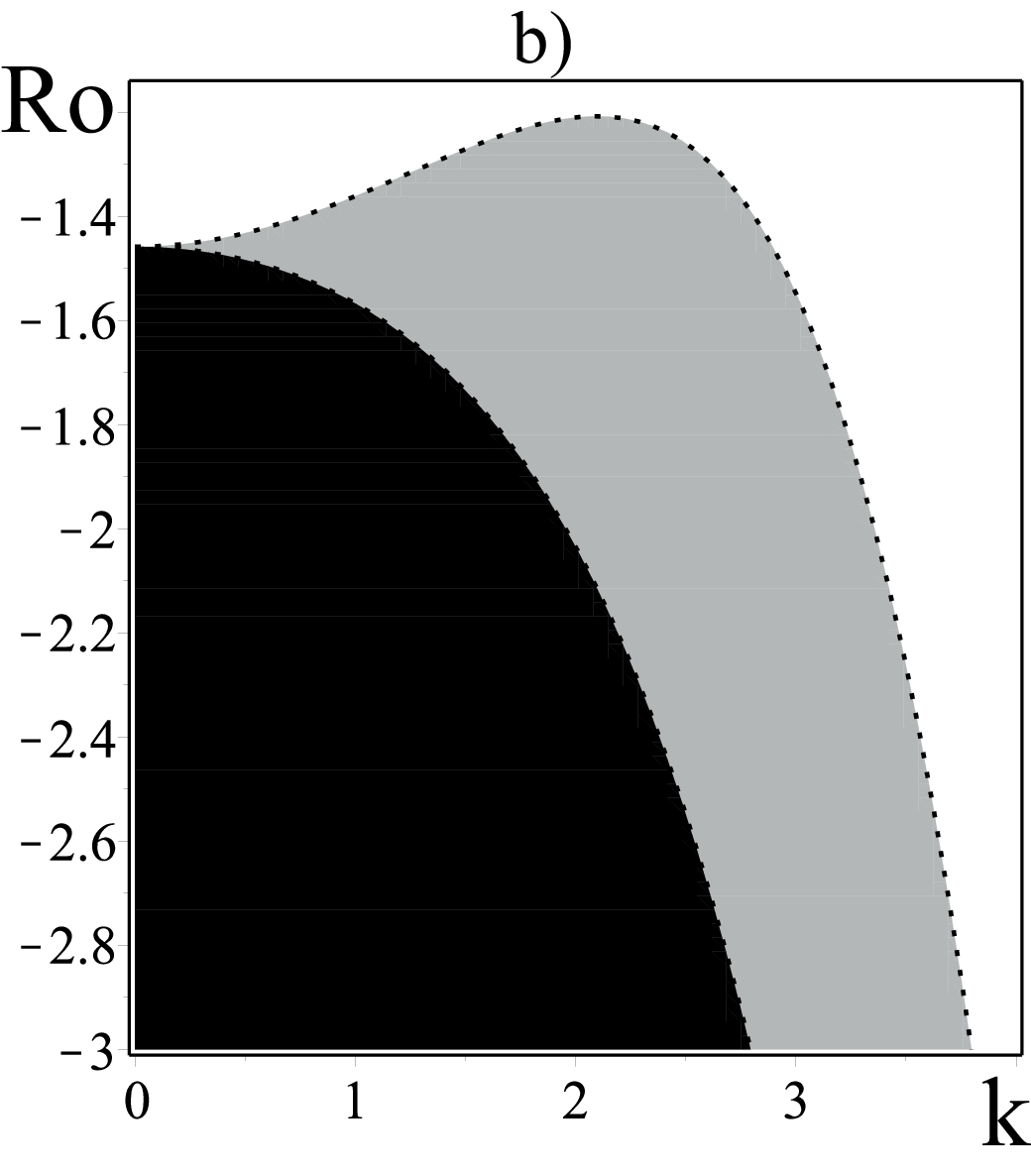}
	\includegraphics[width=5.5 cm, height=5.5 cm]{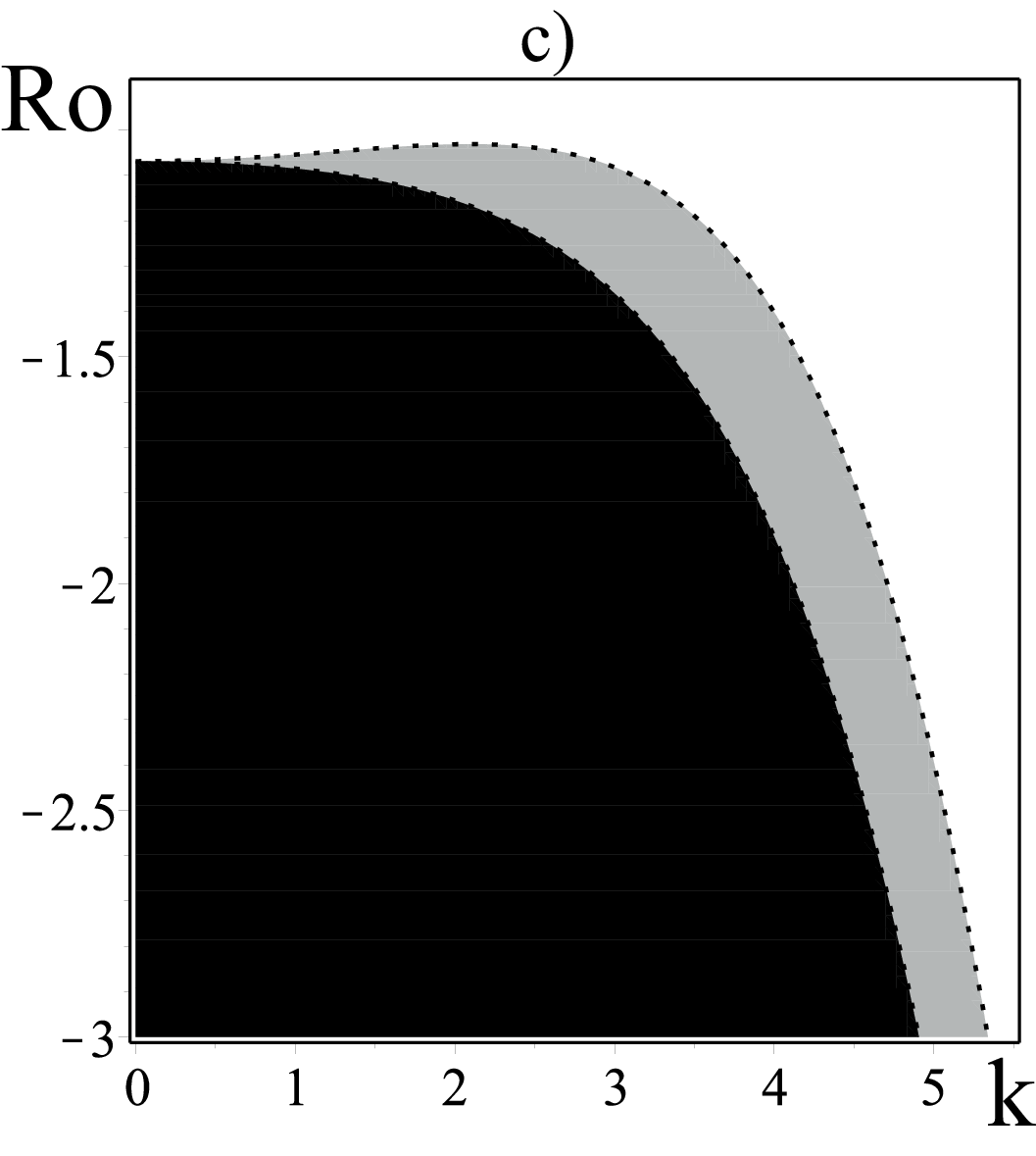}\\
\caption{\small The region of azimuthal MRI in a "pure" fluid is shown in black, and in a nanofluid -- in gray. The plots are constructed for the following Taylor numbers: a) $\textrm{Ta}=100$, b) $\textrm{Ta}=300$, c) $\textrm{Ta}=2000$, with fixed parameters $\textrm{Q}_{\varphi}=10$, $\textrm{Rb}=-1$, $\textrm{R}_n=0.122$, and $L_e=5000$.}\label{fg6}
\end{figure}	
where the coefficients $a_j$ $(j=0 \ldots 6)$ are given by
\[ a_0={\Pr}^2\textrm{Pm}^2L_ea^2,\; a_1={\Pr}(1+L_e)a^4\textrm{Pm}^2, a_2=a^6[\textrm{Pm}^2+2\textrm{Pm}(1+\textrm{Pm})\Pr(1+L_e)+\]
\[+{\Pr}^2L_e(1+\textrm{Pm}^2)+4\textrm{Pm}{\Pr}^2L_e] +\pi^2\textrm{Pm}^2{\Pr}^2\textrm{Ta}(1+\textrm{Ro})L_e-k^2\textrm{R}_nL_e\Pr\textrm{Pm}^2-4\pi^2\textrm{Q}_{\varphi}\textrm{Rb}\textrm{Pm}\textrm{Pr}^2L_e, \]
\[a_3=a^8[2\textrm{Pm}(1+\textrm{Pm})+\Pr(1+L_e)(1+4\textrm{Pm}+\textrm{Pm}^2)]+\pi^2a^2\textrm{Ta}(1+\textrm{Ro})(2\textrm{Pm}{\Pr}^2L_e+\textrm{Pm}^2\Pr(1+L_e))-\]
\[-k^2a^2\textrm{R}_nL_e(\textrm{Pm}^2+\textrm{Pm}\Pr(2+\textrm{Pm}))-4\pi^2a^2\textrm{Q}_{\varphi}\textrm{Rb}({\Pr}^2L_e(1+\textrm{Pm})+\textrm{Pm}\textrm{Pr}(1+L_e)),   \]
\[a_4=a^{10}[2\Pr(1+\textrm{Pm})(1+L_e)+{\Pr}^2L_e]+\pi^2a^4\textrm{Ta}(1+\textrm{Ro})({\Pr}^2L_e+2\textrm{Pm}\textrm{Pr}(1+L_e)+\textrm{Pm}^2)-\]
\[-k^2a^4\textrm{R}_nL_e(\textrm{Pm}(2+\textrm{Pm})+\Pr(2\textrm{Pm}+1))-\] \[-4\pi^2a^4\textrm{Q}_{\varphi}\textrm{Rb}({\Pr}^2L_e+\textrm{Pr}(1+\textrm{Pm})(1+L_e)+\textrm{Pm}),\]
\[a_5=a^{12}[2(1+\textrm{Pm})+\Pr(1+L_e)]+\pi^2a^6\textrm{Ta}(1+\textrm{Ro})({\Pr}(1+L_e)+2\textrm{Pm})-\]
\[-k^2a^6\textrm{R}_nL_e(\textrm{Pr}+2\textrm{Pm}+1)-4\pi^2a^6\textrm{Q}_{\varphi}\textrm{Rb}({\Pr}(1+L_e)+1+\textrm{Pm}), \]
\[a_6=a^{14}+\pi^2a^8\textrm{Ta}(1+\textrm{Ro})-k^2a^8\textrm{R}_nL_e-4\pi^2a^8\textrm{Q}_{\varphi}\textrm{Rb} .\]
Here, $\textrm{Q}_{\varphi} = \mu_e H_{0\varphi}^2 h^4 / (4\pi \rho_0 R_0^2 \nu \eta)$ is the azimuthal Chandrasekhar number.
The reality of the coefficients $a_j$ in equation (\ref{eq52}) allows us to apply the Lienard-Chipart asymptotic stability criterion, from which the positivity of the coefficients $a_j > 0$ and the Hurwitz determinants $\Delta_3, \Delta_5 > 0$ follows. From the explicit form of the coefficients $a_j$, it is evident that non-uniform rotation with positive Rossby numbers ($\textrm{Ro} > 0$) has a stabilizing effect, while the concentration of nanoparticles (terms with the concentration Rayleigh number $\textrm{R}_n$) has a destabilizing influence. The azimuthal magnetic field $H_{0\varphi}$ exerts both stabilizing and destabilizing effects depending on the sign of the magnetic Rossby number $\textrm{Rb}$. The condition $a_6 > 0$ yields the following stability criterion:
\begin{equation} \label{eq53}
\textrm{Ro}>-1-\frac{a^6}{\pi^2\textrm{Ta}}+\frac{k^2\textrm{R}_nL_e}{\pi^2\textrm{Ta}}+\frac{4\textrm{Q}_{\varphi}\textrm{Rb}}{\textrm{Ta}}=\textrm{Ro}_{\textrm{cr}}  \end{equation}
\begin{figure}
  \centering
\includegraphics[width=5.5 cm, height=5.5 cm]{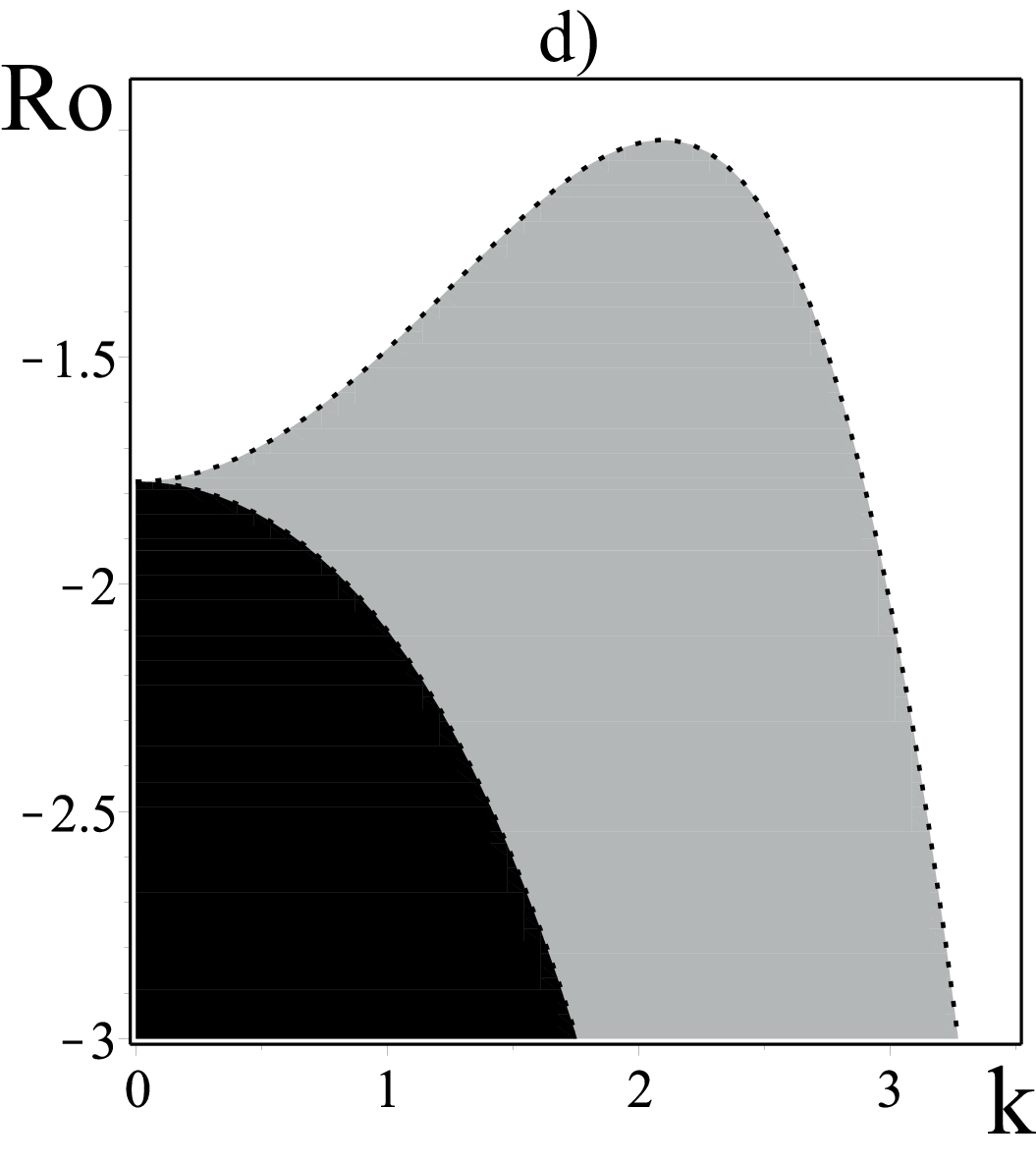}
	\includegraphics[width=5.5 cm, height=5.5 cm]{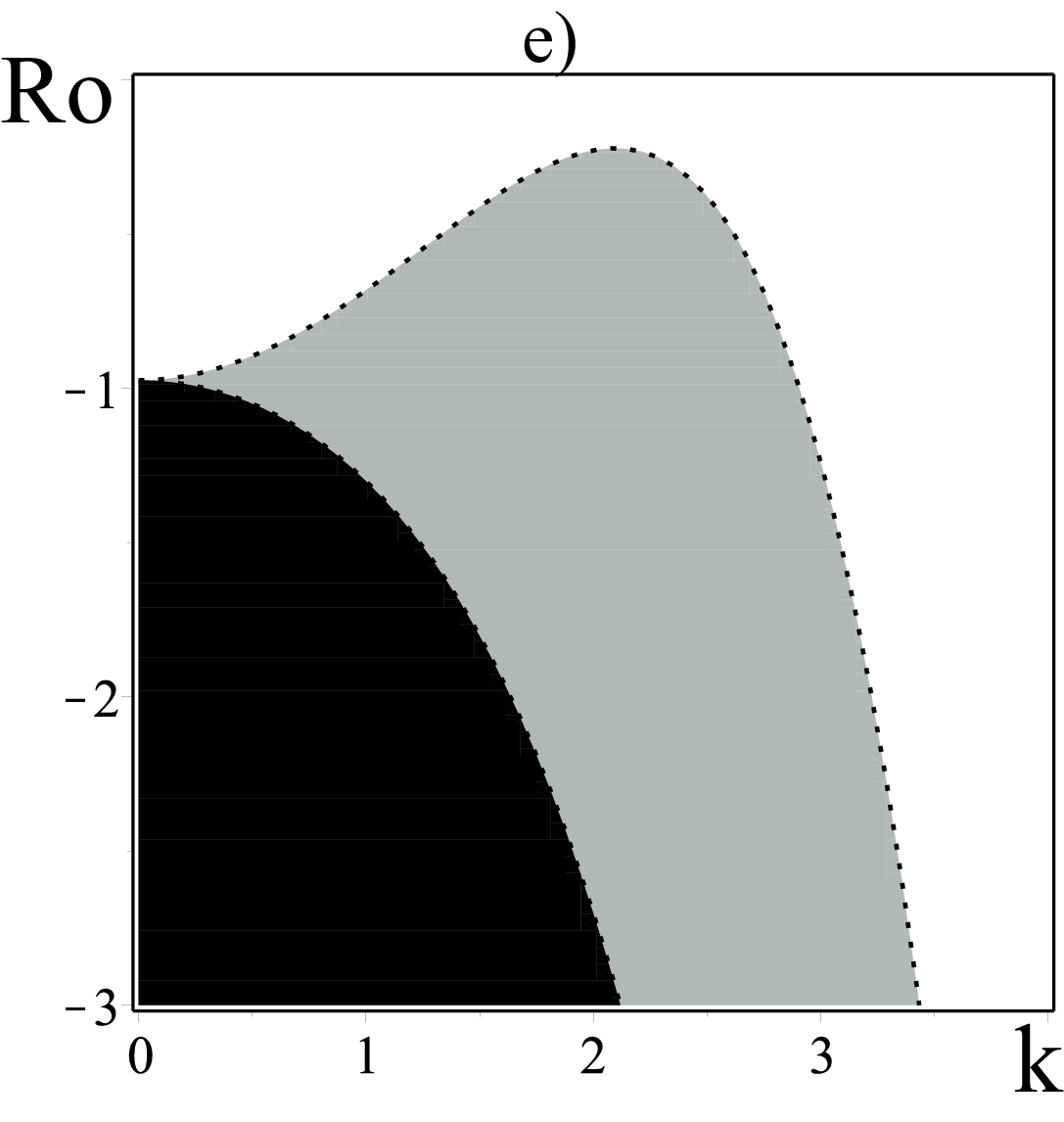}
	\includegraphics[width=5.5 cm, height=5.5 cm]{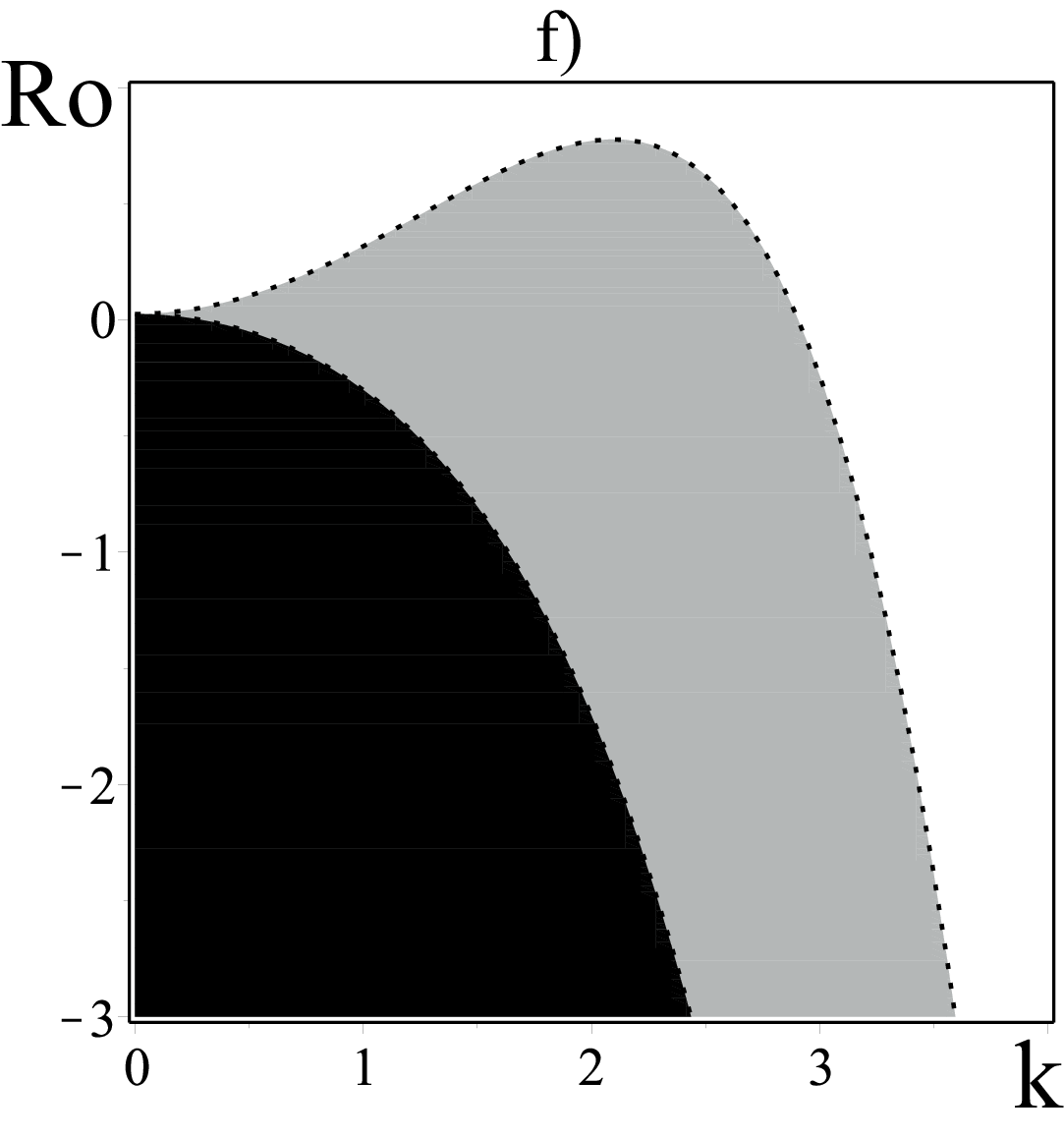}\\
	\caption{\small The black color shows the regions in which the azimuthal MRI arises in a pure fluid, and the gray color, in a
nanofluid. The graphs are plotted for azimuthal Chandrasekhar numbers: d) $\textrm{Q}_{\varphi}=10$, e) $\textrm{Q}_{\varphi}=50$, f) $\textrm{Q}_{\varphi}=100$ for fixed parameters $\textrm{Ta}=100$, $\textrm{Rb}=1/2$, $\textrm{R}_n=0.122, L_e=5000$.}\label{fg7}
\end{figure}
or, in dimensional variables,
\[\textrm{Ro}>-1-\frac{\omega_\nu^2}{4\alpha^2\Omega^2}+\textrm{Rb}\frac{\omega_{A\varphi}^2}{\Omega^2}\frac{\omega_\nu}{\omega_\eta}+(1-\alpha^2)\frac{\textrm{R}_nL_e\omega_\nu^2}{4\alpha^2\Omega^2(|{\bf k}|h)^4}. \]
For the case of a "pure" electrically conducting fluid, the stability criterion (\ref{eq53}) at $\textrm{R}_n = 0$ reduces to a simpler criterion obtained in \cite{26s}. In the absence of the azimuthal Alfven frequency ($\omega_{A\varphi} = 0$) and for $\textrm{R}_n = 0$, the stability criterion (\ref{eq53}) agrees with the result of \cite{24s}.
It is evident that a rotating flow of an ideal nanofluid under the condition $\omega_\nu = \omega_\eta$ and in the limit $\omega_\nu \rightarrow 0$, subjected to an azimuthal magnetic field, is stable with respect to axisymmetric perturbations if the following inequality is satisfied:
\begin{equation} \label{eq54} \textrm{Ro}>-1+\textrm{Rb}\frac{\omega_{A\varphi}^2}{\Omega^2} \end{equation} 
The flow of an ideal "pure" fluid was considered in \cite{18s}, where the kinetic and magnetic energies are equal to each other:
\[
\frac{\rho_0(\Omega R)^2}{2} = \frac{H_{0\varphi}^2}{4\pi}
\]
In \cite{27s}, an exact stationary solution of ideal magnetohydrodynamics was found:
\[
\Omega = \frac{H_{0\varphi}}{R\sqrt{2\pi\rho_0}}, \quad P = \textrm{const}
\]
which is referred to as Chandrasekhar's equipartition. It was also proven there that this flow is marginally stable.
From Chandrasekhar's equipartition, it follows that 
\begin{equation} \label{eq55}
\omega_{A\varphi} = \frac{\mu_e H_{0\varphi}^2}{4\pi\rho_0 R_0^2} = \Omega, \quad \textrm{Ro} = \textrm{Rb} = -1
\end{equation}
The particular case of Chandrasekhar's equipartition (\ref{eq55}) satisfies the inequality (\ref{eq54}). Therefore, the stability criteria for ideal nanofluids and "pure" fluids coincide with (\ref{eq54}).

We now proceed to investigate the development of azimuthal MRI in a nanofluid under axisymmetric perturbations for Rossby numbers $\textrm{Ro} < \textrm{Ro}_{\textrm{cr}}$. To this end, we numerically 
\begin{figure}
  \centering
  \includegraphics[width=5.5 cm, height=5.0 cm]{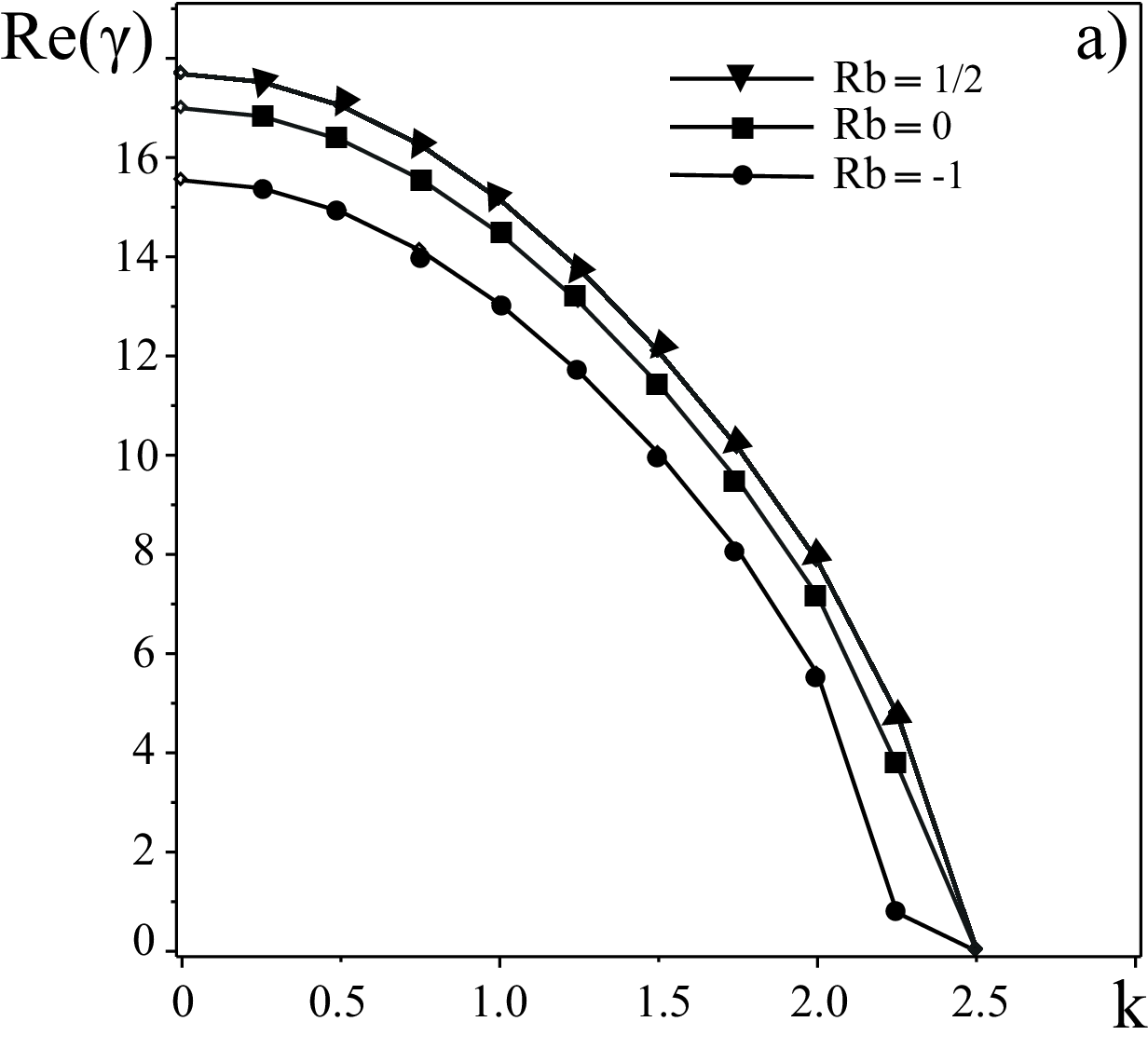}
	\includegraphics[width=5.5 cm, height=5.0 cm]{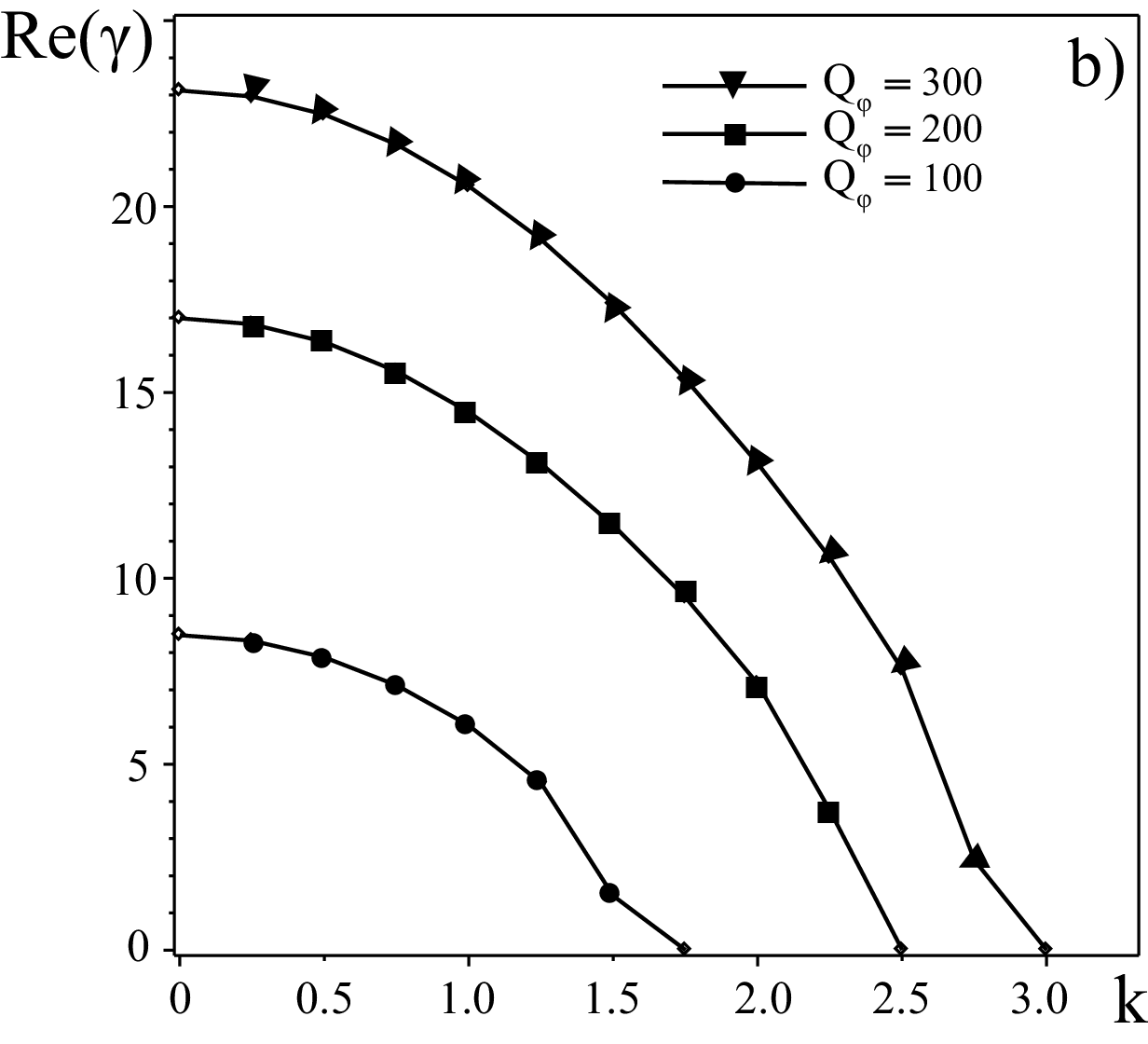}
	\includegraphics[width=5.5 cm, height=5.0 cm]{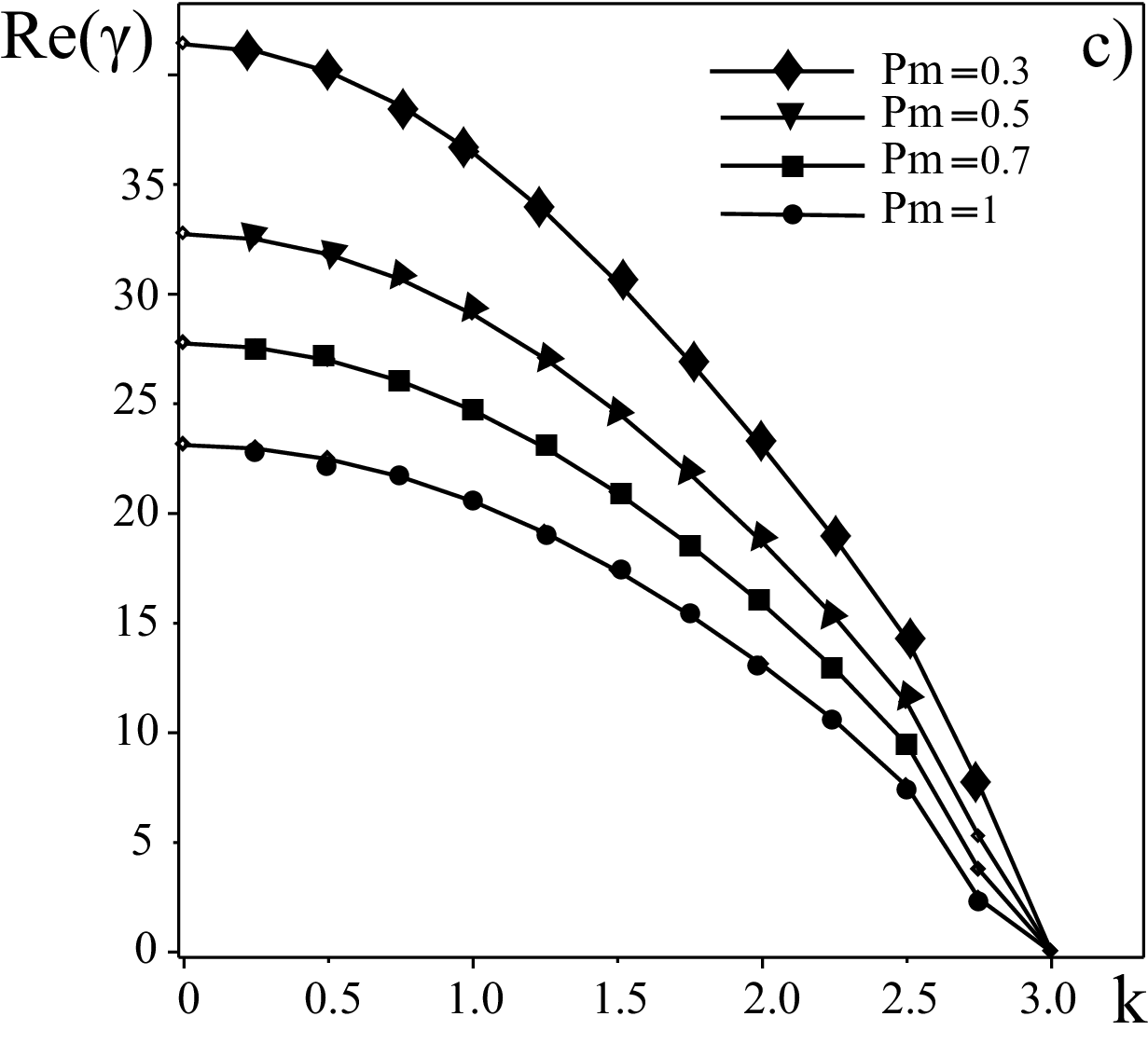} \\
\caption{\small Dependence of the growth rate ($\textrm{Re}(\gamma)>0$) of the AMRI in a nanofluid on the radial wavenumber $\textrm{k}$. Panels (a), (b), and (c) respectively illustrate the effects of: inhomogeneous azimuthal magnetic field for $\textrm{Rb} = -1, 0, 1/2$; varying strength of the azimuthal magnetic field for $\textrm{Q}_{\varphi} = 100, 200, 300$; and magnetic Prandtl number for $\textrm{Pm} = 0.3, 0.5, 0.7, 1$ on the development of AMRI.}\label{fg8}
\end{figure}
construct instability regions for various values of the rotation parameter $\textrm{Ta}$ (Taylor number), the azimuthal magnetic field $\textrm{Q}_{\varphi}$ (azimuthal Chandrasekhar number), and the magnetic Rossby number $\textrm{Rb}$.
Figs.~\ref{fg6}a-\ref{fg6}c show the instability domains of the azimuthal MRI in the $(\textrm{k},\textrm{Ro})$ plane for different Taylor numbers $\textrm{Ta} = 100, 300, 2000$, with fixed parameters $\textrm{Rb} = -1$, $\textrm{Q}_{\varphi} = 10$, $\textrm{R}_n = 0.122$, and $L_e = 5000$. It is evident from Fig.~\ref{fg6} that the presence of nanoparticles enhances the instability region compared to the case of a "pure" electrically conducting fluid.
Figs.~\ref{fg7}d-\ref{fg7}f present the instability domains for the azimuthal MRI in the presence of a positively sheared magnetic field profile ($\textrm{Rb} = 1/2$) in the $(\textrm{k},\textrm{Ro})$ plane for different azimuthal Chandrasekhar numbers $\textrm{Q}_{\varphi} = 10, 50, 100$, while keeping the other parameters fixed: $\textrm{Ta} = 100$, $\textrm{R}_n = 0.122$, and $L_e = 5000$. As seen in Fig.~\ref{fg7}, the nanoparticle concentration also promotes a broader instability region compared to the "pure" fluid case. Moreover, increasing the azimuthal magnetic field strength (i.e., $\textrm{Q}_{\varphi}$) shifts the instability boundary toward positive Rossby numbers ($\textrm{Ro} > 0$).

To investigate the development of azimuthal magnetorotational instability (AMRI) in nanofluids under axisymmetric perturbations for Rossby numbers $\textrm{Ro} < \textrm{Ro}_{\textrm{cr}}$, we perform a numerical analysis of the dispersion relation (\ref{eq52}). In particular, we compute the growth rate $\textrm{Re}(\gamma) > 0$ as a function of the radial wavenumber $\textrm{k}$ for representative nanofluid parameters.
Fig. \ref{fg8}a shows the influence of different radial profiles of the azimuthal magnetic field characterized by magnetic Rossby numbers $\textrm{Rb} = -1, 0, 1/2$ on the AMRI growth rate for fixed parameters: $\textrm{Ta} = 2000$, $\textrm{Ro} = -1.2$, $\textrm{Q}_{\varphi} = 10$, and $\textrm{Pm} = 1$. The results demonstrate that the instability grows faster for positive magnetic Rossby numbers $\textrm{Rb} > 0$, indicating a destabilizing effect of outwardly increasing azimuthal magnetic fields.
We next assess the impact of increasing azimuthal magnetic field strength, quantified by Chandrasekhar numbers $\textrm{Q}_{\varphi} = 100, 200, 300$, on the AMRI growth rate for $\textrm{Ta} = 300$, $\textrm{Ro} = -1$, $\textrm{Rb} = 1/2$, and $\textrm{Pm} = 1$. As shown in Figure \ref{fg8}b, stronger azimuthal fields lead to higher growth rates, thus promoting the onset of instability.
Finally, Fig. \ref{fg8}c presents the variation of the AMRI growth rate $\textrm{Re} \gamma(\textrm{k})$ with the magnetic Prandtl number $\textrm{Pm} = 0.3, 0.5, 0.7, 1$, at fixed $\textrm{Q}_{\varphi} = 300$, $\textrm{Ta} = 300$, $\textrm{Ro} = -1$, and $\textrm{Rb} = 1/2$. The results reveal that lower values of $\textrm{Pm}$ (i.e., $\textrm{Pm} < 1$) significantly enhance the instability growth rate, suggesting that nanofluids with low magnetic diffusivity are more susceptible to the development of AMRI.

Thus, AMRI in nanofluids is triggered by an increasing azimuthal magnetic field of the form \( H_{0\varphi} = C R^{\alpha} \) with a positive profile (\( \alpha > 1 \), \( \textrm{Rb} > 0 \)) for magnetic Prandtl numbers \( \textrm{Pm} \leq 1 \).

\section{Helical MRI in thin layers of nanofluid}

In the case of $\textrm{Ra} = N_A = 0$, equation (\ref{eq45}) yields the dispersion relation for the helical MRI in thin layers of nanofluid:
\begin{equation} \label{eq56}
\mathscr P(\gamma)\equiv a_0\gamma^7+a_1\gamma^6+a_2\gamma^5+a_3\gamma^4+a_4\gamma^3+a_5\gamma^2+a_6\gamma+a_7=0,
\end{equation}
where the coefficients $a_j$ $(j=0 \ldots 7)$ are given by
\[a_0=A_0,\; a_1=A_1,\; a_2=A_2-k^2\textrm{R}_nL_e\Pr\textrm{Pm}^3,\; a_3=A_3-k^2a^2\textrm{R}_nL_e(\textrm{Pm}^3+\Pr\textrm{Pm}^2(3+\textrm{Pm})),\]
\[ a_4=A_4-C_0-k^2\textrm{R}_nL_e(\Pr\textrm{Pm}^2(a^4+\pi^2\textrm{Q})+a^4\textrm{Pm}(\Pr+\textrm{Pm})(3+\textrm{Pm})),\]
\[ a_5=A_5-C_1-k^2\textrm{R}_nL_e(\Pr(2a^2\textrm{Pm}(a^4+\pi^2\textrm{Q})+a^6(1+\textrm{Pm}))+\]
\[+a^2\textrm{Pm}^2(a^4+\pi^2\textrm{Q})+a^6\textrm{Pm}(3+\textrm{Pm})),\]
\[ a_6=A_6-C_2-k^2\textrm{R}_nL_e(2a^4\textrm{Pm}(a^4+\pi^2\textrm{Q})+a^8(1+\textrm{Pm})+a^4(a^4+\pi^2\textrm{Q})\Pr),\]
\[a_7=A_7-C_3-k^2a^6\textrm{R}_nL_e(a^4+\pi^2\textrm{Q}).\]
Here we introduced the following notation for $A_n$ $(n=0,\ldots 7)$  and $C_m$ $(m=0,\ldots 3)$:
\[A_0=a^2{\Pr}^2\textrm{Pm}^3L_e,\; A_1=a^4\textrm{Pm}^2\Pr[2\Pr L_e(1+\textrm{Pm})+\Pr L_e+\textrm{Pm}(1+L_e)], \]
\[A_2=a^6[\textrm{Pm}^2(\textrm{Pm}+\textrm{Pr}(1+L_e))+2\textrm{Pm}(1+\textrm{Pm})\Pr(\Pr L_e+\textrm{Pm}(1+L_e))+\]
\[+(a^6(1+4\textrm{Pm}+\textrm{Pm}^2)+2a^2\textrm{Pm}\pi^2\textrm{Q}+\pi^2\textrm{Pm}^2\textrm{Ta}(1+\textrm{Ro})-4\pi^2\textrm{Q}\xi^2\textrm{RbPm})\textrm{Pm}{\Pr}^2L_e, \]
\[A_3=a^8[\textrm{Pm}^2+2\textrm{Pm}(1+\textrm{Pm})(\textrm{Pm}+\Pr(1+L_e))]+a^2\Pr(a^6(1+4\textrm{Pm}+\textrm{Pm}^2)+2a^2\textrm{Pm}\pi^2\textrm{Q}+\]
\[+\pi^2\textrm{Pm}^2\textrm{Ta}(1+\textrm{Ro})-4\pi^2\textrm{Q}\xi^2\textrm{RbPm})(\Pr L_e+\textrm{Pm}(1+L_e))+(2a^4(1+\textrm{Pm})(a^4+\pi^2\textrm{Q})+\]
\[+2a^2\pi^2\textrm{Ta}(1+\textrm{Ro})\textrm{Pm}-4\pi^2a^2\textrm{Q}\xi^2\textrm{Rb}(1+\textrm{Pm}))\textrm{Pm}{\Pr}^2L_e, \]
\[A_4=2a^{10}\textrm{Pm}(1+\textrm{Pm})+a^4(a^6(1+4\textrm{Pm}+\textrm{Pm}^2)+2a^2\textrm{Pm}\pi^2\textrm{Q}+\pi^2\textrm{Pm}^2\textrm{Ta}(1+\textrm{Ro})-4\pi^2\textrm{Q}\xi^2\textrm{RbPm})\times\]
\[\times(\textrm{Pm}+\textrm{Pr}(1+L_e))+a^2\Pr(2a^4(1+\textrm{Pm})(a^4+\pi^2\textrm{Q})+2a^2\pi^2\textrm{Ta}(1+\textrm{Ro})\textrm{Pm}-4\pi^2a^2\textrm{Q}\xi^2\textrm{Rb}(1+\textrm{Pm}))\times\]
\[\times(\Pr L_e+\textrm{Pm}(1+L_e))+(a^2(a^4+\pi^2\textrm{Q})^2+\pi^2a^4\textrm{Ta}(1+\textrm{Ro})+\pi^4\textrm{PmRoTaQ}-4\pi^2\textrm{Q}\xi^2\textrm{Rb}(a^4+\pi^2\textrm{Q})-\]
\[-4\pi^4\textrm{Q}^2\xi^2)\textrm{Pm}{\Pr}^2L_e,  \]
\[A_5=a^6(a^6(1+4\textrm{Pm}+\textrm{Pm}^2)+2a^2\textrm{Pm}\pi^2\textrm{Q}+\pi^2\textrm{Pm}^2\textrm{Ta}(1+\textrm{Ro})-4\pi^2\textrm{Q}\xi^2\textrm{RbPm})+\]
\[+a^4(2a^4(1+\textrm{Pm})(a^4+\pi^2\textrm{Q})+2a^2\pi^2\textrm{Ta}(1+\textrm{Ro})\textrm{Pm}-4\pi^2a^2\textrm{Q}\xi^2\textrm{Rb}(1+\textrm{Pm}))(\textrm{Pm}+\textrm{Pr}(1+L_e))+\]
\[+a^2\Pr(a^2(a^4+\pi^2\textrm{Q})^2+\pi^2a^4\textrm{Ta}(1+\textrm{Ro})+\pi^4\textrm{PmRoTaQ}-4\pi^2\textrm{Q}\xi^2\textrm{Rb}(a^4+\pi^2\textrm{Q})-\]
\[-4\pi^4\textrm{Q}^2\xi^2)(\Pr L_e+\textrm{Pm}(1+L_e)),\]
\[A_6=a^6(2a^4(1+\textrm{Pm})(a^4+\pi^2\textrm{Q})+2a^2\pi^2\textrm{Ta}(1+\textrm{Ro})\textrm{Pm}-4\pi^2a^2\textrm{Q}\xi^2\textrm{Rb}(1+\textrm{Pm}))+a^4(\textrm{Pm}+\textrm{Pr}(1+L_e))\times \]
\begin{figure}
  \centering
  \includegraphics[width=5.5 cm, height=5.5 cm]{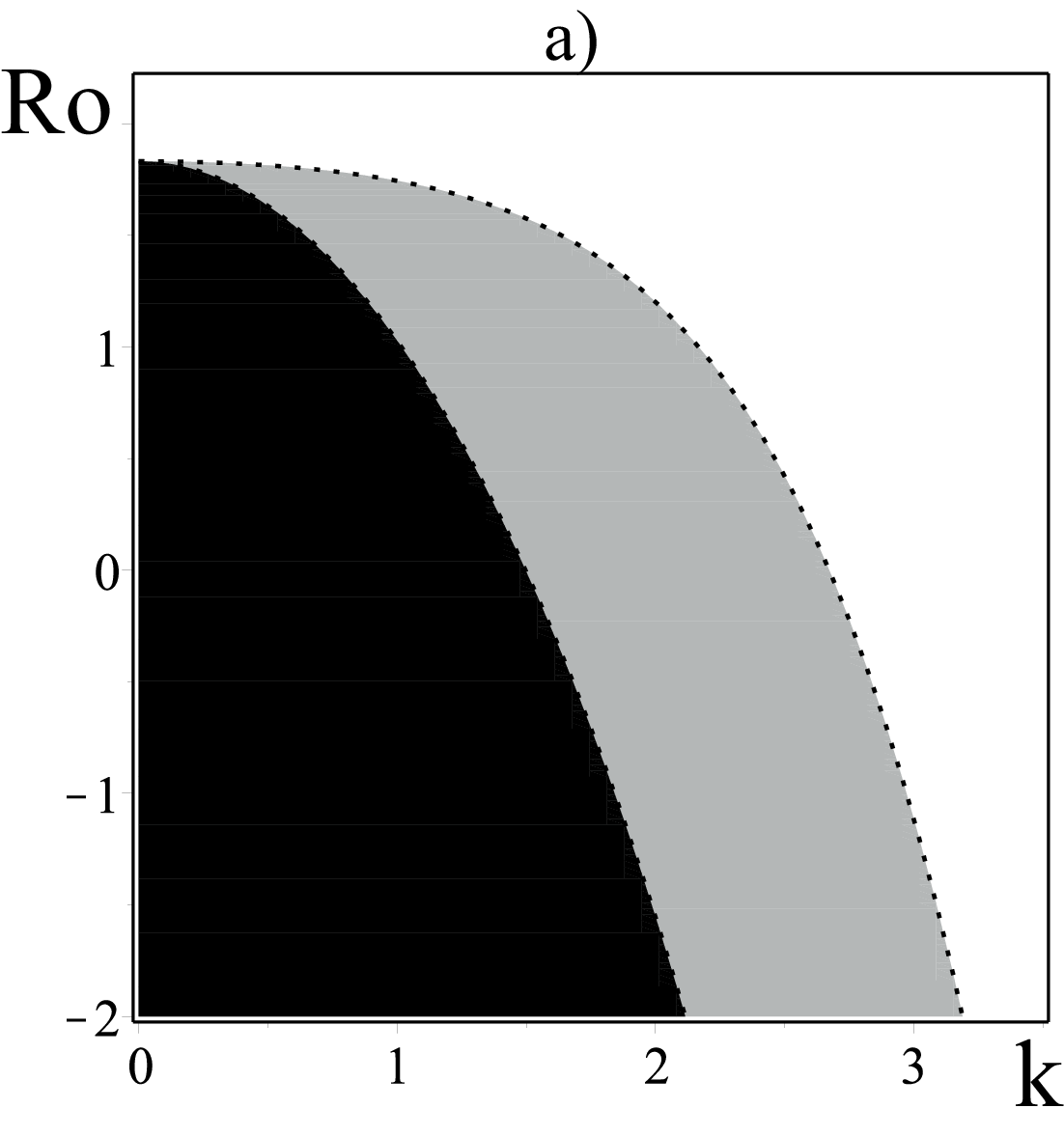}
	\includegraphics[width=5.5 cm, height=5.5 cm]{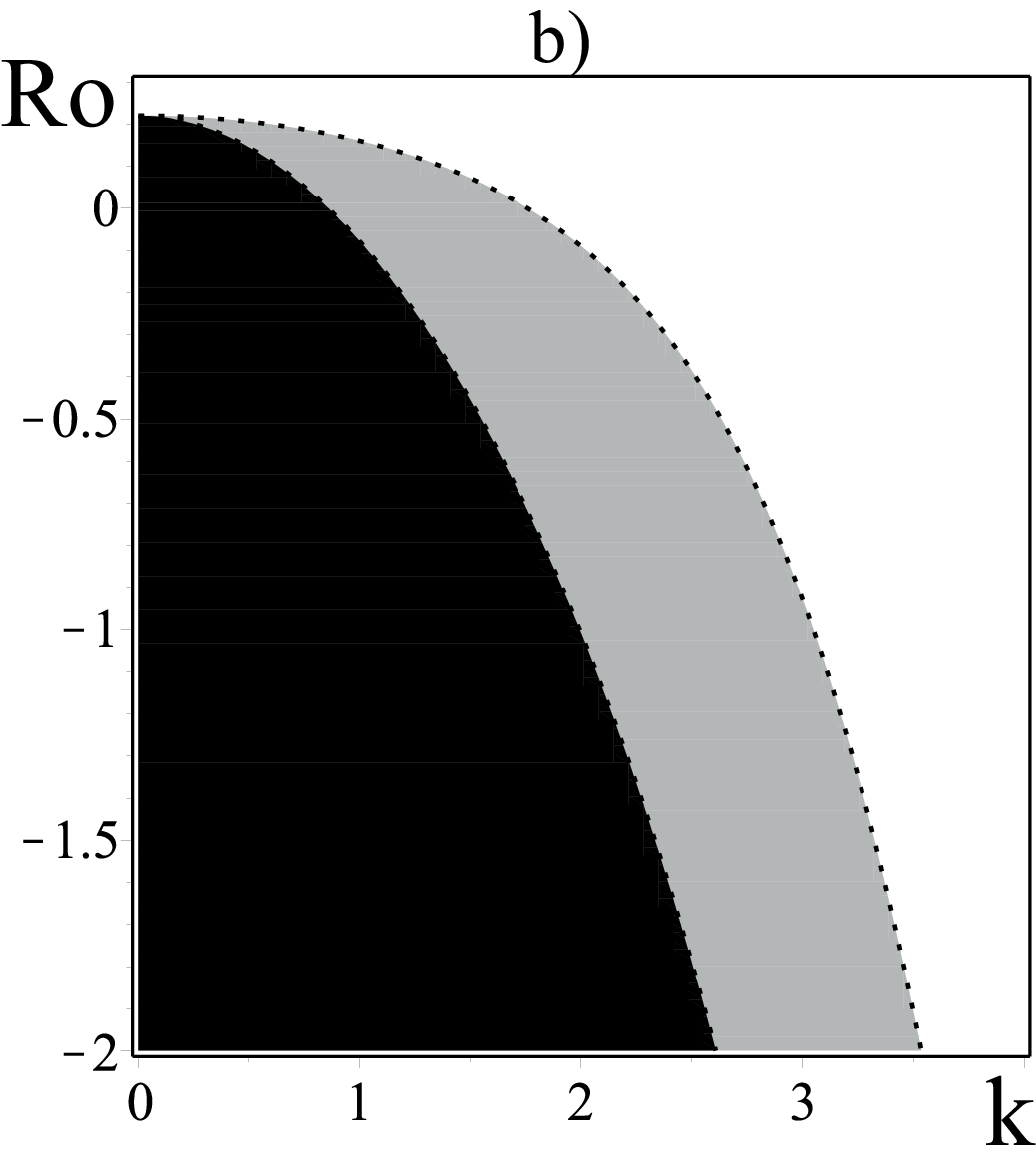}
	\includegraphics[width=5.5 cm, height=5.5 cm]{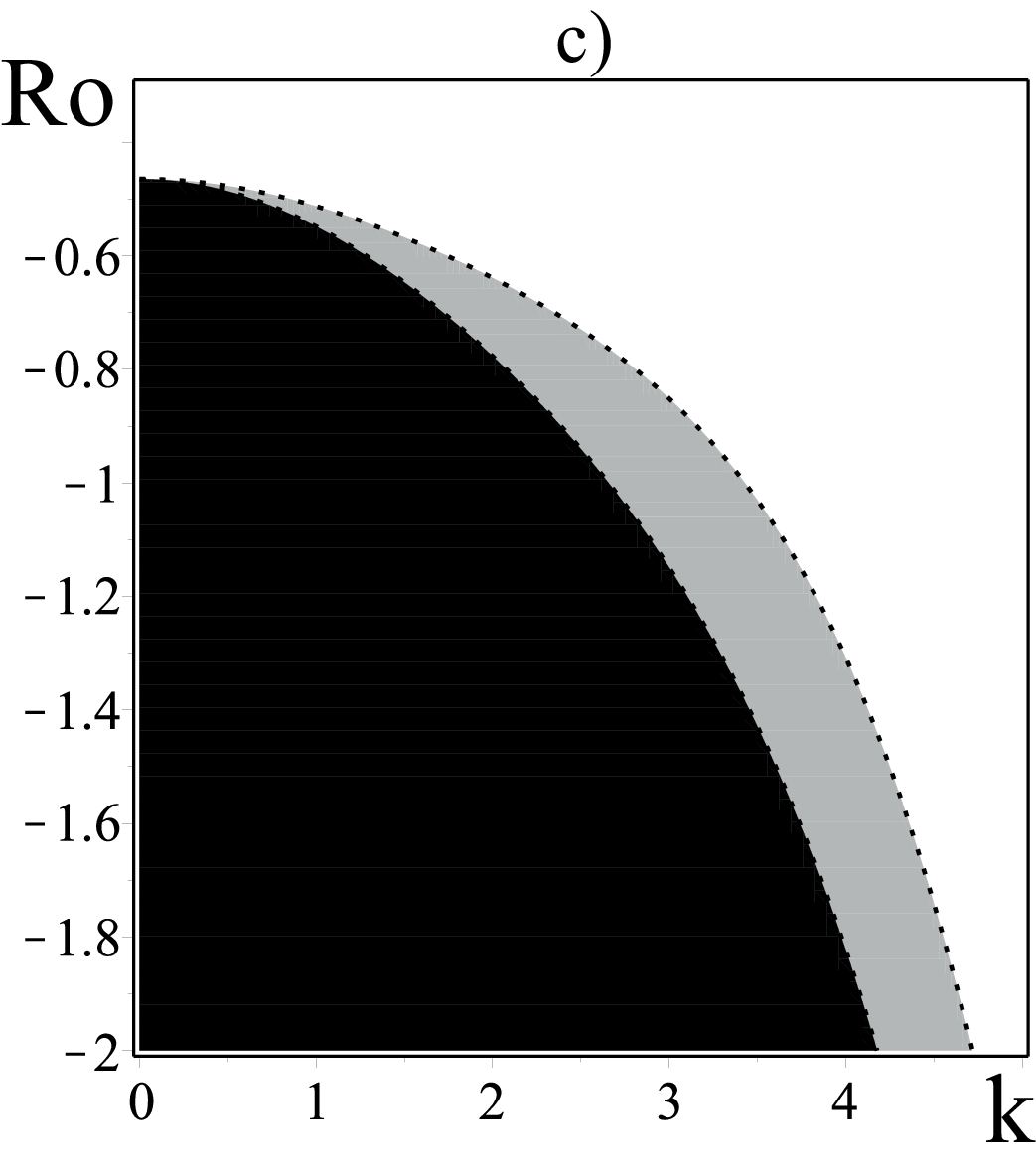}\\
	\caption{\small The black region indicates the domain where helical MRI arises in a "pure" fluid, while the gray region corresponds to the case of a nanofluid. The plots are constructed for the following Taylor numbers: a) $\textrm{Ta}=100$, b) $\textrm{Ta}=300$, c) $\textrm{Ta}=2000$, with fixed parameters $\textrm{Q}=10$, $\textrm{Q}_{\varphi}=100$, $\textrm{Pm}=0.7$, $\textrm{Rb}=1/2$, $\textrm{R}_n=0.122$, $L_e=5000$, $N_B=7.5 \cdot 10^{-4}$.}\label{fg9}
\end{figure}
\[\times(a^2(a^4+\pi^2\textrm{Q})^2+\pi^2a^4\textrm{Ta}(1+\textrm{Ro})+\pi^4\textrm{PmRoTaQ}-4\pi^2\textrm{Q}\xi^2\textrm{Rb}(a^4+\pi^2\textrm{Q})-4\pi^4\textrm{Q}^2\xi^2),\]
\[A_7=a^6(a^2(a^4+\pi^2\textrm{Q})^2+\pi^2a^4\textrm{Ta}(1+\textrm{Ro})+\pi^4\textrm{PmRoTaQ}-4\pi^2\textrm{Q}\xi^2\textrm{Rb}(a^4+\pi^2\textrm{Q})-4\pi^4\textrm{Q}^2\xi^2),\]
\[C_0=4\pi^4\textrm{Q}\xi\sqrt{\textrm{Ta}}N_B\Pr\textrm{Pm}^2,\]
\[ C_1=2\pi^4a^2\textrm{QPm}\xi\sqrt{\textrm{Ta}}\left[N_B\Pr(4+\textrm{Ro}(1-\textrm{Pm}))+\frac{2N_B}{L_e}\textrm{Pm}\right],\]
\[C_2=2\pi^4a^4\textrm{Q}\xi\sqrt{\textrm{Ta}}\left[N_B\Pr(2+\textrm{Ro}(1-\textrm{Pm}))+\frac{N_B}{L_e}\textrm{Pm}(4+\textrm{Ro}(1-\textrm{Pm}))\right],\]
\[C_3=2\pi^4a^6\textrm{Q}\xi\sqrt{\textrm{Ta}}\frac{N_B}{L_e}(2+\textrm{Ro}(1-\textrm{Pm})).\]
In the limiting cases, the dispersion relation (\ref{eq56}) reduces to that of the standard MRI (\ref{eq46}) when $H_{0\varphi} = 0$, and to the azimuthal MRI dispersion relation (\ref{eq53}) when $H_{0z} = 0$. 
To analyze the stability of the system, we apply the classical Lienard-Chipart criterion \cite{23s} to the dispersion relation (\ref{eq56}) with real coefficients $a_j$ $(j = 0, \ldots, 7)$. Based on the explicit expressions for these coefficients, we conclude that axisymmetric perturbations can be destabilized by differential rotation with a negative Rossby profile $(\textrm{Ro} < 0)$, a helical magnetic field with a positive azimuthal inhomogeneity profile $(\textrm{Rb} > 0)$, the presence of nanoparticles $(\textrm{R}_n \neq 0)$, as well as the combined effect of the helical magnetic field and nanoparticle generation $(N_B \neq 0)$, especially when the magnetic Prandtl number deviates from unity $(\textrm{Pm} \neq 1)$. 
Since the onset of instability occurs via the neutral point $\gamma = 0$, the necessary and sufficient condition for the stability of the rotating nanofluid with respect to axisymmetric perturbations is given by:
\begin{equation} \label{eq57}
\textrm{Ro}>\frac{-a^2(a^4+\pi^2\textrm{Q})^2-\pi^2a^4\textrm{Ta}+k^2(a^4+\pi^2\textrm{Q})\textrm{R}_nL_e}{\pi^2\textrm{Ta}(a^4+\pi^2\textrm{QPm})-2\pi^4\textrm{Q}\xi\sqrt{\textrm{Ta}}\frac{N_B}{L_e}(1-\textrm{Pm})}+$$
$$+\frac{4\textrm{Q}\xi^2(\textrm{Rb}(a^4+\pi^2\textrm{Q})+\pi^2\textrm{Q})+4\pi^2\textrm{Q}\xi\sqrt{\textrm{Ta}}\frac{N_B}{L_e}}{\textrm{Ta}(a^4+\pi^2\textrm{QPm})-2\pi^2\textrm{Q}\xi\sqrt{\textrm{Ta}}\frac{N_B}{L_e}(1-\textrm{Pm})}=\textrm{Ro}_{\textrm{cr}},\end{equation}
\begin{figure}
  \centering
  \includegraphics[width=5.5 cm, height=5.5 cm]{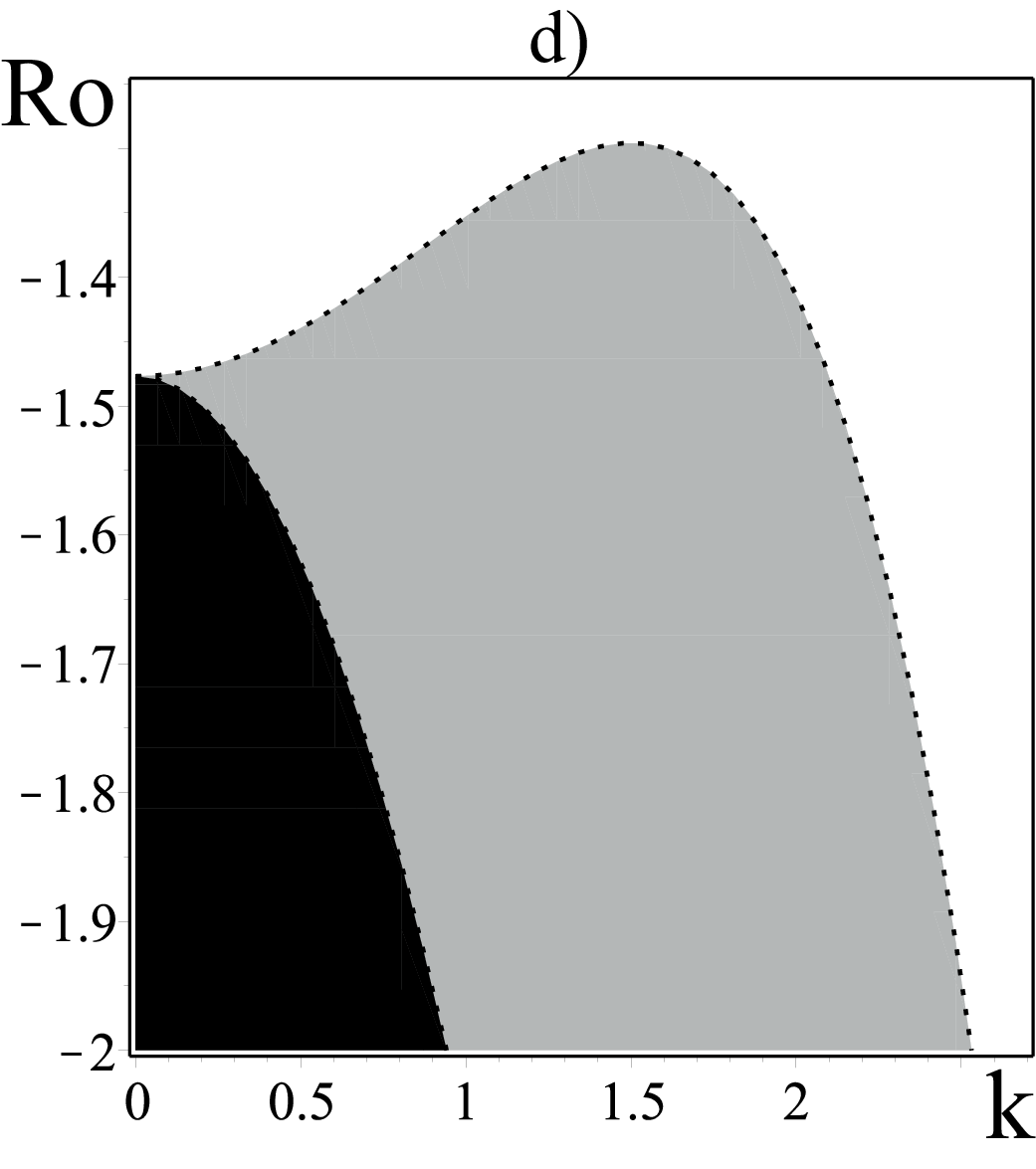}
	\includegraphics[width=5.5 cm, height=5.5 cm]{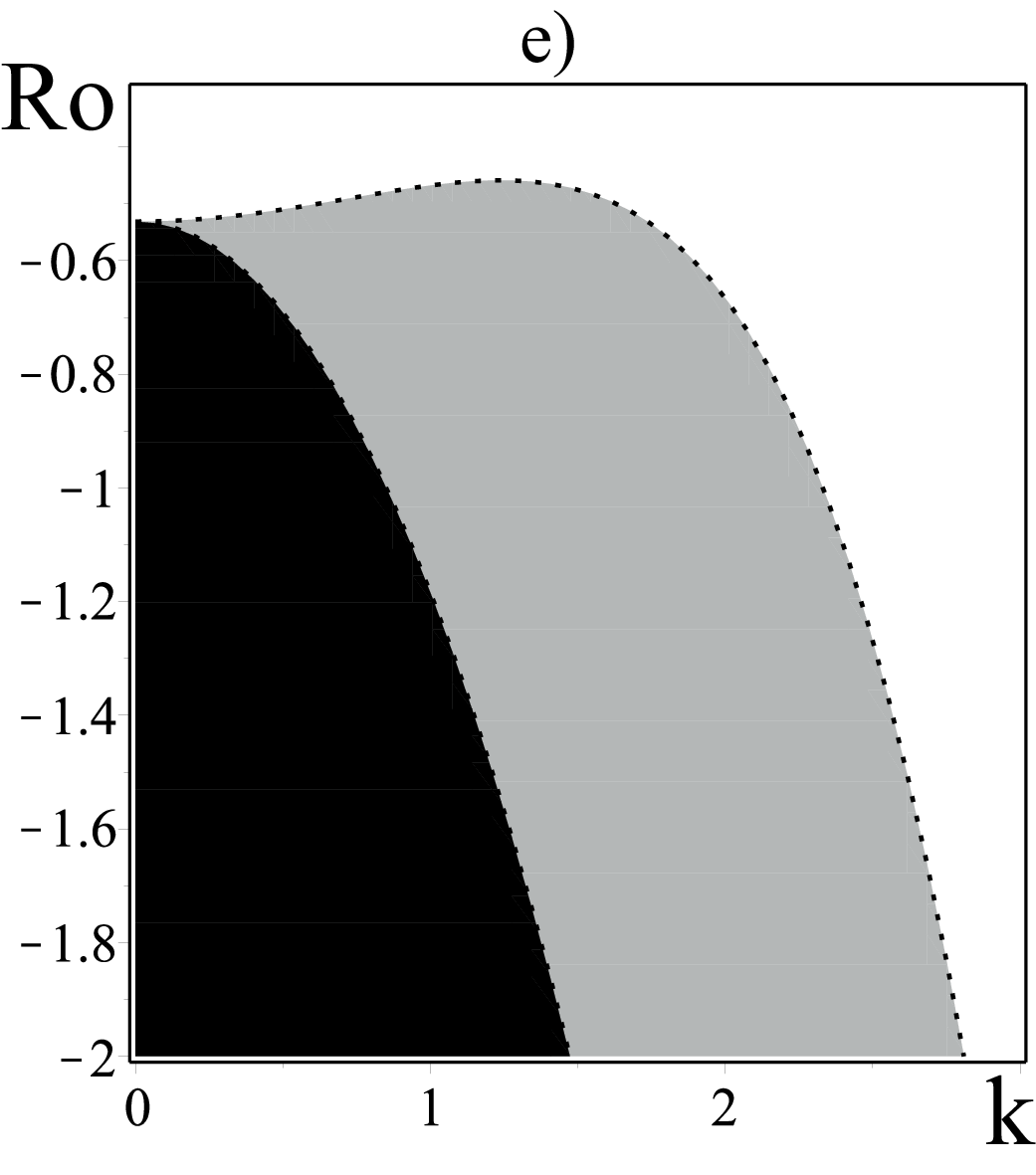}
	\includegraphics[width=5.5 cm, height=5.5 cm]{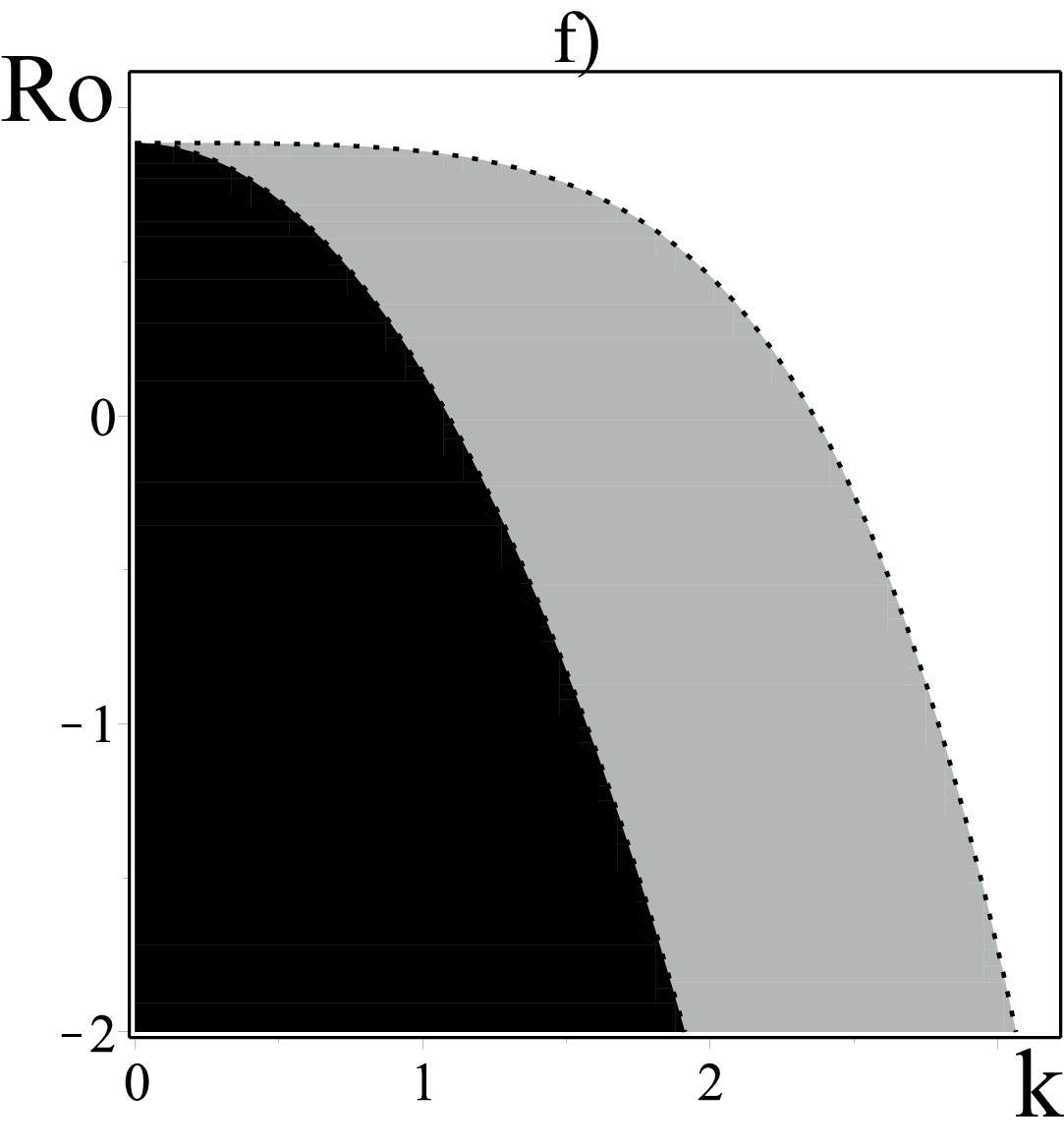}\\
	\caption{\small The black region indicates the domain where helical MRI arises in a "pure" fluid, while the gray region corresponds to the case of a nanofluid. The plots are constructed for azimuthal Chandrasekhar numbers: d) $\textrm{Q}_{\varphi}=30  $, e) $\textrm{Q}_{\varphi}=50 $, f) $\textrm{Q}_{\varphi}=80 $, with fixed parameters $\textrm{Q}=10, \textrm{Ta}=100, \textrm{Pm}=0.7, \textrm{Rb}=1/2, \textrm{R}_n=0.122, L_e=5000, N_B=7.5 \cdot 10^{-4} $.}\label{fg10}
\end{figure}
or, in dimensional variables,
\[\textrm{Ro}>\frac{-(\omega_A^2+\omega_\nu\omega_\eta)^2-4\alpha^2\Omega^2\omega_\eta^2+(1-\alpha^2)(\omega_A^2+\omega_\nu\omega_\eta)\textrm{R}_nL_e\frac{\omega_\nu\omega_\eta}{(|{\bf k}|h)^4}}{4\alpha^2\Omega^2(\omega_A^2+\omega_\eta^2)-4\alpha^3\Omega\omega_A\omega_{A\varphi}\frac{\omega_\eta}{(|{\bf k}|h)}\frac{N_B}{L_e}(1-\textrm{Pm})}+\]
\[+\frac{\omega_{A\varphi}^2(\textrm{Rb}(\omega_A^2+\omega_\nu\omega_\eta)+\omega_A^2)+2\alpha\Omega\omega_A\omega_{A\varphi}\frac{\omega_\eta}{(|{\bf k}|h)}\frac{N_B}{L_e}}{\Omega^2(\omega_A^2+\omega_\eta^2)-\alpha\Omega\omega_A\omega_{A\varphi}\frac{\omega_\eta}{(|{\bf k}|h)}\frac{N_B}{L_e}(1-\textrm{Pm})}=\textrm{Ro}_{\textrm{cr}}. \]
The stability condition (\ref{eq57}) encompasses the previously derived criteria for the standard MRI (when $H_{0\varphi}=0$) and for the azimuthal MRI (when $H_{0z}=0$), discussed in the preceding sections. 
In the special case where $\textrm{Ro} = 0$ and $\textrm{Ta} = 0$, the necessary and sufficient condition for the stability of the nanofluid with respect to axisymmetric perturbations imposes a restriction on the profile of the azimuthal magnetic field inhomogeneity:
\begin{equation} \label{eq58}
\textrm{Rb}> \frac{a^2(a^4+\pi^2\textrm{Q})^2-4\pi^4\textrm{Q}^2\xi^2-k^2\textrm{R}_nL_e(a^4+\pi^2\textrm{Q})}{4\pi^2\textrm{Q}\xi^2(a^4+\pi^2\textrm{Q})}=\textrm{Rb}_{\textrm{cr}}
\end{equation}
or, in dimensional variables,
\[\textrm{Rb}>\frac{(\omega_A^2+\omega_\nu\omega_\eta)^2-4\alpha^2\omega_A^2\omega_{A\varphi}^2-(1-\alpha^2)\frac{\omega_\nu\omega_\eta}{(|{\bf k}|h)^4}\textrm{R}_nL_e(\omega_A^2+\omega_\nu\omega_\eta)}{4\alpha^2\omega_{A\varphi}^2(\omega_A^2+\omega_\nu\omega_\eta)}=\textrm{Rb}_{\textrm{cr}} . \]
We now determine the region of helical MRI development in a thin layer of nanofluid that occurs for Rossby numbers $\textrm{Ro} < \textrm{Ro}_{\textrm{cr}}$.

Using numerical analysis and expression (\ref{eq57}) for the critical Rossby number, we identify the instability domains of helical MRI (HMRI) in both the "pure" conducting fluid and the nanofluid. In Figs.~\ref{fg9}a)--c), the instability regions for the nanofluid are highlighted in gray in the $(\textrm{k}, \textrm{Ro})$ plane for various values of the Taylor number $\textrm{Ta} = 100, 300, 2000$. The remaining nanofluid parameters are kept fixed: $\textrm{Q} = 10$, $\textrm{Q}_{\varphi} = 100$, $\textrm{Pm} = 0.7$, $\textrm{Rb} = 1/2$, $\textrm{R}_n = 0.122$, $L_e = 5000$, and $N_B = 7.5 \cdot 10^{-4}$. 
\begin{figure}
  \centering
\includegraphics[width=5.5 cm, height=5.0 cm]{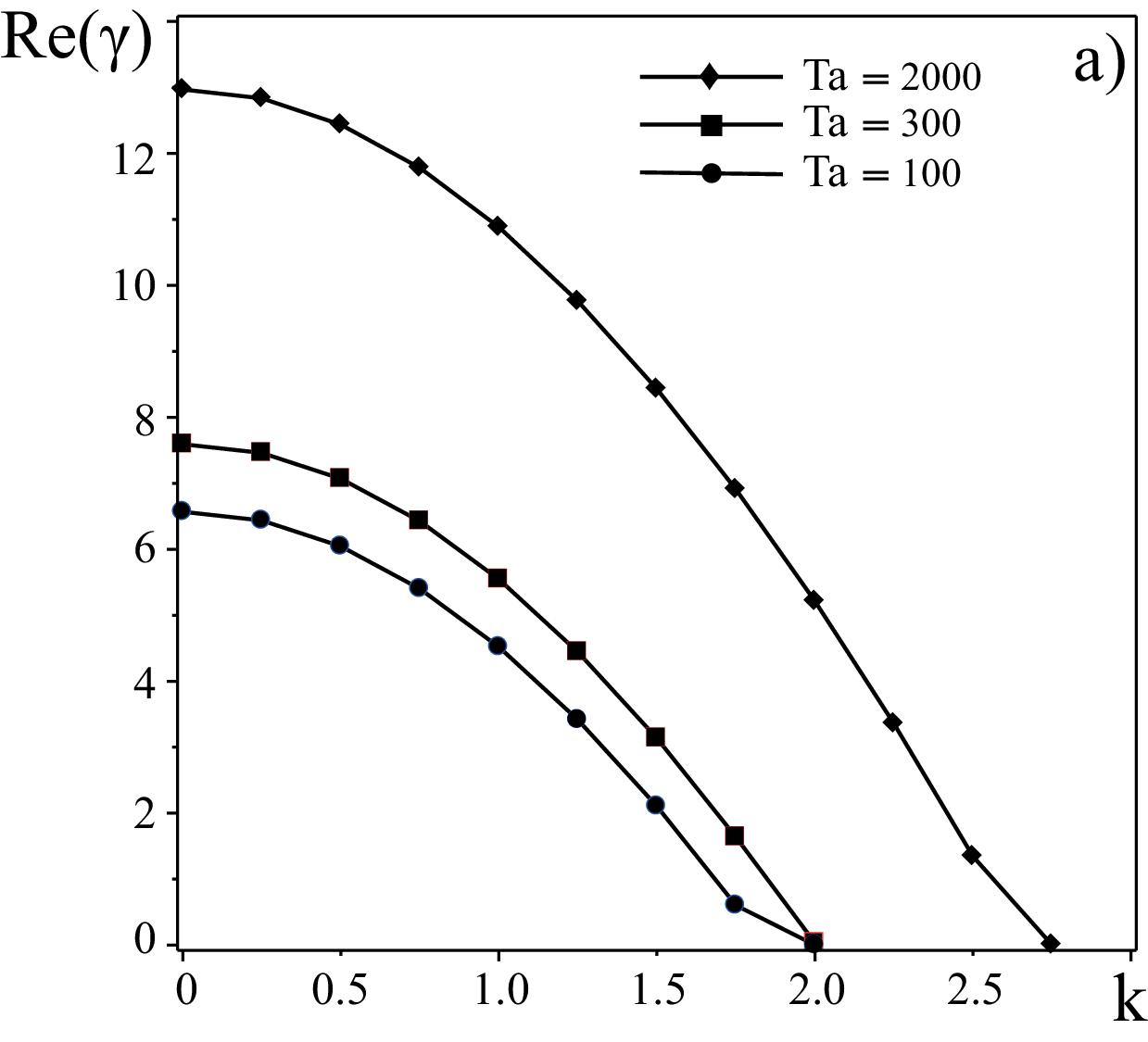}
	\includegraphics[width=5.5 cm, height=5.0 cm]{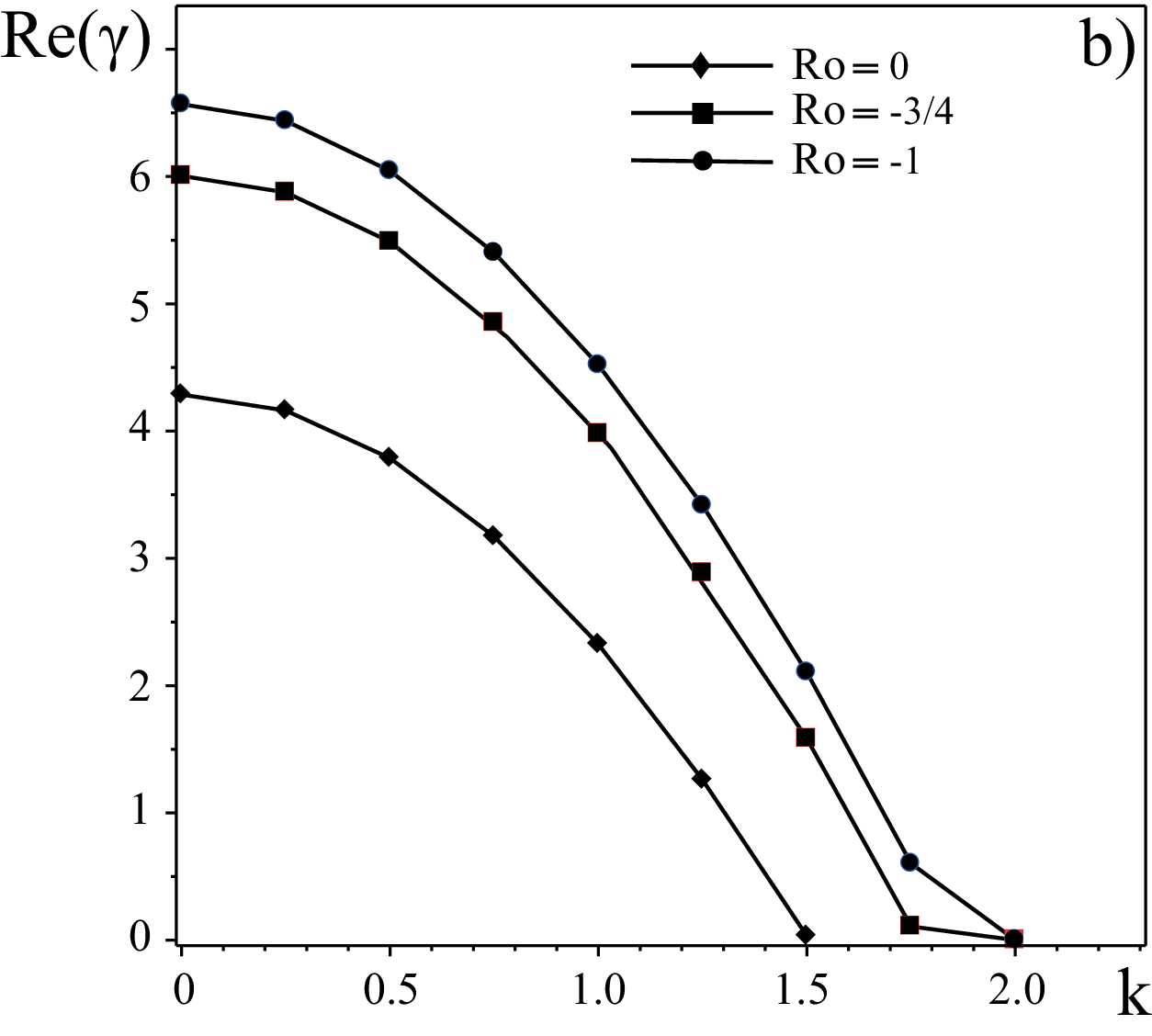}
	\includegraphics[width=5.5 cm, height=5.0 cm]{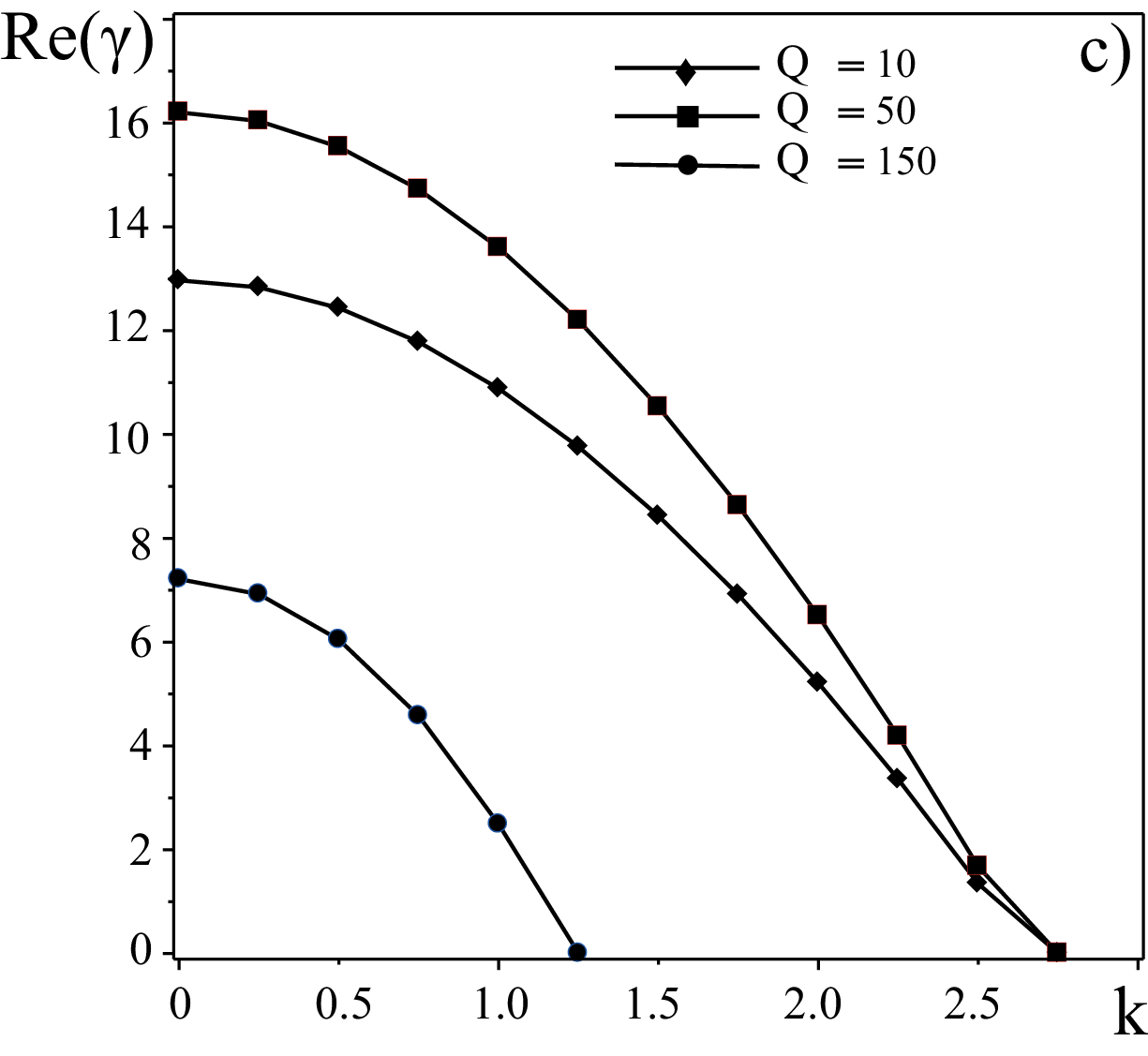}\\
\caption{ \small Dependence of the MRI increment ($\textrm{Re}(\gamma)>0$) in a nanofluid on the radial wave number $\textrm{k}$. The plots illustrate the influence of: a) rotation effect for $\textrm{Ta}=100, 300, 2000$; b) Rossby number $\textrm{Ro}=0, -3/4, -1$; c) axial magnetic field strength $\textrm{Q}=10, 50, 150$. }\label{fg11}
\end{figure}
From Figs.~\ref{fg9}a)--c), it can be observed that the presence of nanoparticles leads to an expansion of the instability region compared to the "pure" conducting fluid, whose instability domains are shown in black. In Figs.~\ref{fg10}d)--f), the HMRI regions are presented for a positive magnetic field gradient ($\textrm{Rb} = 1/2$) in the $(\textrm{k}, \textrm{Ro})$ plane for different values of the Chandrasekhar number $\textrm{Q}_{\varphi} = 30, 50, 80$, with the other parameters fixed: $\textrm{Q} = 10$, $\textrm{Ta} = 100$, $\textrm{Pm} = 0.7$, $\textrm{Rb} = 1/2$, $\textrm{R}_n = 0.122$, $L_e = 5000$, and $N_B = 7.5 \cdot 10^{-4}$. 
As seen in Figs.~\ref{fg10}d)-f), the presence of nanoparticles also enhances the instability region in comparison to the "pure" conducting fluid. Moreover, as the strength of the azimuthal magnetic field (i.e., $\textrm{Q}_{\varphi}$) increases -- similar to the case of azimuthal MRI -- the instability boundary shifts toward higher (positive) values of the Rossby number ($\textrm{Ro} > 0$).

We now proceed to the numerical analysis of the dispersion relation (\ref{eq56}). Fig.~\ref{fg11} presents the dependence of the increment of axisymmetric magneto-rotational instability (MRI) on the radial wave numbers $\textrm{k}$ for various values of the physical parameters of the nanofluid. From the plots in Fig.~\ref{fg11}a), one can observe that the growth rate of perturbations increases with the Taylor number $\textrm{Ta}=100, 300, 2000$, indicating the intensifying effect of rotation. The other parameters of the nanofluid are held constant: $\textrm{Q}=10$, $\textrm{Q}_{\varphi}=100$, $\textrm{Pm}=0.7$, $\textrm{Pr}=5$, $\textrm{Rb}=1/2$, $\textrm{Ro}=-1$, $\textrm{R}_n=0.122$, $L_e=5000$, $N_B=7.5 \cdot 10^{-4}$. For the same set of parameters, Fig.~\ref{fg11}b) illustrates the MRI increment for a rotating nanofluid with Taylor number $\textrm{Ta}=100$ and various Rossby numbers $\textrm{Ro}=0, -3/4, -1$. It follows that the growth rate of axisymmetric perturbations is higher for negative Rossby numbers ($\textrm{Ro}<0$) compared to the case of uniform rotation ($\textrm{Ro}=0$): 
\[\gamma(k)|_{\textrm{Ro}=-1} > \gamma(k)|_{\textrm{Ro}=-3/4} > \gamma(k)|_{\textrm{Ro}=0}.\]
Figure~\ref{fg11}c) shows the growth rates of MRI for different values of the axial magnetic field $\textrm{Q}=10, 50, 150$, assuming a Rayleigh-type rotation profile ($\textrm{Ro}=-1$) and Taylor number $\textrm{Ta}=2000$. It is evident that increasing the axial magnetic field strength $H_{0z}$ may lead either to an increase in the instability increment (from $\textrm{Q}=10$ to $\textrm{Q}=50$) or to its decrease (from $\textrm{Q}=50$ to $\textrm{Q}=150$). A similar behavior was previously observed for standard MRI.

\begin{figure}
  \centering
\includegraphics[width=5.5 cm, height=5.0 cm]{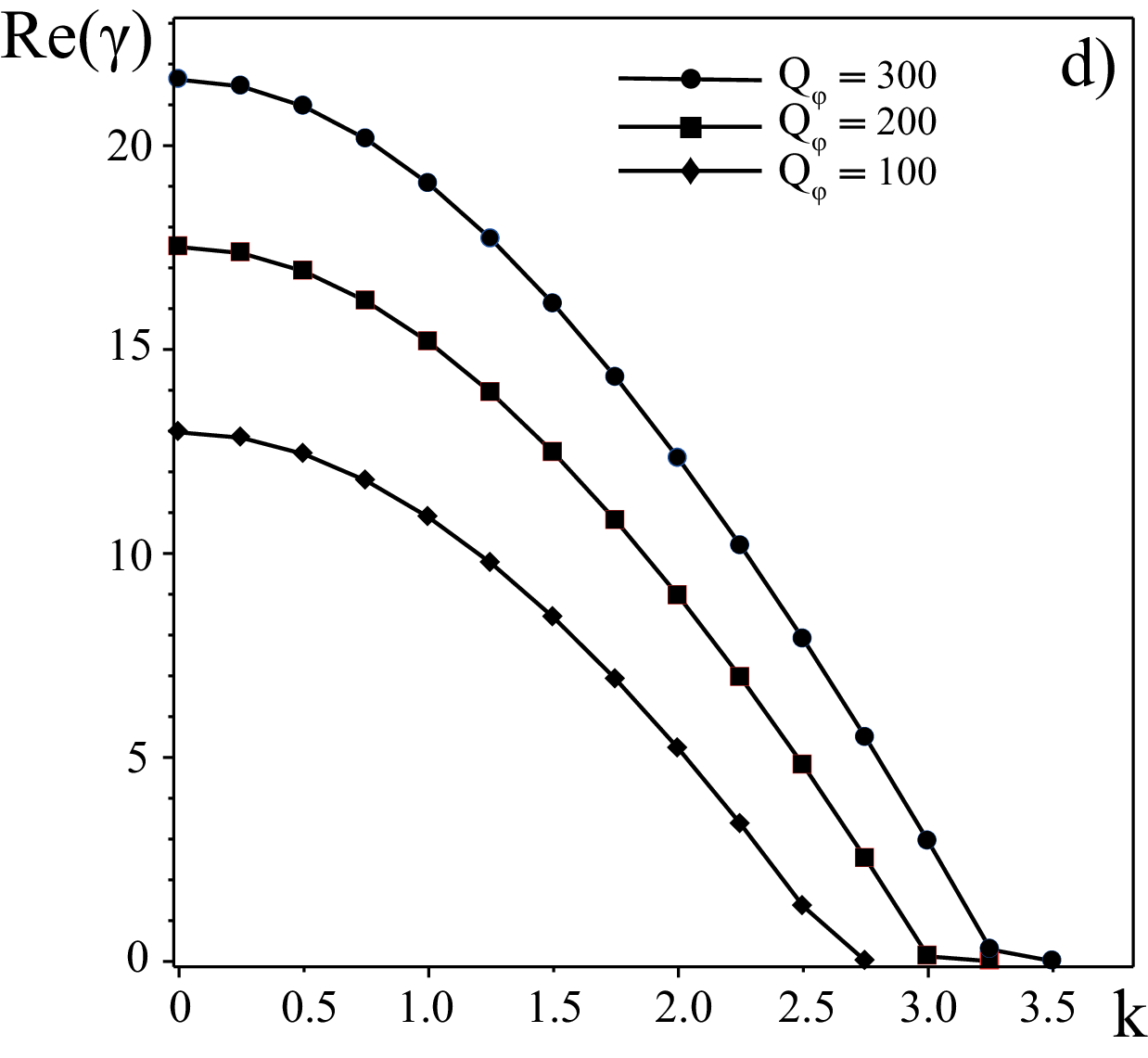}
	\includegraphics[width=5.5 cm, height=5.0 cm]{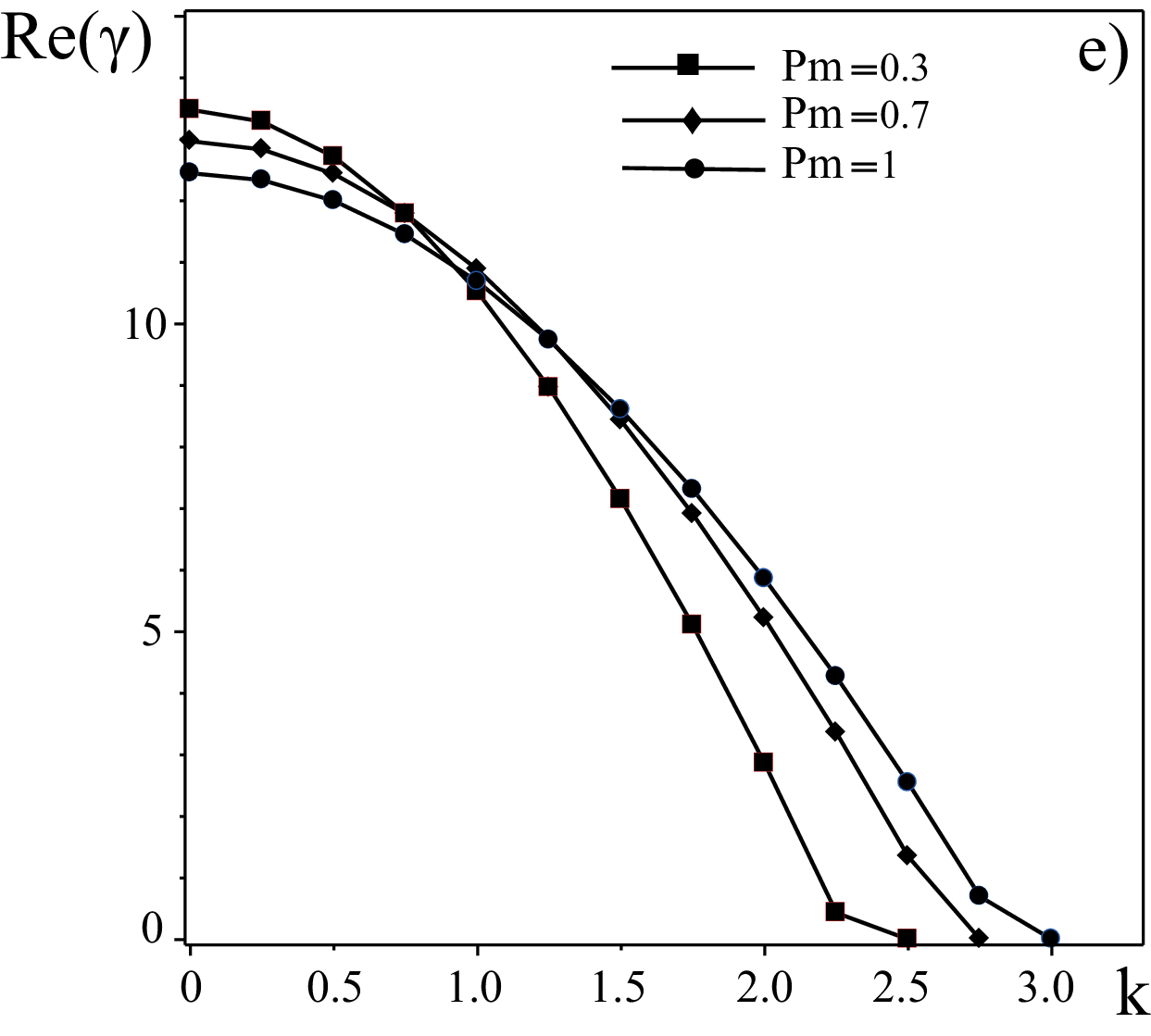}
	\includegraphics[width=5.5 cm, height=5.0 cm]{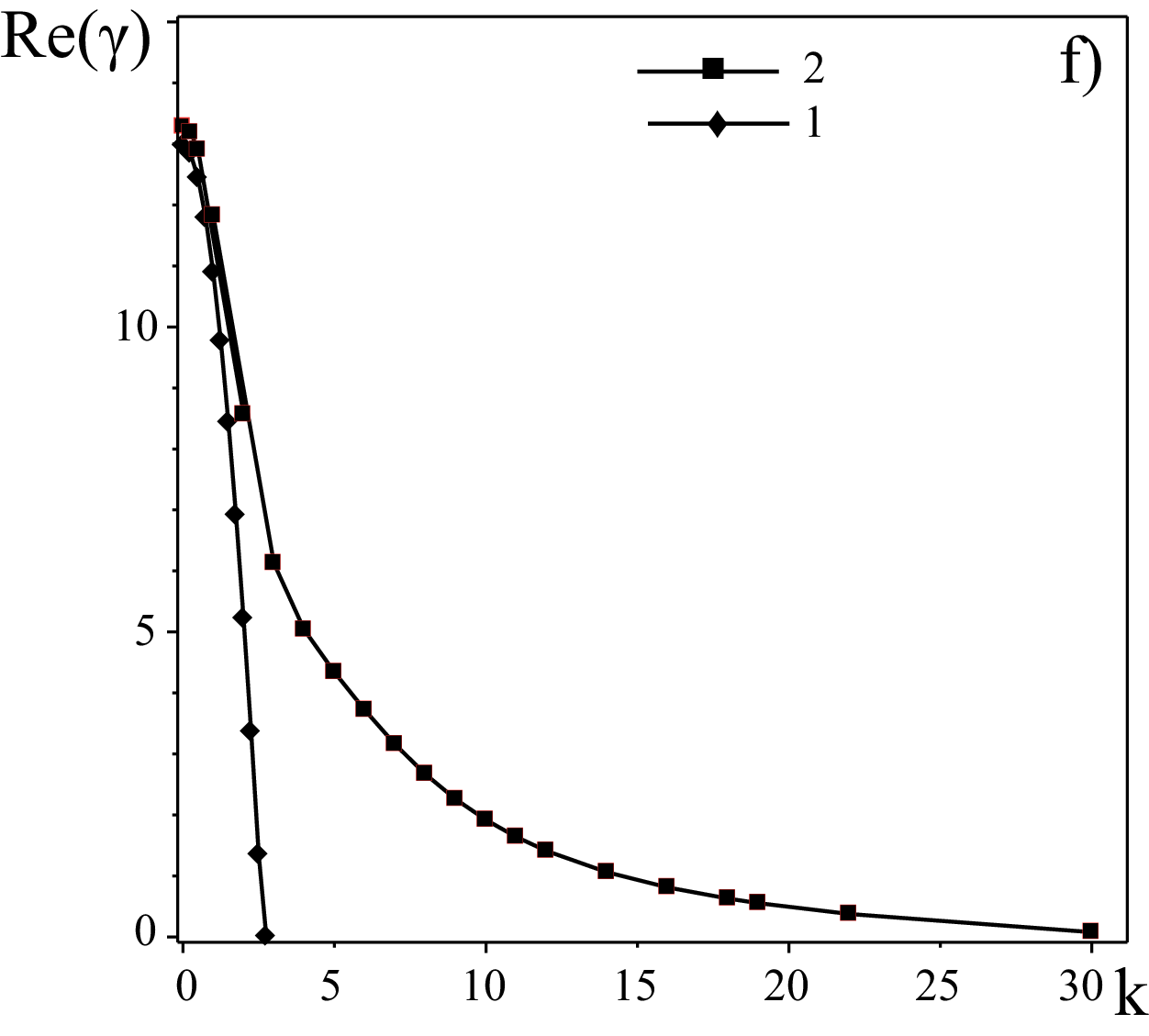}\\
\caption{ \small Dependence of the increment ($\textrm{Re}(\gamma)>0$) of helical MRI in a nanofluid on the radial wave number $\textrm{k}$. The plots illustrate the influence of: d) azimuthal magnetic field $\textrm{Q}_{\varphi}=100, 200, 300$; e) magnetic Prandtl number $\textrm{Pm}=0.3, 0.7, 1$; f) nanoparticle concentration on the helical MRI.}\label{fg12}
\end{figure}

The influence of the azimuthal magnetic field $\textrm{Q}_{\varphi}=100, 200, 300$ on MRI, for parameters $\textrm{Ta}=2000$ and $\textrm{Ro}=-1$, is shown in Fig.~\ref{fg12}d). As in the case of azimuthal MRI (AMRI), the growth rate of perturbations increases with the strength of the azimuthal magnetic field $H_{0\varphi}$.
In Fig.~\ref{fg12}e), the MRI increment is plotted for various magnetic Prandtl numbers $\textrm{Pm}=0.3, 0.7, 1$. The results suggest that for $\textrm{Pm}<1$, the increase in the growth rate $\textrm{Re}\,\gamma(\textrm{k})$ occurs in the long-wavelength (small $\textrm{k}$) part of the perturbation spectrum.
The effect of nanoparticle concentration on MRI is presented in Fig.~\ref{fg12}f). Curve 1 corresponds to the following nanofluid parameters: $\textrm{Q}=10$, $\textrm{Q}_{\varphi}=100$, $\textrm{Ta}=2000$, $\textrm{Pm}=0.7$, $\textrm{Pr}=5$, $\textrm{Rb}=1/2$, $\textrm{Ro}=-1$, $\textrm{R}_n=0.122$, $L_e=5000$, $N_B=7.5 \cdot 10^{-4}$. 
To model an increased nanoparticle concentration (i.e., an increased volume fraction at the upper boundary $\varphi_u$), the following parameters are adjusted: $\textrm{R}_n=1200$, $N_B=750$, $L_e=1000$, while the other dimensionless numbers $(\textrm{Q},\textrm{Q}_{\varphi},\textrm{Ta},\textrm{Pm},\textrm{Pr}, \textrm{Rb}, \textrm{Ro})$ remain unchanged. Curve 2 in Fig.~\ref{fg12}f) corresponds to the case with enhanced nanoparticle concentration. It is clear that under such conditions, the MRI increment is higher, with $\gamma_2(0)=13.28 > \gamma_1(0)=12.9$, and the instability starts to develop in the short-wavelength (large $\textrm{k}$) region of the spectrum.

As a result of increased nanoparticle concentration, the growth rates of helical MRI rise due to the combined effect of nanoparticle generation ($N_B \gg 1$) and the presence of a helical magnetic field, particularly for magnetic Prandtl numbers $\textrm{Pm} < 1$.

\section{Steady-state convection regime in an axial magnetic field}

From the dispersion relation (\ref{eq45}), we determine the critical Rayleigh number $\textrm{Ra}_{st}$ for steady-state convection $(\gamma = 0)$ in the presence of an axial magnetic field:
\begin{equation} \label{eq59}
  \textrm{Ra}_{st}=\frac{a^6}{k^2}+\frac{\pi^2 a^2\textrm{Q}}{k^2}+\frac{\pi^2a^4\textrm{Ta}}{k^2 (a^4 + \pi^2\textrm{Q})}+\frac{\pi^2\textrm{TaRo}(a^4+\pi^2\textrm{QPm})}{k^2(a^4+\pi^2\textrm{Q})}-\textrm{R}_n(L_e+N_A)
\end{equation}
The minimum value of the critical Rayleigh number is found from the condition $\partial\textrm{Ra}_{st}/\partial k = 0$ and corresponds to the wave numbers $k = k_c$ satisfying the following equation:
\begin{equation} \label{eq60}
\frac{{2k_c^2-\pi^2}}{{k_c}}-\frac{{\pi^4\textrm{Q}}}{{k_c (\pi^2+k_c^2 )^2 }}+\frac{{2\pi^2 k_c\textrm{Ta}(1+\textrm{Ro})}}{(\pi^2+k_c^2 )\left( (\pi^2+k_c^2)^2+\pi^2\textrm{Q}\right)} -$$
$$-\frac{\pi^2\textrm{Ta}((\pi^2+k_c^2)^2+\pi^2\textrm{Q}+2k_c^2(\pi^2+k_c^2))}{k_c((\pi^2+k_c^2)^2+\pi^2\textrm{Q})^2}-$$
$$-\frac{\pi^2\textrm{TaRo}((\pi^2+k_c^2)^2+\pi^2\textrm{QPm})((\pi^2+k_c^2)^2+\pi^2\textrm{Q}+2k_c^2(\pi^2+k_c^2))}{k_c(\pi^2+k_c^2 )^2((\pi^2+k_c ^2)^2+\pi^2\textrm{Q})^2}= 0 
\end{equation}	
\begin{figure}
  \centering
\includegraphics[width=6.5 cm, height=6.0 cm]{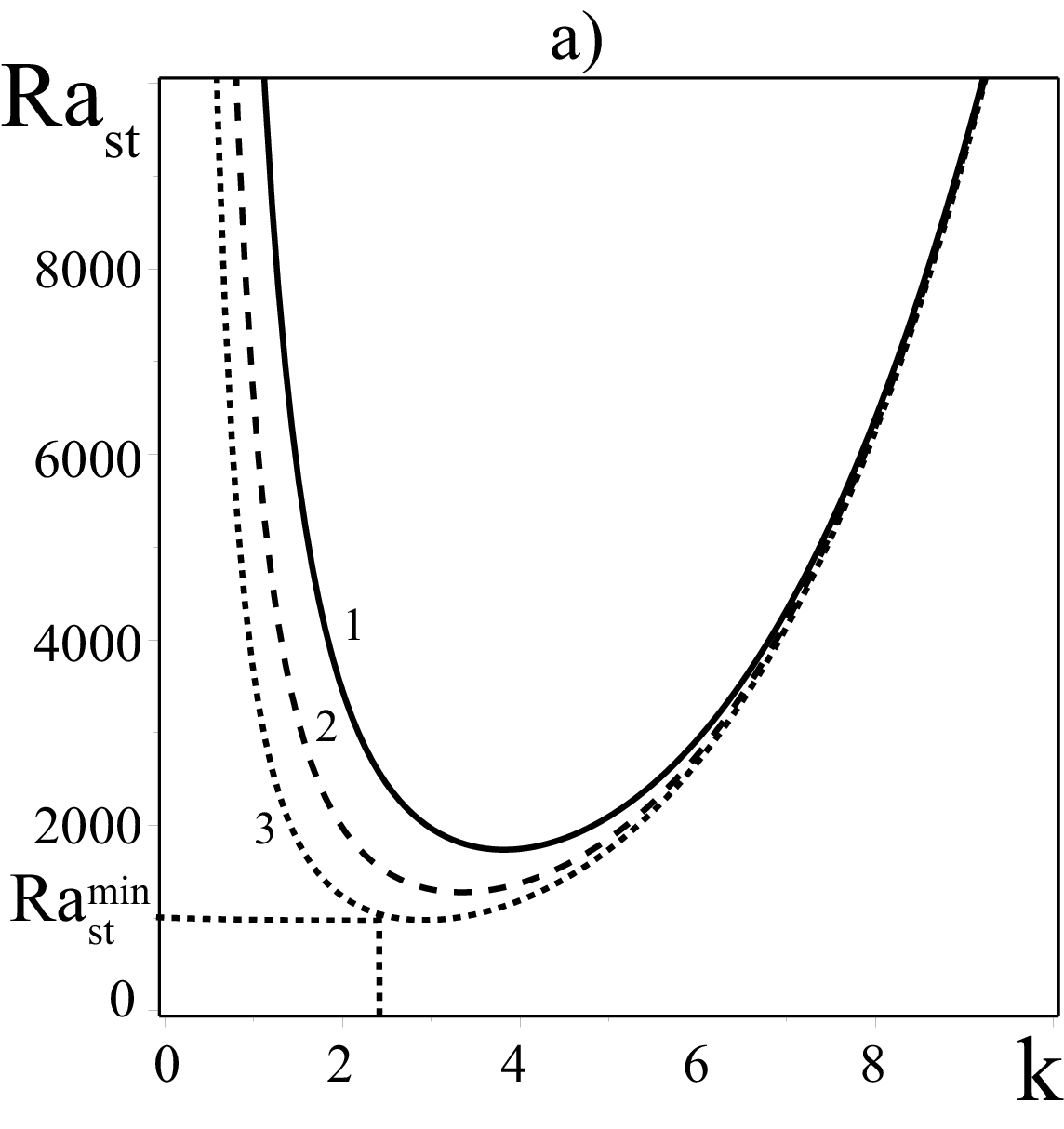}
	\includegraphics[width=6.5 cm, height=6.0 cm]{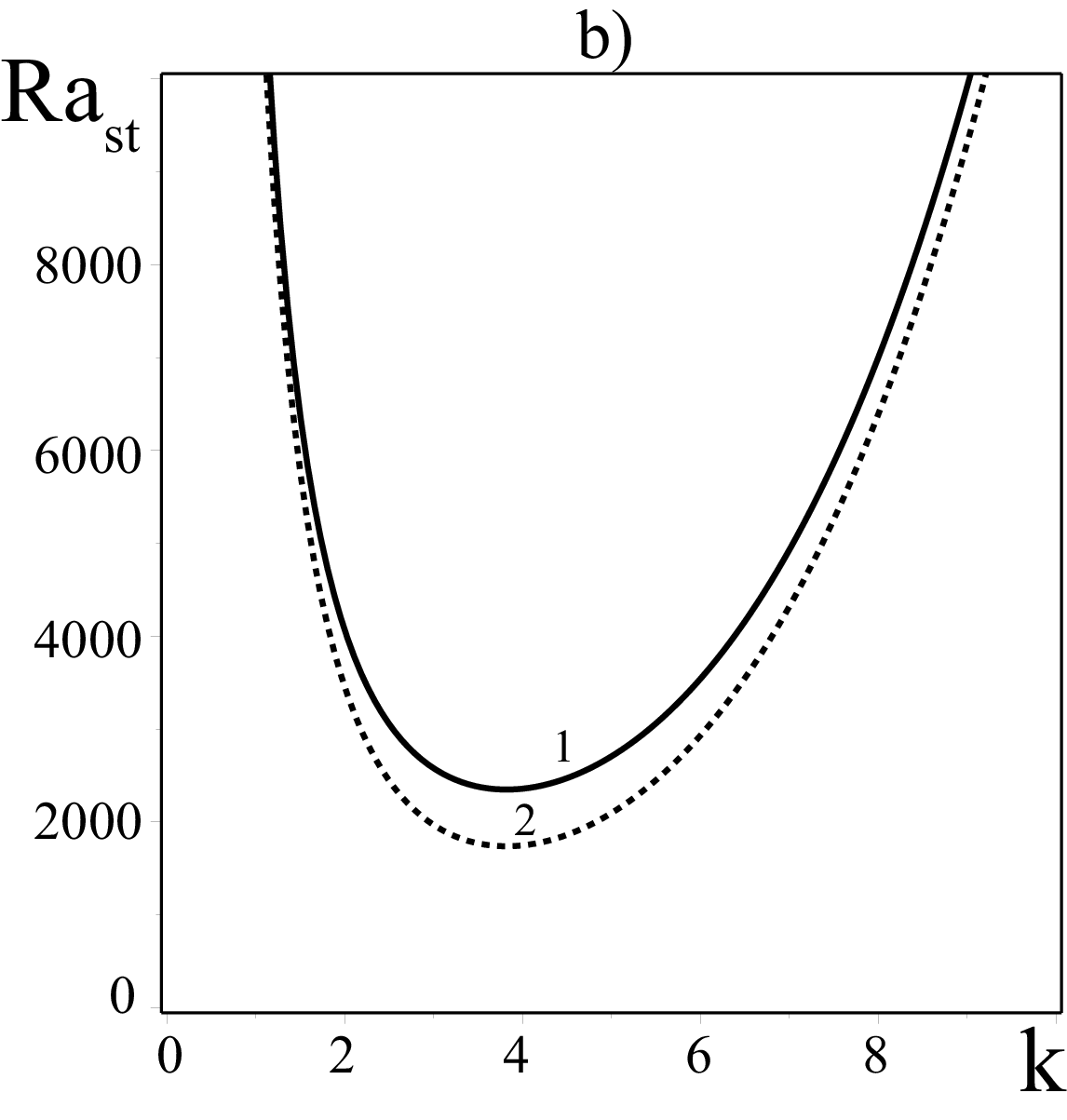}\\
\caption{\small Dependence of the stationary Rayleigh number $\textrm{Ra}_{st}$ on the wave number $k$ for a differentially rotating nanofluid in an axial magnetic field with fixed parameters: $\textrm{Ta} = 300$, $\textrm{Q} = 50$, $\textrm{Pm} = 1$, $\textrm{R}_n = 0.122$, $L_e = 5000$, $N_A = 5$.  (a) Curve~1 corresponds to $\textrm{Ro} = 2$, curve~2 to $\textrm{Ro} = 0$, and curve~3 to $\textrm{Ro} = -1$; (b) Curve~1 represents a "pure" electrically conducting fluid, and curve~2 a conducting nanofluid. Both curves (1, 2) are plotted for $\textrm{Ro} = 2$.}\label{fg13}
\end{figure}
As evident from equation (\ref{eq60}), the critical wave number is independent of the nanofluid parameters.
In Fig.~\ref{fg13}a, the minimum value of the critical Rayleigh number $\textrm{Ra}_{st}^{\textrm{min}}$ corresponds to the point on the neutral stability curve that separates the stable and unstable perturbation domains. It is clear that for positive Rossby number profiles $(\textrm{Ro} \geqslant 0)$, the minimum critical Rayleigh number is higher than for negative rotation profiles, such as the Rayleigh-type profile $(\textrm{Ro} = -1)$. This implies that negative differential rotation reduces the threshold for the onset of instability compared to both uniform $(\textrm{Ro} = 0)$ and positively sheared $(\textrm{Ro} = 2)$ rotation.  
Fig.~\ref{fg13}b demonstrates that increasing the nanoparticle concentration leads to a further reduction in the threshold for steady convection. The curves (1, 2) are plotted for $\textrm{Ro} = 2$, although the general trend remains valid across a wide range of Rossby numbers.
In several limiting cases, expression (\ref{eq59}) reproduces well-known classical results. When nanoparticles are absent $(\textrm{R}_n = 0)$, equation (\ref{eq59}) yields the critical Rayleigh number for steady convection in a differentially rotating fluid $(\textrm{Ro} \neq 0)$ subjected to an axial magnetic field, as shown in \cite{10s}. Setting $\textrm{Ro} = 0$ recovers the result of \cite{9s}. For the case $\textrm{Ro} = 0$ and $\textrm{R}_n = 0$, we retrieve the classical result by Chandrasekhar \cite{6s}. Finally, in the absence of both rotation and magnetic field $(\textrm{Ta} = \textrm{Q} = 0)$, equation (\ref{eq59}) reduces to the well-established result from \cite{28s}.

In order to analyze the influence of rotation, Rossby number, magnetic field, Lewis number, modified diffusion coefficient, and nanoparticle concentration, we evaluate the respective derivatives
\[ \frac{d \textrm{R}_1}{d \textrm{T}_1},\; \frac{d\textrm{R}_1}{d\textrm{Ro}},\; \frac{d\textrm{R}_1}{d \textrm{Q}_1},\; \frac{d\textrm{R}_1}{dL_e}, \; \frac{d \textrm{R}_1}{dN_A},\;\frac{d \textrm{R}_1}{d\textrm{R}_n} \]
expressed in terms of Chandrasekhar variables:
\[ \textrm{R}_1=\frac{\textrm{Ra}_{st}}{\pi^4},\quad \textrm{T}_1=\frac{\textrm{Ta}}{\pi^4}, \quad \textrm{Q}_1=\frac{\textrm{Q}}{\pi^2}, \quad \textrm{x}=\frac{k^2}{\pi^2}.\]
For these variables, expression (\ref{eq59}) takes the form
\begin{equation} \label{eq61}	
\textrm{R}_{1}=\frac{(1+\textrm{x})((1+\textrm{x})^2+\textrm{Q}_1)^2+(1+\textrm{x})^2(1+\textrm{Ro})\textrm{T}_1+\textrm{RoPm}\textrm{Q}_1\textrm{T}_1}{\textrm{x}((1+\textrm{x})^2+\textrm{Q}_1)}-\textrm{R}_n(L_e+N_A) .
\end{equation}
From this expression, we find the derivatives
\begin{equation} \label{eq62}		
	\frac{d \textrm{R}_1}{d \textrm{T}_1}=\frac{(1+\textrm{x})^2}{\textrm{x}((1+\textrm{x})^2+\textrm{Q}_1)}+\frac{\textrm{Ro}((1+\textrm{x})^2+ \textrm{Q}_1\textrm{Pm})}{\textrm{x} ((1+\textrm{x})^2+\textrm{Q}_1)}
\end{equation}
\begin{equation}\label{eq63}
\frac{d\textrm{R}_1}{d\textrm{Ro}}=\frac{\textrm{T}_1 ((1+\textrm{x})^2+\textrm{Q}_1\textrm{Pm}) }{\textrm{x}(1+\textrm{x})^2+\textrm{Q}_1)}
\end{equation}
\begin{equation}\label{eq64}
\frac{d \textrm{R}_1}{d \textrm{Q}_1}=\frac{1+\textrm{x}}{\textrm{x}}-\frac{(1+\textrm{x})^2\textrm{T}_1(1+\textrm{Ro}(1-\textrm{Pm}))}{\textrm{x}((1+\textrm{x})^2+\textrm{Q}_1)^2},
\end{equation}	
\begin{equation}\label{eq65}	
\frac{d \textrm{R}_1}{d L_e}=\frac{d \textrm{R}_1}{d N_A}=-\textrm{R}_n
\end{equation}	
\begin{equation}\label{eq66}
\frac{d \textrm{R}_1}{d \textrm{R}_n}=-(L_e+N_A)	
\end{equation}	
According to equation~(\ref{eq62}), the quantity $d\textrm{R}_1/d\textrm{T}_1$ is positive under differential rotation with non-negative Rossby numbers ($\textrm{Ro} \geqslant 0$), indicating that such rotation has a stabilizing effect on stationary convection. In contrast, when $\textrm{Ro} < 0$, differential rotation can exert a destabilizing influence, as evidenced by the negative derivative $d\textrm{R}_1/d\textrm{T}_1 < 0$.

Equation~(\ref{eq63}) reveals that increasing the rotation parameter $\textrm{T}_1$ affects stationary convection depending on the rotation profile, specifically the sign of the Rossby number. For positive $\textrm{Ro}$, the effect remains stabilizing, since the derivative $d\textrm{R}_1/d\textrm{Ro} > 0$. Conversely, for negative $\textrm{Ro}$, the influence becomes destabilizing due to the reversed sign, with $d\textrm{R}_1/d\textrm{Ro} < 0$.

As shown in equation~(\ref{eq64}), the axial magnetic field (expressed through the Chandrasekhar number $\textrm{Q}_1$) can have either a stabilizing or destabilizing effect on stationary convection. A destabilizing effect arises under the following condition:
\[\textrm{Ro}(\textrm{Pm} - 1) < 1\]
This inequality is satisfied for $\textrm{Ro} \geqslant 0$ with $\textrm{Pm} < 1$, for $\textrm{Ro} < 0$ with $\textrm{Pm} > 1$, and for any value of $\textrm{Ro}$ when $\textrm{Pm} = 1$.

Equation~(\ref{eq65}) demonstrates that both the Lewis number $L_e$ and the modified diffusion coefficient $N_A$ have a destabilizing effect on stationary convection when the nanoparticle Rayleigh number is positive ($\textrm{R}_n > 0$). Since for most nanofluids the sum $L_e + N_A$ is typically positive~\cite{9s}, equation~(\ref{eq66}) implies that the concentration-based Rayleigh number generally contributes to destabilization. Equations~(\ref{eq65}) and~(\ref{eq66}) are consistent with the findings in~\cite{9s}, which examined the convective instability of a uniformly rotating nanofluid layer in the presence of a constant vertical magnetic field.

\section{Stationary convection regime in a helical magnetic field}

\begin{figure}
  \centering
	\includegraphics[width=5.5 cm, height=5.0 cm]{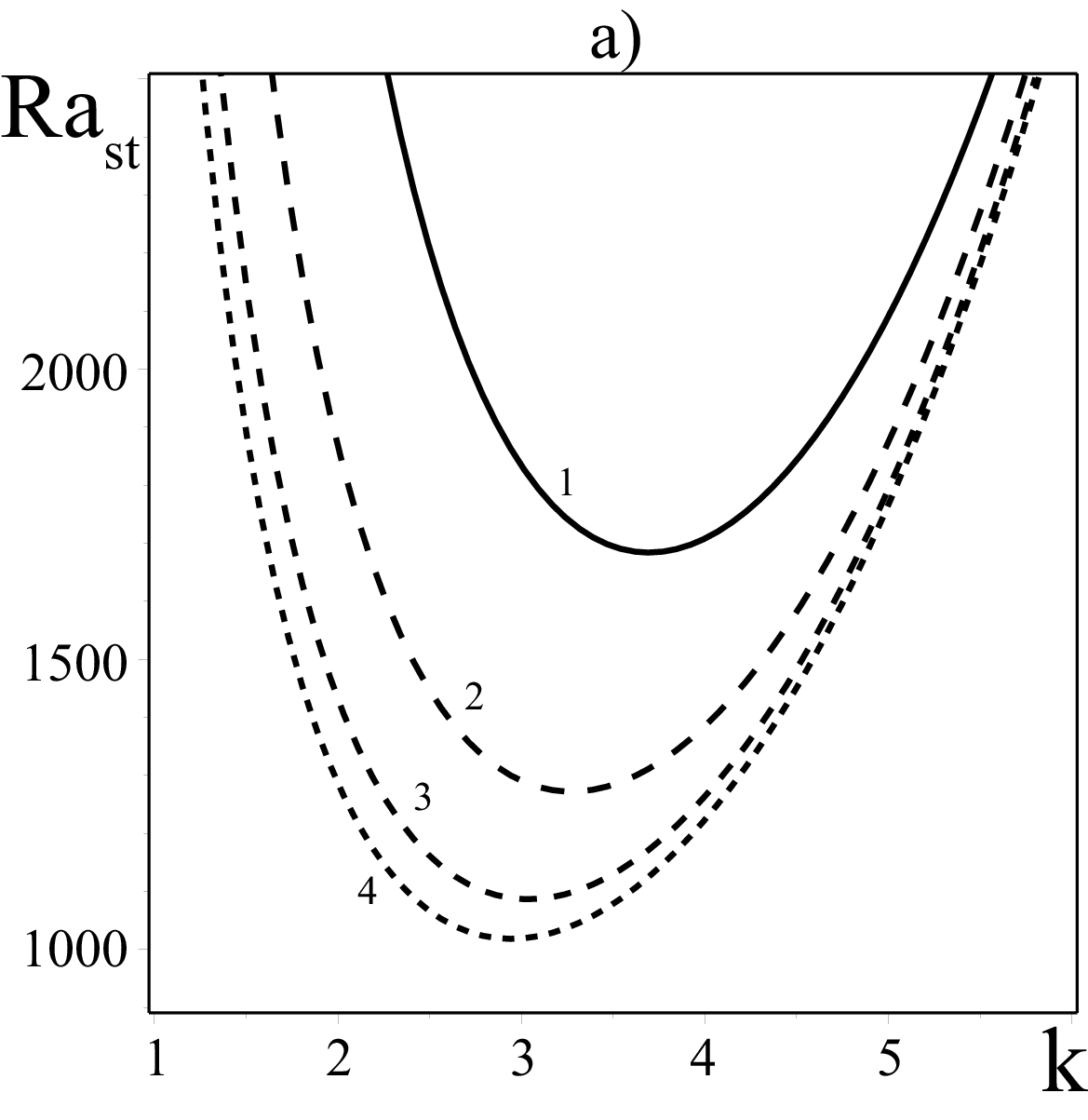}
	\includegraphics[width=5.5 cm, height=5.0 cm]{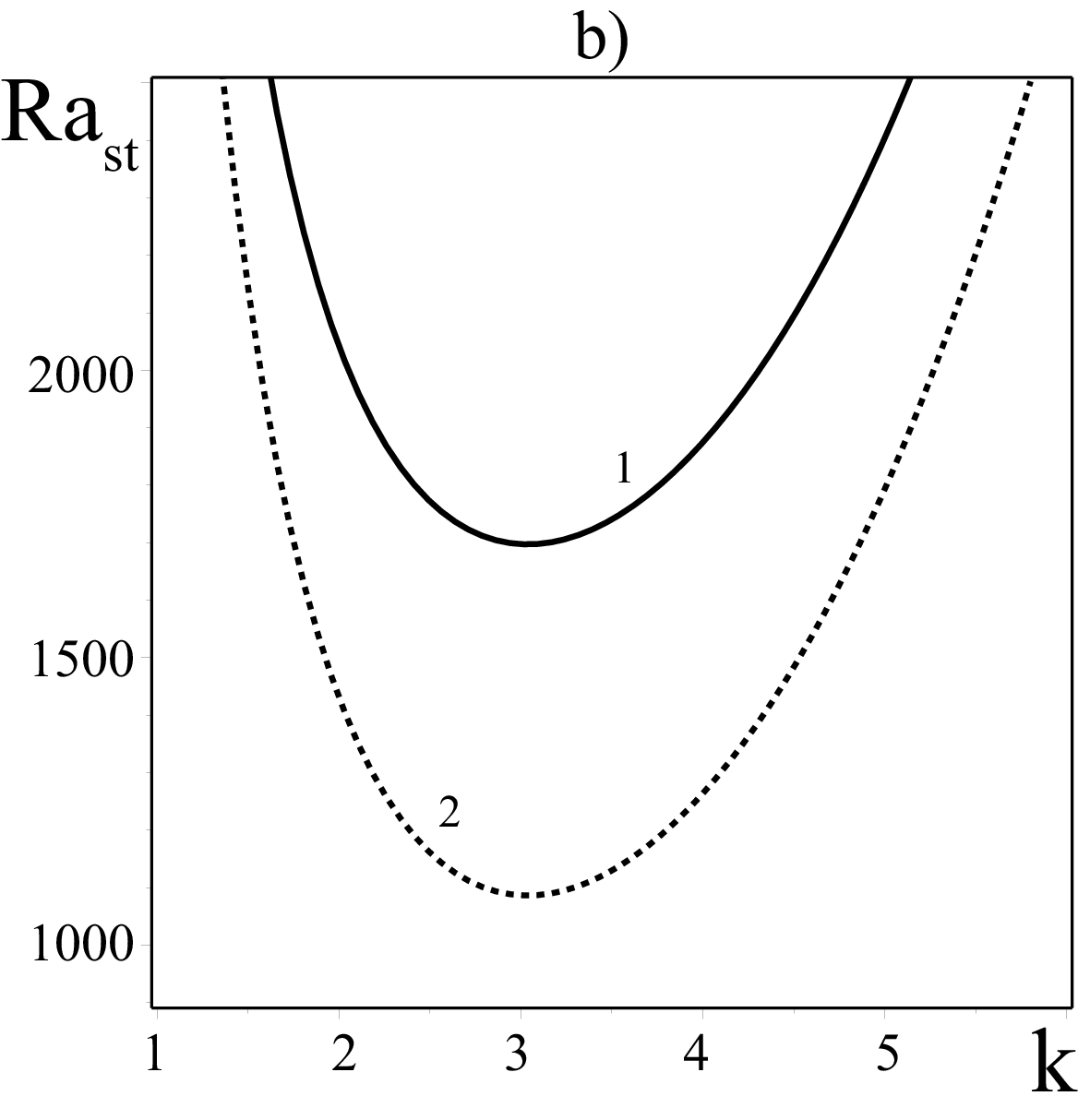}
	\includegraphics[width=5.5 cm, height=5.0 cm]{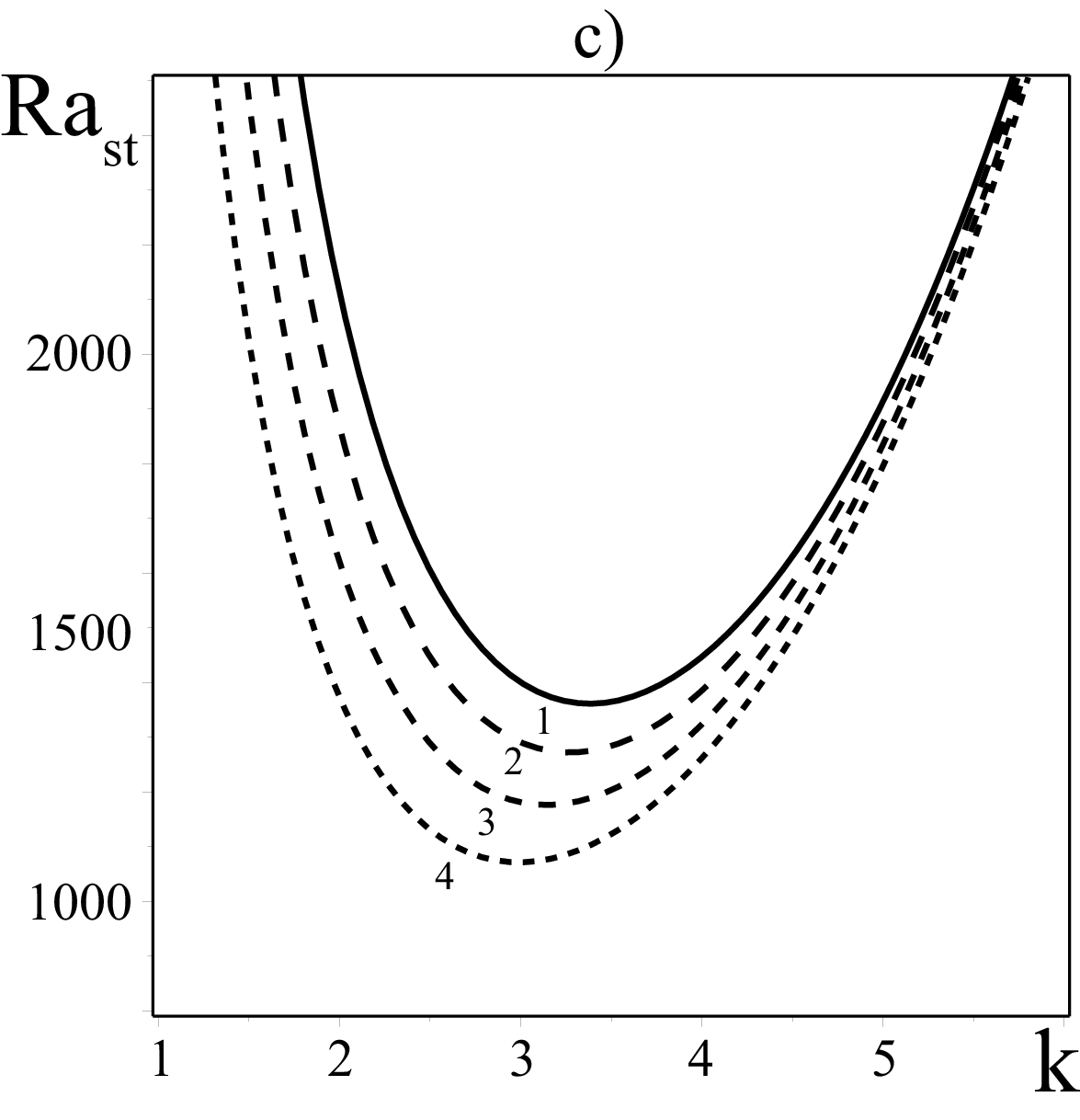}\\
\caption{\small Dependence of the stationary Rayleigh number $\textrm{Ra}_{st}$ on the wavenumber $k$ for a differentially rotating nanofluid in a helical magnetic field: (a) $\textrm{Rb} = -1/2$ and $\xi = 1$: curve~1 -- $\textrm{Ro} = 2$, curve~2 -- $\textrm{Ro} = 0$, curve~3 -- $\textrm{Ro} = -3/4$, curve~4 -- $\textrm{Ro} = -1$; (b) $\textrm{Ro} = 0$ and $\xi = 1$: curve~1 -- $\textrm{Rb} = -1$, curve~2 -- $\textrm{Rb} = -1/2$, curve~3 -- $\textrm{Rb} = 0$, curve~4 -- $\textrm{Rb} = 1/2$; (c) $\textrm{Ro} = -3/4$ and $\xi = 1$: curve~1 -- $\textrm{Rb} = -1$, curve~2 -- $\textrm{Rb} = -1/2$, curve~3 -- $\textrm{Rb} = 0$, curve~4 -- $\textrm{Rb} = 1/2$.}\label{fg14}
\end{figure}

In a similar manner to the case of an axial magnetic field, the critical Rayleigh number $\textrm{Ra}_{st}$ corresponding to the onset of stationary convection ($\gamma = 0$) in a helical magnetic field can be determined from the dispersion relation~(\ref{eq45}):
\begin{equation}\label{eq67}
\textrm{Ra}_{st}=\left[\textrm{Ra}_{st}^{(0)}-\textrm{R}_n(L_e+N_A)+\frac{2\pi^4\textrm{Q}\xi\sqrt{\textrm{Ta}}(2+\textrm{Ro}(1-\textrm{Pm}))N_B(N_A-1)}{k^2(a^4+\pi^2\textrm{Q})L_e}\right]\times $$
$$\times\left[1-\frac{2\pi^4\textrm{Q}\xi\sqrt{\textrm{Ta}}(2+\textrm{Ro}(1-\textrm{Pm}))(N_A-N_B)}{a^2(a^4+\pi^2\textrm{Q})^2L_e}\right]^{-1}=D_1(k)\cdot D_2^{-1}(k),
\end{equation}	
where $\textrm{Ra}_{st}^{(0)}$ is the critical Rayleigh number for stationary convection in a pure fluid under a helical magnetic field, which coincides with the result reported in~\cite{12s}:
$$\textrm{Ra}_{st}^{(0)}= \frac{a^6}{k^2} + \frac{\pi^2 a^2\textrm{Q}}{k^2} + \frac{\pi^2a^4 \textrm{Ta}}{k^2 (a^4 + \pi^2\textrm{Q})} + \frac{\pi^2  \textrm{TaRo}(a^4+\pi^2\textrm{QPm})-4\pi^4\xi^2 \textrm{Q}^2}{k^2(a^4+\pi^2\textrm{Q})}-\frac{4\pi^2}{k^2}\cdot\xi^2\textrm{QRb} .$$
The minimum value of the critical Rayleigh number is determined from the condition $\partial\textrm{Ra}_{st}/\partial k = 0$ and corresponds to the wavenumbers $k = k_c$ that satisfy the following equation:
\begin{equation}\label{eq68}
 \left[\left(\frac{\partial \textrm{Ra}_{st}^{(0)}}{\partial k}\right)_{k=k_c}-4\pi^4\textrm{Q}\xi\sqrt{\textrm{Ta}}(2+\textrm{Ro}(1-\textrm{Pm}))N_B(N_A-1)\cdot\frac{(\pi^2+k_c^2)^2+\pi^2\textrm{Q}+2k_c^2(\pi^2+k_c^2)}{k_c^3 ((\pi^2+k_c^2)^2+\pi^2\textrm{Q})^2L_e}\right]\times $$
$$\times D_2^{-1}(k_c)-D_2^{-2}(k_c)\cdot 4k_c\pi^4\textrm{Q}\xi\sqrt{\textrm{Ta}}(2+\textrm{Ro}(1-\textrm{Pm}))(N_A-N_B)\times $$
$$\times \frac{(\pi^2+k_c^2)^2+\pi^2\textrm{Q}+2(\pi^2+k_c^2)}{(\pi^2+k_c^2)((\pi^2+k_c^2)^2+\pi^2\textrm{Q})^3L_e }\cdot D_1(k_c)=0,
\end{equation}	
where
\[\left(\frac{\partial \textrm{Ra}_{st}^{(0)}}{\partial k}\right)_{k=k_c}=\frac{{2k_c ^2-\pi^2 }}{{k_c }}-\frac{{\pi ^4\textrm{Q}}}{{k_c (\pi^2+k_c^2)^2 }}+\frac{{2\pi^2 k_c\textrm{Ta}(1+\textrm{Ro})}}{(\pi^2+k_c^2)\left( (\pi^2+k_c^2)^2+\pi^2\textrm{Q}\right)} - \]
\[-\frac{\pi^2 \textrm{Ta}((\pi^2+k_c^2)^2+\pi^2\textrm{Q}+2k_c^2(\pi^2+ k_c^2))}{k_c((\pi^2+k_c^2)^2  + \pi ^2\textrm{Q})^2}-\]
\[-\frac{\pi ^2 \textrm{TaRo}((\pi^2  + k_c^2 )^2  + \pi ^2\textrm{QPm})((\pi^2+ k_c^2 )^2  + \pi ^2\textrm{Q}+2k_c^2(\pi ^2  + k_c ^2))}{k_c(\pi ^2  + k_c^2)^2((\pi^2+k_c^2)^2+\pi^2\textrm{Q})^2}+\]
\[+4\pi^4\xi^2\textrm{Q}^2\cdot\frac{2((\pi^2+ k_c^2)^2+\pi^2\textrm{Q})+4k_c^4(\pi^2+ k_c^2)}{k_c^3((\pi^2+ k_c^2)^2+\pi^2\textrm{Q})^2}+4\pi^2\xi^2\textrm{QRb}\cdot\frac{2}{k_c^3}. \]
If the azimuthal magnetic field is absent $(\xi = 0)$, then expression~(\ref{eq68}) coincides with~(\ref{eq60}). 
Therefore, the critical wavenumber $k_c$ depends on the nanofluid parameters only in the presence of a helical magnetic field.
Fig.~\ref{fg14} shows the dependence of the stationary Rayleigh number $\textrm{Ra}_{st}$ 
on the wavenumber $k$ for various profiles of differential rotation $(\textrm{Ro})$ and magnetic field $(\textrm{Rb})$. The numerical results presented in Fig.~\ref{fg14} were obtained for the following fixed parameters: $\textrm{Ta}=300$, $\textrm{Q}=50$, $\textrm{Pm}=0.7$, $\textrm{R}_n=0.122$, $L_e=5000$, $N_A=5$, $N_B=7.5\cdot 10^{-4}$. The minimum value of the stationary Rayleigh number $\textrm{Ra}_{st}^{\textrm{min}}$ in the plots corresponds to a point on the neutral stability curve separating stable and unstable perturbation regimes.

Fig.~\ref{fg14}a presents the case of a homogeneous azimuthal magnetic field $(\textrm{Rb} = -1/2)$ with the parameter $\xi = 1$. It is observed that with an increasing positive Rossby number profile $\textrm{Ro}$ (curve~1: $\textrm{Ro}=2$, curve~2: $\textrm{Ro}=0$), the minimum value $\textrm{Ra}_{st}^{\textrm{min}}$ also increases, indicating a higher threshold for the onset of instability. On the other hand, for negative rotation profiles Keplerian $(\textrm{Ro} = -3/4)$ (curve~3) and Rayleigh $(\textrm{Ro} = -1)$ (curve~4) a decrease in the critical Rayleigh number is observed, indicating a lower threshold for the development of instability compared to the cases of homogeneous $(\textrm{Ro} = 0)$ and differential $(\textrm{Ro} = 2)$ rotation.

\begin{figure}
  \centering
	\includegraphics[width=5.5 cm, height=5.0 cm]{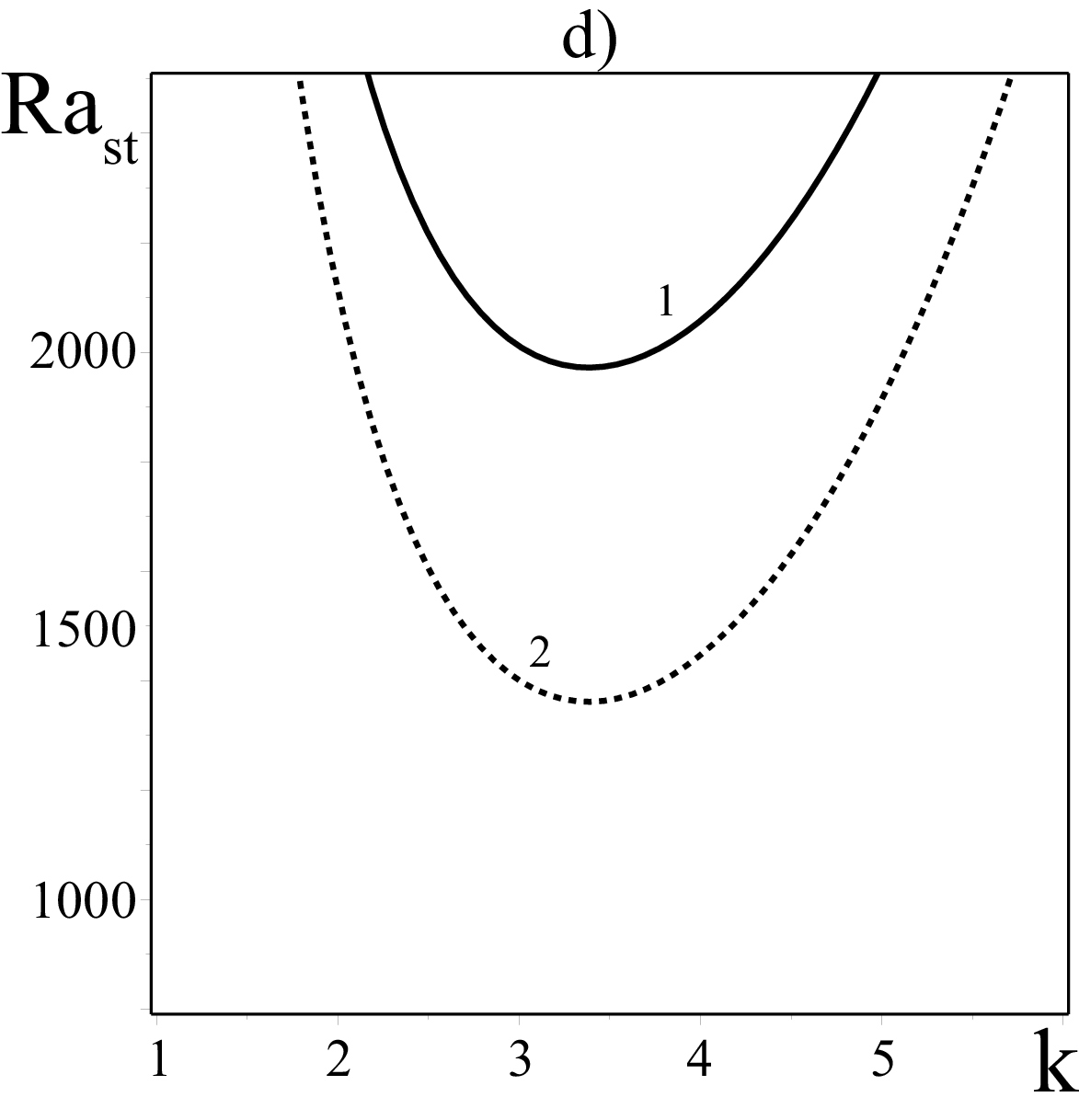}
	\includegraphics[width=5.5 cm, height=5.0 cm]{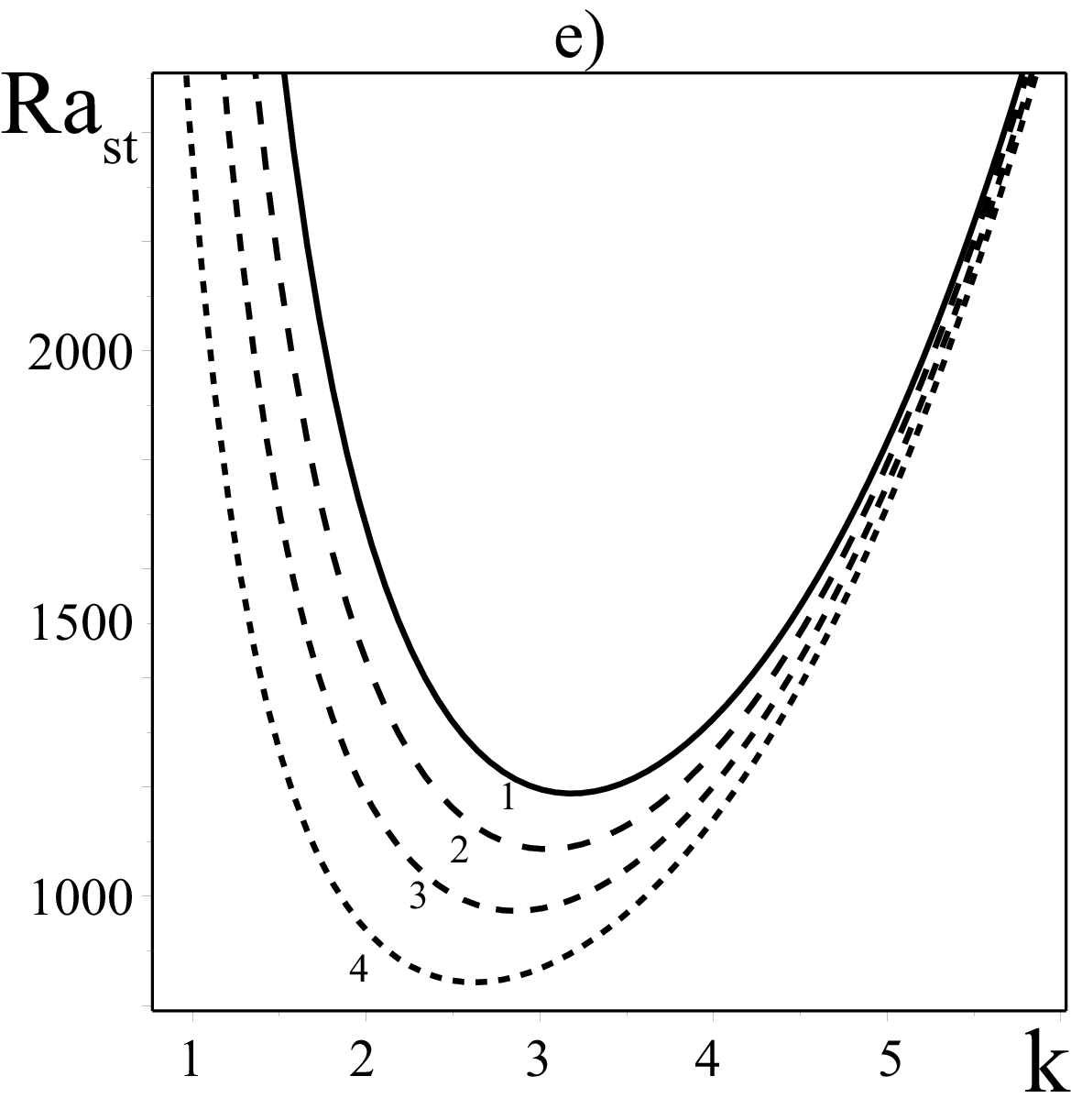}
	\includegraphics[width=5.5 cm, height=5.0 cm]{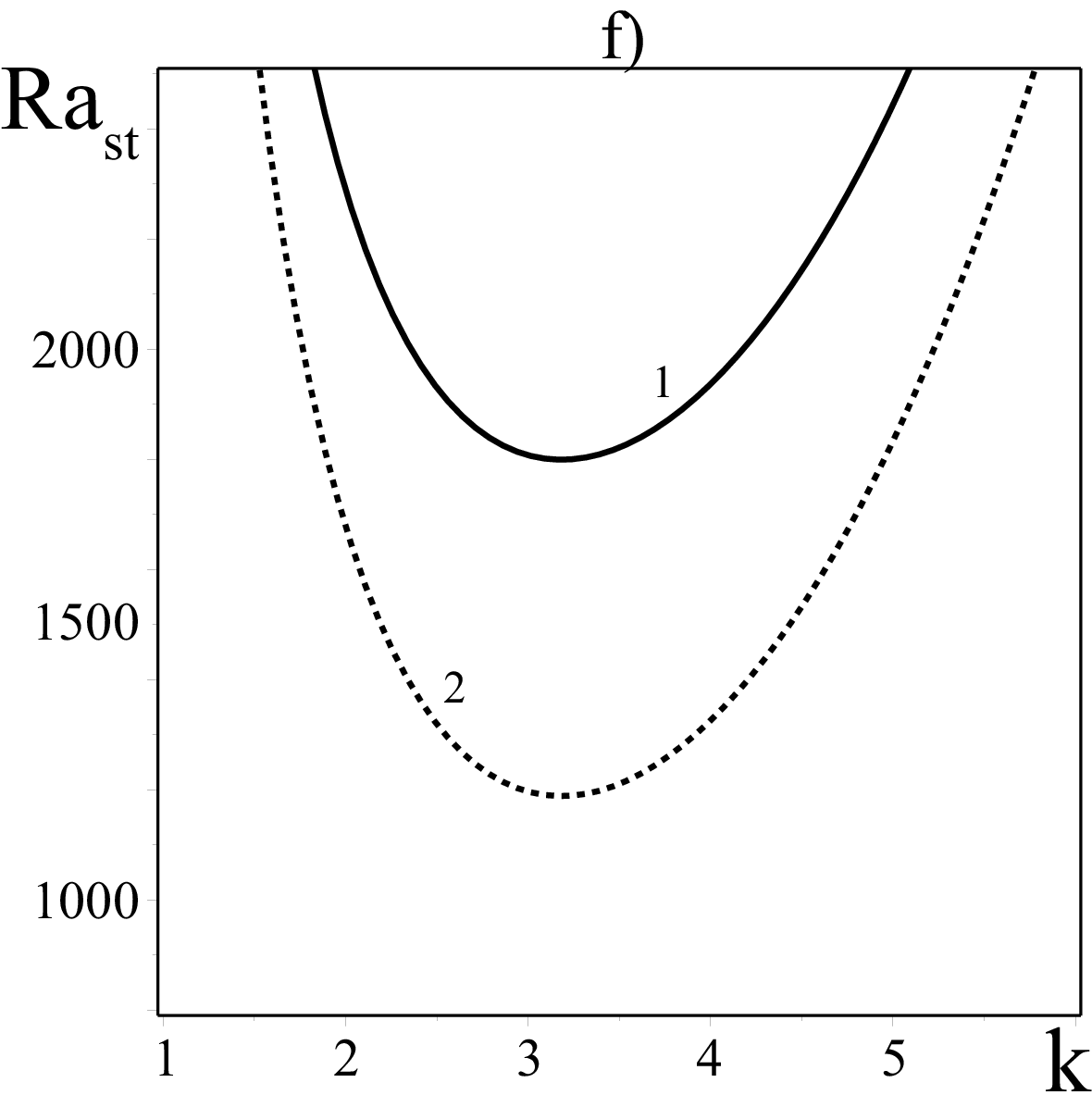}\\
\caption{\small Dependence of the stationary Rayleigh number $\textrm{Ra}_{st}$ on the wavenumber $k$: (a) $\textrm{Rb} = -1/2$, $\textrm{Ro} = -3/4$, $\xi = 1$; (b) $\textrm{Ro} = 0$, $\textrm{Rb} = -1$, $\xi = 1$; (c) $\textrm{Ro} = -3/4$, $\textrm{Rb} = -1$, $\xi = 1$. In all panels, curve~1 corresponds to a "pure" electrically conducting fluid, and curve~2 corresponds to a conducting nanofluid.}\label{fg15}
\end{figure}

Fig.~\ref{fg15}a demonstrates that the presence of nanoparticles (curve~2) promotes a reduction in the stationary convection threshold compared to the "pure" fluid (curve~1). Here, curves~(1,2) are plotted for a Rossby number $\textrm{Ro} = -3/4$, but the conclusions remain valid for arbitrary values of $\textrm{Ro}$.

We now analyze the influence of the azimuthal magnetic field on stationary convection by fixing the Rossby number $\textrm{Ro}$ and varying the magnetic Rossby number $(\textrm{Rb} = -1, -1/2, 0, 1/2)$ with $\xi = 1$. Fig.~\ref{fg14}b presents the results for the case of uniform rotation $\textrm{Ro} = 0$. It can be seen that for positive values $\textrm{Rb} \geqslant 0$ (curve~3: $\textrm{Rb} = 0$, curve~4: $\textrm{Rb} = 1/2$), the critical Rayleigh numbers are lower than those for negative values $\textrm{Rb} < 0$ (curve~1: $\textrm{Rb} = -1$, curve~2: $\textrm{Rb} = -1/2$). 
Thus, depending on the profile of the azimuthal magnetic field inhomogeneity $H_{0\phi}(R) = C R^{\alpha}$ ($C = \textrm{const}$, $\alpha$ being an arbitrary real number), the instability threshold can either increase $(\textrm{Rb} < 0)$ or decrease $(\textrm{Rb} > 0)$. 
In Fig.~\ref{fg15}b, curve~1 corresponds to a "pure" electrically conducting fluid, while curve~2 corresponds to a conducting nanofluid. Both curves (1,2) are plotted for $\textrm{Ro} = 0$ and $\textrm{Rb} = -1$. It is evident that the presence of nanoparticles lowers the threshold for stationary convection. This conclusion remains valid for any profile of the azimuthal magnetic field inhomogeneity (i.e., for any magnetic Rossby number $\textrm{Rb}$).
A similar behavior is observed for the Keplerian rotation profile $\textrm{Ro} = -3/4$ (see Fig.~\ref{fg14}c and Fig.~\ref{fg15}c).  
Comparing the results in Fig.~\ref{fg14}b, Fig.~\ref{fg15}b with those in Fig.~\ref{fg14}c, Fig.~\ref{fg15}c, it can be seen that for the rotation profile $\textrm{Ro} = -3/4$, the instability thresholds are lower than in the case of $\textrm{Ro} = 0$ for all considered profiles of the azimuthal magnetic field inhomogeneity.

In what follows, we conduct a comprehensive analysis of how rotation, characterized by the Rossby number, as well as azimuthal and helical magnetic fields, the Lewis number, the modified diffusion coefficient, and the nanoparticle concentration, influence the threshold of stationary convection.  
To facilitate this, we evaluate the relevant partial derivatives
\[ \frac{d \widetilde{\textrm{R}}}{d \widetilde{\textrm{T}}},\; \frac{d\widetilde{\textrm{R}}}{d\textrm{Ro}},\; \frac{d\widetilde{\textrm{R}}}{d \widetilde{\textrm{Q}}},\;\frac{d\widetilde{\textrm{R}}}{d \widetilde{\xi}},\;\frac{d\widetilde{\textrm{R}}}{d \textrm{Rb}},\; \frac{d\widetilde{\textrm{R}}}{dL_e}, \; \frac{d \widetilde{\textrm{R}}}{dN_A},\;\frac{d \widetilde{\textrm{R}}}{d\textrm{R}_n} .\]
Here we introduced the Chandrasekhar variables
\[\widetilde{\textrm{R}}=\frac{\textrm{Ra}_{st}}{\pi^4},\quad \widetilde{\textrm{T}}=\frac{\textrm{Ta}}{\pi^4}, \quad \widetilde{\textrm{Q}}=\frac{\textrm{Q}}{\pi^2}, \quad \textrm{x}=\frac{k^2}{\pi^2}, \quad \widetilde \xi= \frac{\xi}{\pi}. \]
In these variables, the expression for the critical Rayleigh number (Eq.~\ref{eq67}) takes the form:
\[\widetilde{\textrm{R}}=\left[\widetilde{\textrm{R}}^{(0)}-\textrm{R}_n(L_e+N_A)+\frac{2\pi^3\widetilde{\textrm{Q}}\widetilde{\xi}\sqrt{\widetilde{\textrm{T}}}(2+\textrm{Ro}(1-\textrm{Pm}))N_B(N_A-1)}{\textrm{x}((1+\textrm{x})^2+\widetilde{\textrm{Q}})L_e}\right]\times \]
\[\times\left[1-\frac{2\widetilde{\textrm{Q}}\widetilde{\xi}\sqrt{\widetilde{\textrm{T}}}(2+\textrm{Ro}(1-\textrm{Pm}))(N_A-N_B)}{\pi(1+\textrm{x})((1+\textrm{x})^2+\widetilde{\textrm{Q}})^2L_e}\right]^{-1}=\widetilde{D}_1(k)\cdot \widetilde{D}_2^{-1}(k),\]
where \[ \widetilde{\textrm{R}}^{(0)}=\frac{(1+\textrm{x})((1+\textrm{x})^2+\widetilde{\textrm{Q}})^2+\widetilde{\textrm{T}}((1+\textrm{x})^2(1+\textrm{Ro})+\textrm{RoPm}\widetilde{\textrm{Q}})-4\widetilde{\xi}^2\widetilde{\textrm{Q}}^2}{\textrm{x}((1+\textrm{x})^2+\widetilde{\textrm{Q}})}-\frac{4\widetilde{\xi}^2\widetilde{\textrm{Q}}}{\textrm{x}}\textrm{Rb} .  \]
Using these expressions, we can easily calculate the required derivatives
\begin{equation}\label{eq69}
 \frac{d \widetilde{\textrm{R}}}{d \widetilde{\textrm{T}}}=\left[\frac{d\widetilde{\textrm{R}}^{(0)}}{d\widetilde{\textrm{T}}}+\frac{\pi^3\widetilde{\textrm{Q}}\widetilde{\xi}(2+\textrm{Ro}(1-\textrm{Pm}))N_B(N_A-1)}{\textrm{x}((1+\textrm{x})^2+\widetilde{\textrm{Q}})L_e\sqrt{\widetilde{\textrm{T}}}}\right]\cdot \widetilde{D}_2^{-1}+\widetilde{D}_2^{-2}\cdot\widetilde{D}_1\times$$
$$\times\frac{\widetilde{\textrm{Q}}\widetilde{\xi}(2+\textrm{Ro}(1-\textrm{Pm}))(N_A-N_B)}{\pi(1+\textrm{x})((1+\textrm{x})^2+\widetilde{\textrm{Q}})^2L_e\sqrt{\widetilde{\textrm{T}}}},\; \frac{d \widetilde{\textrm{R}}^{(0)}}{d\widetilde{\textrm{T}}}=\frac{(1+\textrm{x})^2}{\textrm{x}((1+\textrm{x})^2+\widetilde{\textrm{Q}})}+\frac{\textrm{Ro}((1+\textrm{x})^2+\widetilde{\textrm{Q}} \textrm{Pm})}{\textrm{x}((1+\textrm{x})^2+\widetilde{\textrm{Q}})} ,
\end{equation}
\begin{equation}\label{eq70}
\frac{d\widetilde{\textrm{R}}}{d\textrm{Ro}}=\left[\frac{d\widetilde{\textrm{R}}^{(0)}}{d\textrm{Ro}}+\frac{2\pi^3\widetilde{\textrm{Q}}\widetilde{\xi}\sqrt{\widetilde{\textrm{T}}}(1-\textrm{Pm})N_B(N_A-1)}{\textrm{x}((1+\textrm{x})^2+\widetilde{\textrm{Q}})L_e}\right]\cdot \widetilde{D}_2^{-1}+\widetilde{D}_2^{-2}\cdot\widetilde{D}_1\times$$
$$\times\frac{2\widetilde{\textrm{Q}}\widetilde{\xi}\sqrt{\widetilde{\textrm{T}}}(1-\textrm{Pm})(N_A-N_B)}{\pi(1+\textrm{x})((1+\textrm{x})^2+\widetilde{\textrm{Q}})^2L_e},\;\frac{d \widetilde{\textrm{R}}^{(0)}}{d \textrm{Ro}}=\frac{\widetilde{\textrm{T}} ((1+\textrm{x})^2+\widetilde{\textrm{Q}} \textrm{Pm}) }{\textrm{x}((1+\textrm{x})^2+\widetilde{\textrm{Q}})},
\end{equation}
\begin{equation}\label{eq71}
\frac{d\widetilde{\textrm{R}}}{d\widetilde{\textrm{Q}}}=\left[\frac{d\widetilde{\textrm{R}}^{(0)}}{d\widetilde{\textrm{Q}}}+\frac{2\pi^3\widetilde{\xi}\sqrt{\widetilde{\textrm{T}}}(2+\textrm{Ro}(1-\textrm{Pm}))N_B(N_A-1)(1+\textrm{x})^2}{\textrm{x}((1+\textrm{x})^2+\widetilde{\textrm{Q}})L_e}\right]\cdot \widetilde{D}_2^{-1}+\widetilde{D}_2^{-2}\cdot\widetilde{D}_1\times$$
$$\times\frac{2\widetilde{\xi}\sqrt{\widetilde{\textrm{T}}}(2+\textrm{Ro}(1-\textrm{Pm}))(N_A-N_B)((1+\textrm{x})^2-\widetilde{\textrm{Q}})}{\pi(1+\textrm{x})((1+\textrm{x})^2+\widetilde{\textrm{Q}})^2L_e},$$
$$\frac{d \widetilde{\textrm{R}}^{(0)}}{d \widetilde{\textrm{Q}}}=\frac{1+\textrm{x}}{\textrm{x}}-\frac{(1+\textrm{x})^2\widetilde T(1+\textrm{Ro}(1-\textrm{Pm})) }{\textrm{x}((1+\textrm{x})^2+\widetilde{\textrm{Q}})^2}-\frac{4\widetilde \xi^2}{\textrm{x}}\cdot \left(\textrm{Rb}+\frac{\widetilde{ \textrm{Q}}(2(1+\textrm{x})^2+\widetilde{\textrm{Q}})}{(1+\textrm{x})^2+\widetilde{\textrm{Q}})^2}\right),
\end{equation}
\begin{equation}\label{eq72}
\frac{d\widetilde{\textrm{R}}}{d\widetilde{\xi}}=\left[\frac{d\widetilde{\textrm{R}}^{(0)}}{d\widetilde{\xi}}+
\frac{2\pi^3\widetilde{\textrm{Q}}\sqrt{\widetilde{\textrm{T}}}(2+\textrm{Ro}(1-\textrm{Pm}))N_B(N_A-1)}{\textrm{x}((1+\textrm{x})^2+\widetilde{\textrm{Q}})L_e}\right]\cdot \widetilde{D}_2^{-1}+\widetilde{D}_2^{-2}\cdot\widetilde{D}_1\times $$
$$\times\frac{2\widetilde{\textrm{Q}}\sqrt{\widetilde{\textrm{T}}}(2+\textrm{Ro}(1-\textrm{Pm}))(N_A-N_B)}{\pi(1+\textrm{x})((1+\textrm{x})^2+\widetilde{\textrm{Q}})^2L_e},\;\frac{d\widetilde{\textrm{R}}^{(0)}}{d\widetilde\xi}=-\frac{8\widetilde\xi \widetilde{\textrm{Q}}}{\textrm{x}}\cdot \frac{\widetilde{\textrm{Q}}(1+\textrm{Rb})+\textrm{Rb}(1+\textrm{x})^2 }{(1+\textrm{x})^2+\widetilde{\textrm{Q}}},
\end{equation}
\begin{equation}\label{eq73}
\frac{d\widetilde{\textrm{R}}}{d \textrm{Rb}}=\frac{d\widetilde{\textrm{R}}^{(0)}}{d \textrm{Rb}}=-\frac{4\widetilde \xi^2 \widetilde Q}{\textrm{x}} 
\end{equation}
\begin{equation}\label{eq74}
\frac{d\widetilde{\textrm{R}}}{dL_e}=\left[-\textrm{R}_n-\frac{2\pi^3\widetilde{\textrm{Q}}\widetilde{\xi}\sqrt{\widetilde{\textrm{T}}}(2+\textrm{Ro}(1-\textrm{Pm}))N_B(N_A-1)}{\textrm{x}((1+\textrm{x})^2+\widetilde{\textrm{Q}})L_e^2}\right]\cdot \widetilde{D}_2^{-1}-\widetilde{D}_2^{-2}\cdot\widetilde{D}_1\times$$
$$\times\frac{2\widetilde{\textrm{Q}}\widetilde{\xi}\sqrt{\widetilde{\textrm{T}}}(2+\textrm{Ro}(1-\textrm{Pm}))(N_A-N_B)}{\pi(1+\textrm{x})((1+\textrm{x})^2+\widetilde{\textrm{Q}})^2L_e^2},\end{equation}
\begin{equation}\label{eq75}
\frac{d \widetilde{\textrm{R}}}{dN_A}=\left[-\textrm{R}_n+\frac{2\pi^3\widetilde{\textrm{Q}}\widetilde{\xi}\sqrt{\widetilde{\textrm{T}}}(2+\textrm{Ro}(1-\textrm{Pm}))N_B}{\textrm{x}((1+\textrm{x})^2+\widetilde{\textrm{Q}})L_e}\right]\cdot \widetilde{D}_2^{-1}+\widetilde{D}_2^{-2}\cdot\widetilde{D}_1\times$$
$$\times\frac{2\widetilde{\textrm{Q}}\widetilde{\xi}\sqrt{\widetilde{\textrm{T}}}(2+\textrm{Ro}(1-\textrm{Pm}))}{\pi(1+\textrm{x})((1+\textrm{x})^2+\widetilde{\textrm{Q}})^2L_e},\end{equation}
\begin{equation}\label{eq76}
\frac{d \widetilde{\textrm{R}}}{d\textrm{R}_n}=-(N_A+L_e)\cdot\widetilde{D}_2^{-1}.
\end{equation}
It is readily seen that expressions (\ref{eq69})--(\ref{eq76}) contain terms of the form $\sim \widetilde{\textrm{Q}}\widetilde{\xi}N_B(N_A-1)/L_e$, which arise due to the combined effect of nanoparticle generation $(N_B)$ and the helical magnetic field $(\widetilde{\textrm{Q}}\widetilde{\xi})$.  
For given nanofluid parameters $(N_A, N_B, L_e)$, the contribution of these terms is relatively small.  
However, terms of the form $\sim \widetilde{D}_1 \widetilde{\textrm{Q}}\widetilde{\xi}N_B(N_A-1)/L_e$ contribute to the destabilization of convection when the concentration Rayleigh number is positive $(\textrm{R}_n > 0)$.  
Therefore, in all expressions (\ref{eq69})-(\ref{eq75}), the destabilizing factor is the nanoparticle concentration, represented by the parameter $\textrm{R}_n$.

From Eqs.~(\ref{eq69})-(\ref{eq70}), it follows that differential rotation has a stabilizing effect $(d \widetilde{\textrm{R}}^{(0)}/d \widetilde{\textrm{T}} > 0,\ d \widetilde{\textrm{R}}^{(0)}/d \textrm{Ro} > 0)$ for positive Rossby numbers $(\textrm{Ro} > 0)$.  
In contrast, for negative Rossby numbers $(\textrm{Ro} < 0)$, differential rotation may exert a destabilizing influence $(d \widetilde{\textrm{R}}^{(0)}/d \widetilde{\textrm{T}} < 0,\ d \widetilde{\textrm{R}}^{(0)}/d \textrm{Ro} < 0)$.

From Eq.~(\ref{eq71}), it is evident that the quantity $d \widetilde{\textrm{R}}^{(0)}/d \widetilde{\textrm{Q}}$ can be either positive or negative, indicating that the axial magnetic field (represented by the Chandrasekhar number $\widetilde{\textrm{Q}}$) can have a stabilizing or destabilizing effect on stationary convection.  
The destabilizing effect arises under specific conditions for the profiles of differential rotation (Rossby number $\textrm{Ro}$) and azimuthal magnetic field inhomogeneity (magnetic Rossby number $\textrm{Rb}$):
\[ \textrm{Ro}(\textrm{Pm}-1)<1, \; \textrm{Rb}>\frac{\widetilde{\textrm{Q}}(2(1+\textrm{x})^2+\widetilde{\textrm{Q}})}{(1+\textrm{x})^2+\widetilde{\textrm{Q}})^2} (\textrm{at}~ \textrm{Rb}>0),\; \textrm{Rb}<\frac{\widetilde{\textrm{Q}}(2(1+\textrm{x})^2+\widetilde{\textrm{Q}})}{(1+\textrm{x})^2+\widetilde{\textrm{Q}})^2} (\textrm{at}~ \textrm{Rb}<0).   \]
The first inequality holds for positive Rossby numbers $(\textrm{Ro} > 0)$ and magnetic Prandtl numbers $(\textrm{Pm} < 1)$, or for $\textrm{Ro} < 0$ and $\textrm{Pm} > 1$.

We analyze the influence of the inhomogeneous azimuthal magnetic field on stationary convection using Eqs.~(\ref{eq72}) and (\ref{eq73}).  
Due to the inhomogeneity of the azimuthal magnetic field, the Lorentz force introduces a magnetic field gradient, which induces a drift in the fluid flow.  The direction of this drift, and consequently its effect on the stability of convective flows, depends on the sign of the magnetic field gradient (i.e., the magnetic Rossby number $\textrm{Rb}$).  
For positive $\textrm{Rb} > 0$ (for a magnetic field profile $H_{0\varphi} = C R^{\alpha}$ with $\alpha > 1$), convection is destabilized.  
In contrast, for negative $\textrm{Rb} < 0$ $(\alpha < 1)$, convection is stabilized.  
This phenomenon is observed in the stationary instability (see Fig.~13).
An inhomogeneous azimuthal magnetic field with a magnetic Rossby number $\textrm{Rb} = -1$ has a stabilizing effect, as indicated by $d \widetilde{R} / d \widetilde{\xi} > 0$. For $\textrm{Rb} \geqslant 1$, the inhomogeneous azimuthal magnetic field exerts a destabilizing effect, as shown by $d \widetilde{R} / d \widetilde{\xi} < 0$. From Eq.~(\ref{eq72}), it follows that a homogeneous azimuthal magnetic field with $\textrm{Rb} = -1/2$ can either stabilize or destabilize stationary convection depending on the other system parameters.  
Eq.~(\ref{eq73}) illustrates that both stabilizing $(\textrm{Rb} < 0)$ and destabilizing $(\textrm{Rb} > 0)$ effects depend on the sign of the magnetic Rossby number, which in turn is determined by the profile of the inhomogeneous azimuthal magnetic field $H_{0\phi}(R)$.
From Eqs.~(\ref{eq74})-(\ref{eq75}), it is evident that the Lewis number $L_e$ and the modified diffusion coefficient $N_A$ have a destabilizing effect when the nanoparticle Rayleigh number $\textrm{R}_n > 0$. For most types of nanofluids, the sum of the Lewis number and the modified diffusion coefficient is always positive, i.e., $L_e + N_A > 0$ \cite{9s}. Therefore, from Eq.~(\ref{eq76}), it follows that the concentration Rayleigh number always exerts a destabilizing effect.

In contrast to stationary convection in the axial magnetic field, the influence of the rotation effects, Rossby number, azimuthal magnetic field, and spiral magnetic field on the stationary non-homogeneously rotating convection of nanofluids in a spiral magnetic field occurs with a destabilizing contribution from the nanoparticle concentration.

\section {Conclusion}
In this study, a linearized system of magnetohydrodynamic equations is formulated to investigate hydrodynamic instabilities in a non-uniformly rotating nanofluid exposed to a spiral magnetic field, with imposed constant temperature gradients and a uniform nanoparticle concentration in the presence of gravity. 
In the framework of the local WKB approximation, various types of hydromagnetic instabilities are analyzed in the absence of a temperature gradient but accounting for the gradient of nanoparticle concentration. When a non-uniformly rotating nanofluid is subjected solely to an axial magnetic field $H_{0z} \neq 0$, the standard magnetorotational instability (SMRI) arises. If the nanofluid is exposed only to a non-uniform azimuthal magnetic field $H_{0\varphi}(R) \neq 0$, an azimuthal MRI (AMRI) develops. In the presence of a helical magnetic field ${\bf H}_0 = H_{0\varphi}(R)\,{\bf e}_{\varphi} + H_{0z}\,{\bf e}_z$, a helical MRI (HMRI) is excited. For each type of MRI, dispersion relations were derived in the $(k, \textrm{Ro})$ plane for different values of the rotation parameter (Taylor number) $\textrm{Ta}$ and the azimuthal Chandrasekhar number $\textrm{Q}_{\varphi}$. Critical Rossby numbers $(\textrm{Ro}_{cr}, \textrm{Rb}_{cr})$, characterizing threshold values for non-uniform rotation profiles ($\textrm{Ro}$) and non-uniform azimuthal magnetic fields (magnetic Rossby numbers $\textrm{Rb}$), were also obtained.
It was found that the instability regions in the $(k, \textrm{Ro})$ plane for all types of magnetorotational instability (MRI) in nanofluids are significantly larger than those in a pure fluid.

The influence of temperature gradients and nanoparticle concentration on stationary convection in axial and helical magnetic fields is analyzed, taking into account the effects of non-uniform rotation (Rossby number $\textrm{Ro}$) and spatially varying azimuthal magnetic fields (magnetic Rossby number $\textrm{Rb}$). For both axial and helical magnetoconvection, explicit expressions for the critical Rayleigh number $\textrm{Ra}_{st}$ are derived, and corresponding neutral stability curves are constructed. The analysis reveals that the presence of nanoparticles leads to a noticeable reduction in the threshold value of the stationary critical Rayleigh number $\textrm{Ra}_{st}^{\min}$, indicating a destabilizing effect in both configurations.

\end{document}